\documentclass[journal,final]{IEEEtran}
\title{Model-Based and Graph-Based \\ Priors for Group Testing}
\author{Ivan Lau, Jonathan Scarlett, and Yang Sun\thanks{I.~Lau is with the Department of Electrical and Computer Engineering, Rice University. J.~Scarlett and Y.~Sun are with the Department of Computer Science, National University of Singapore (NUS).  J.~Scarlett is also with the Department of Mathematics, NUS, and the Institute of Data Science, NUS. e-mails: \url{ivan.lau@rice.edu}; \url{scarlett@comp.nus.edu.sg}; \url{yang.sun@u.nus.edu}.}}
\date{National University of Singapore}

\usepackage[utf8]{inputenc}
\usepackage{amsmath}                            
\usepackage{amssymb} 
\usepackage{amsfonts}
\usepackage{amsthm}    
\usepackage{mathtools}
\usepackage{graphicx}
\usepackage{import}
\usepackage{lipsum}

\usepackage[thinc]{esdiff}
\usepackage[dvipsnames]{xcolor}
\usepackage{color}
\definecolor{darkblue}{rgb}{0,0,0.6}
\usepackage[breaklinks, pdftex, ocgcolorlinks,colorlinks=true, citecolor=Maroon, filecolor=darkblue, linkcolor=darkblue, urlcolor=violet, linktocpage=true]{hyperref}
\usepackage{chngcntr}
\usepackage{apptools}
\usepackage{subfig}
\AtAppendix{\counterwithin{theorem}{subsection}}
\usepackage{cancel}
\usepackage[normalem]{ulem}

\DeclarePairedDelimiterX{\infdivx}[2]{(}{)}{%
  #1\;\delimsize\|\;#2%
}

\theoremstyle{definition}
\newtheorem{theorem}{Theorem}[section]

\newtheorem{lemma}[theorem]{Lemma}
\newtheorem{corollary}[theorem]{Corollary}

\newcommand{\bone}{\boldsymbol{1}}
\newcommand{\Cless}{\mathcal{C}_{\rm noiseless}}
\newcommand{\Cnoisy}{\mathcal{C}_{\rm noisy}}
\newcommand{\Clessr}{\mathcal{C}_{\rm noiseless}^{[0,1]}}
\newcommand{\Cnoisyr}{\mathcal{C}_{\rm noisy}^{[0,1]}}

\newcommand{\Ntil}{\widetilde{N}}
\newcommand{\Shat}{\widehat{S}}

\newcommand{\sdif}{s_{\mathrm{dif}}}

\newcommand{\seq}{s_{\mathrm{eq}}}

\newcommand{\sbar}{\overline{s}}

\newcommand{\Z}{\mathbb{Z}}
\newcommand{\R}{\mathbb{R}}

\newcommand{\Xv}{\mathsf{X}}
\newcommand{\Yv}{\mathbf{Y}}
\newcommand{\uv}{\mathbf{u}}

\newcommand{\deltatau}{\delta'_{\tau}}
\newcommand{\Itau}{I_{\tau}}

\newcommand{\Ntaudefault}{N_{\tau}^{\text{default}}}
\newcommand{\Ntau}{N_{\tau}}

\newcommand{\Xsdif}{\mathbf{X}_{\sdif}}
\newcommand{\XXsdif}{\mathsf{X}_{\sdif}}

\newcommand{\Xseq}{\mathbf{X}_{\seq}}
\newcommand{\XXseq}{\mathsf{X}_{\seq}}

\newcommand{\dmax}{d_{\mathrm{max}}}
\newcommand{\pe}{\mathbb{P}_{d_{\max}}\left[\text{err}\right]}
\newcommand{\PP}{\mathbb{P}}

\newcommand{\E}{\mathbb{E}}

\definecolor{ao}{rgb}{0.0, 0.5, 0.0}

\usepackage{hyperref}
\usepackage{cancel}

\let\originalleft\left
\let\originalright\right
\renewcommand{\left}{\mathopen{}\mathclose\bgroup\originalleft}
\renewcommand{\right}{\aftergroup\egroup\originalright}

\newcommand{\Sc}{\mathcal{S}}

\theoremstyle{definition}

\newcommand{\argmin}{\operatornamewithlimits{argmin}}
\newcommand{\argmax}{\operatornamewithlimits{argmax}}

\begin{document}

\maketitle

% {\bf \color{blue} TODO:
% \begin{itemize}
%     \item (Jon) Possibly add some discussion on how the theory can extend to settings where $k$ is random
%     \item (Yang) Discuss green parts 
%     \item \magenta{(Jon and Ivan) change ``algorithm'' to ``decoder'' for all converse corollaries?}
%     \item \magenta{(Jon) Check Ivan's update (the magenta part) for Corollary 2.10 and Appendices C.2, D, E.2}
% \end{itemize}
% }

\begin{abstract}
    The goal of the group testing problem is to identify a set of defective items within a larger set of items, using suitably-designed tests whose outcomes indicate whether any defective item is present. In this paper, we study how the number of tests can be significantly decreased by leveraging the structural dependencies between the items, i.e., by incorporating prior information.  To do so, we pursue two different perspectives: (i) As a generalization of the uniform combinatorial prior, we consider the case that the defective set is uniform over a \emph{subset} of all possible sets of a given size, and study how this impacts the information-theoretic limits on the number of tests for approximate recovery; (ii) As a generalization of the i.i.d.~prior, we introduce a new class of priors based on the Ising model, where the associated graph represents interactions between items.  We show that this naturally leads to an Integer Quadratic Program decoder, which can be converted to an Integer Linear Program and/or relaxed to a non-integer variant for improved computational complexity, while maintaining strong empirical recovery performance. 
\end{abstract}

\section{Introduction} \label{sec:intro}
%!TEX root = main.tex

The concept of group testing was first introduced by Robert Dorfman \cite{dorfman1943detection} in 1943 to provide a more efficient method of screening for syphilis during conscription for World War II. It has since gained significant attention due to its wide-ranging applications in areas such as biology, communications, and information technology, as well as connections with problems in data science and theoretical computer science (see \cite[Section 1.7]{aldridge2019group} and the references therein).
Recently, it has also found utility in testing for COVID-19 \cite{Ald21}.
% \underline{\bf TODO: pick some references out of 200++} {\bf \color{blue} [JS: We don't need more references in this particular location]}

The general goal of group testing is to identify a set of ``defective'' items of size $k$ within a larger set of items of size $n$, based on the outcomes of $t$ suitably-designed tests, where each test outcome indicates whether the test includes at least one defective item. The current state-of-the-art group testing schemes can reliably identify the defective set using an asymptotically optimal number of tests \cite{coja2020optimal}. In this paper, we build on a recent line of works demonstrating that we can further decrease the number of tests by leveraging structural dependencies between the items \cite{arasli2021group,gonen2022group,nikpey2022group,goenka2021contact,nikolopoulos2021group,ahn2021adaptive,nikolopoulos2021overlap}.  We study such dependencies from two different perspectives, adopting a natural generalization of the uniform combinatorial prior for the purpose of theoretical analysis, and introducing a new prior based on Ising models as a more practical alternative that generalizes the i.i.d.~prior.

Before outlining the related work and our contributions in more detail, we formally introduce the problem.

\subsection{Problem Setup}
%!TEX root = main.tex

\subsubsection{Observation Model}
\label{sect:observation_model}
We have a set of $n$ items, $V = [n] \coloneqq \{1, \dotsc, n\}$, and an unknown defective set $S \subseteq V$ of size~$k$. The goal is to recover $S$ via a set of tests, where each test contains a subset of $V$, and the test outcome reveals whether any of the items in the test are defective.  We use a binary test matrix $\mathsf{X} \in \{0, 1\}^{t \times n}$ to represent the design of the pooled tests, with
% we write
% $X^{(1)}, X^{(2)}, \dotsc, X^{(T)} \in \{0, 1\}^n$ 
% with
\begin{equation}
    X_{ij} =
    % X^{(i)}_j = 
    \begin{cases}
    1 & \text{the $j$-th item is included in the $i$-th test;}\\
    0 & \text{otherwise,}
    \end{cases}
\end{equation}    
for $i \in \{1, \dotsc , t\}$ and for $j \in V$. 
We focus on the \emph{non-adaptive} setting, in which~$\mathsf{X}$ must be designed prior to observing any outcomes. The test outcomes, denoted by $\mathbf{Y} =(Y_1,\dotsc,Y_t) \in \{0,1\}^t$, may follow one of several observation models.
% dependent on the test matrix $\mathsf{X}$.
% $X^{(i)} \in \{0, 1\}^n$ 
% indicating which of the $n$ items are in the $i$-th test, 
In the \emph{noiseless} group testing model, which is our main focus, the test outcomes are given by
\begin{equation}
    Y_i = \bigvee_{j \in S} X_{ij}
    =
    \begin{cases}
    1 & \text{any defectives in the $i$-th test;}\\
    0 & \text{otherwise.}
    \end{cases}
    \label{constriant_noiseless}
\end{equation}     
In the \emph{noisy} group testing model, the test outcomes might be ``flipped'' by some (random) noise vector $\boldsymbol{\xi} \in \{0,1\}^t$, in which $\xi_{i}=1$ if and only if the 
outcome of the $i$-th test is flipped.
% and $\xi_{i}=0$
% to represent the flips such that the $i$-th test gives the flipped outcome if the $i$-th entry of $\boldsymbol{\xi}$,  otherwise. 
In this model, the corresponding test outcomes are defined as
\begin{equation}
    Y_i
    =
    \begin{cases}
    1 & \text{if $\big( \bigvee_{j \in S} X_{ij} \big) \bigoplus \xi_{i} = 1$;}\\
    0 & \text{otherwise,}
    \end{cases}
    \label{constriant_noise}
\end{equation} 
where $\bigoplus$ represents the logical XOR operator.  We focus primarily on the case that $\boldsymbol{\xi}$ has i.i.d.~${\rm Bernoulli}(\rho)$ entries for some $\rho \in \big(0,\frac{1}{2}\big)$ (i.e., the symmetric noise model).

In addition to the above specific models, we will sometimes work with a general probabilistic model, writing
\begin{equation}
    (Y_{i}|\mathbf{X}_{i}) \sim P_{Y_i|\mathbf{X}_{i;S}},
\end{equation}
where $\mathbf{X}_{i;S}$ is the sub-vector of the $i$-th row of $\mathsf{X}$ containing the entries indexed by $S$.  In all models, we assume conditional independence (given $\mathsf{X}$) among the tests $i = 1, \dotsc, t$.

\subsubsection{Probabilistic Priors}

We consider settings in which the unknown defective set $S$ has some (known) prior distribution, i.e., a distribution over all subsets of $V$.
% Our defective set $S$ is random according to some suitably-chosen prior distribution.  
In the literature, the most common priors are as follows (e.g., see \cite[Page 205]{aldridge2019group}):
\begin{itemize}
    \item \textbf{Combinatorial prior}: The number of defective items $k$ is fixed, and each of the $\binom{n}{k}$ subsets of $V$ of cardinality $k$ are equally likely, i.e., $S$ is uniform over such subsets.
    \item \textbf{i.i.d.~prior}: Each item is defective with some probability $q$, with independence between items.  Hence, the average number of defectives is $\bar{k} = nq$.
\end{itemize}
Note that $k$ and $q$ may scale with $n$; most works consider a sub-linear number of defectives (e.g., $k = \sqrt{n}$ or $q = \frac{1}{\sqrt n}$).

\subsubsection{Test Designs}

As mentioned in Section~\ref{sect:observation_model}, we consider non-adaptive test designs, where the entire test matrix $\mathsf{X}$ must be designed before observing any test outcomes.  This often has significant practical benefits due to allowing parallel testing.  For the more practically-oriented parts of this paper, our focus will be on introducing a suitable prior and designing a decoder for it, but that decoder will be applicable to arbitrary test designs.  That is, the decoder design and the non-adaptive test design will be decoupled.

In contrast, for our theoretical analysis, we will focus on $i.i.d.$~Bernoulli testing, in which each item is placed in a given test independently with probability $\frac{\nu}{k}$ for some constant $\nu>0$.  While there is evidence that slightly more structured random designs can do better \cite{johnson2018performance,coja2020optimal,tan2022performance}, such improvements have been absent for \emph{approximate recovery under sublinear sparsity}, which is our focus. Hence, we believe that the i.i.d.~design is as suitable as any for our purposes.

\subsubsection{Recovery Criteria}
%!TEX root = main.tex

% One of our goals is to provide a theoretical analysis on the required number of tests for the group testing with prior information.  Specifically, we will consider the uniform prior over a \emph{subset} $\mathcal{S}$ of size-$k$ defective sets, whose size may be much smaller than $\binom{n}{k}$.

%  To avoid the nuisance of $k$ being random, we capture this with ``model-based group testing'', where the space of defective sets, denoted by $\mathcal{S}$, is an arbitrary subset of the entire space\footnote{We are mostly interested in the case where $|\mathcal{S}| \ll \binom{n}{k}$.} of $k$-sized subsets of $[n]$.
% Furthermore, $S$ is uniform on $\mathcal{S}$.

% We focus on the \emph{non-adaptive} and  \emph{noiseless} setting.
% Given the tests design $\mathsf{X}$ and the tests outcomes $\mathbf{Y}$,
% a \emph{decoder} forms an estimate $\Shat = \Shat(n, k, \mathcal{S}, \mathsf{X}, \mathbf{Y})$ of $S$.

Given the test design $\mathsf{X}$ and the test outcomes $\mathbf{Y}$, a \emph{decoder} forms an estimate $\Shat = \Shat(n, k, \mathsf{X}, \mathbf{Y})$ of~$S$.  A variety of recovery criteria exist for measuring the performance.  We focus on \emph{approximate recovery} criteria, allowing a certain number of false positives (i.e., $|\Shat \setminus S|$) and false negatives (i..e, $|S \setminus \Shat|$).  In our theoretical analysis, we will treat these two equally, allowing up to $\dmax = \lfloor \alpha^{\ast} k \rfloor$ of each for some $\alpha^{\ast} \in (0, 1)$.  That is, the error probability is
\begin{equation}
    \pe := \PP\left[d(S,\Shat) > \dmax\right], \label{eq:pe_dmax}
\end{equation}
where
\begin{equation}
    d(S,\Shat) = \max\left\{ \big|S \setminus \Shat\big|, \big|\Shat \setminus S\big| \right\}. \label{eq:dist}
\end{equation}
We are interested in the required number of tests to attain asymptotically 
vanishing error probability, i.e., $\lim_{n \to \infty} \pe = 0$.  Permitting a small error probability is commonly given the terminology {\em probabilistic group testing} or the {\em small-error criterion}.

The approximate recovery criterion is less stringent than the usual exact recovery criterion (which corresponds to setting $d_{\max}=0$ in \eqref{eq:pe_dmax}), and both are of practical interest depending on the application.  We focus only on the approximate recovery criterion, but studying exact recovery may be of interest in future work.  We also note that our goal is to understand \emph{improved decoding algorithms} for exploiting prior information, rather than improved test designs (though both are of interest).

We briefly note that a distinct line of works for the noiseless setting (or non-stochastic noise settings) requires \emph{zero error probability}.  This strict requirement incurs $k^2$ dependence in the number of tests for exact recovery, though approximate recovery is still possible with $O(k \log n)$ tests \cite{Che09}.

\subsection{Related Work}

In addition to the related works already mentioned above, we highlight the following aspects of group testing that are particularly relevant to our paper.

\subsubsection{Information-theoretic limits of group testing}

Among many existing theoretical works on group testing, the information-theoretic limits in
\cite{scarlett2016phase} are particularly relevant (see also \cite{Scarlett2017LimitsSupportRecovery,scarlett2017little}).  In particular, we highlight the following.

\begin{lemma}[{\cite{scarlett2016phase}}]
\label{lem:standard_bounds}
    Let $n$ be the number of items, $k = \Theta\left(n^{\theta}\right)$ for some $\theta \in (0, 1)$ be the size of defective set, and suppose that $S$ is uniform over the set of all $k$-sized subsets of $V$.
    Fix $\alpha^{\ast} \in (0, 1)$, and let $\dmax = \lfloor \alpha^{\ast} k \rfloor$.  For the noiseless group testing model, under i.i.d.~Bernoulli testing with parameter $\frac{\ln 2}{k}$, there exists a decoder outputting $\Shat$ such that as $\pe \to 0$ as $n \to \infty$, provided that\footnote{Note that the dependence on $\alpha^*$ here only enters via the $o(1)$ term.}
    \begin{equation}
        t \ge \left( k\log_2 \frac{n}{k} \right) (1+o(1)). % = \left(1-\theta \right) \left( k\log_2 n \right) (1+o(1))
        \label{eq:test_achiev}
    \end{equation}
    Conversely, for any non-adaptive test design\footnote{In \cite{scarlett2016phase} this was stated specifically for the i.i.d.~design, but the general case is given in \cite[Thm.~1]{scarlett2017little}.} and any decoder $\Shat$, in order to achieve $\pe \to 0$, it
    is necessary that
    % as $n \to \infty$, we have $\pe \to 1$ whenever
    \begin{equation}
            t  \ge (1- \alpha^{\ast})\left( k\log_2 \frac{n}{k} \right) (1-o(1)). \label{eq:test_conv}
    \end{equation}
\end{lemma}

Observe that the upper and lower bounds match as $\alpha^* \to 0$.  More generally, recent results in \cite{truong2020all,niles2021all} demonstrated that \eqref{eq:test_conv} that is tight for general designs (i.e., there exists a matching achievability result), whereas Bernoulli designs cannot do better than \eqref{eq:test_achiev}, so they are mildly suboptimal in terms of the dependence on $\alpha^*$.

We refer the reader to \cite{malyutov1978separating,scarlett2016phase,coja2020optimal,aldridge2019group} and the references therein for other works on the information-theoretic limits of group testing.  
% These works are complementary to those in \cite{scarlett2016phase,Scarlett2017LimitsSupportRecovery}, as they consider group testing of different settings (e.g., adaptive instead of non-adaptive).
These works typically adopt an information-theoretic decoder, roughly amounting to a brute force search over the space of defective sets, which is computationally intractable.
A distinct line of works has sought schemes that require not only order-optimal number of tests $t$, but also computationally feasible decoder, running in time polynomial in~$n$ (e.g., see \cite[Chapters 2 and 3]{aldridge2019group} and the references therein), or even polynomial in $k \log n$ (e.g., see \cite[Table 1]{Pri20} and the references therein).  Our theoretical focus, however, is on information-theoretic limits.

\subsubsection{Decoding techniques} \label{relatedwork_lp}

In the theoretical part of our paper, we will consider an information-theoretic decoder (similar to \cite{scarlett2016phase}) whose definition is quite technical, so its details are deferred to later (see Appendix~\hyperref[sect:inf_theoretic_decoder]{A-II}).\footnote{Throughout the paper, by ``Appendix'' we mean ``Supplementary Material'', attached as a separate document.}  In the practically-oriented part of our paper, we will use techniques based on Linear Programming and Quadratic Programming; here we briefly outline some existing works that adopted similar approaches.

We first discuss the noiseless setting.  Given the test matrix $\mathsf{X}$ and the test outcome vector $\mathbf{Y}$, a natural decoding rule is to choose $\Shat$ to be the smallest set of items consistent with the test results; this was termed the \emph{smallest satisfying set} (SSS) algorithm in \cite{aldridge2014group}.  Since this poses a potentially hard combinatorial optimization problem, it was proposed in \cite{6288622} to relax it to a linear program.  The resulting optimization problems can be written in a unified manner as follows:
\begin{multline}
        {\rm minimize}_{\uv} ~ \sum_{j=1}^{n} u_{j} ~~~
        \text{subject to } \uv \in \Cless \\ \text{ (or $\uv \in \Clessr$ if relaxed)}
    \label{lp_noiseless_intro}
\end{multline}
where
\begin{equation}
    \begin{aligned}
        \Cless = \bigg\{ \uv \in \{0,1\}^n \,:\, & \sum_{j=1}^{n} X_{ij}u_{j} = 0 \text{ when } Y_{i}=0 \\ 
        & \sum_{j=1}^{n} X_{ij}u_{j} \geq 1 \text{ when } Y_{i}=1 \bigg\},
    \end{aligned} \label{eq:C_noiseless}
\end{equation}
and $\Clessr$ is defined analogously with $[0,1]^n$ in place of $\{0,1\}^n$.  Denoting the resulting estimate by $\hat{\uv}$ (with suitable rounding to $\{0,1\}^n$ in the relaxed case), the defective set is estimated according to $\Shat = \{ j \in [n] \,:\, \hat{u}_j = 1\}$.

In the noisy setting, it was proposed in \cite{6288622} to weigh the sparsity-based objective in \eqref{lp_noiseless_intro} with an additional term indicating how many test results are flipped compared to their nominal outcome, leading to the following: 
\begin{multline}
        {\rm minimize}_{\uv,\boldsymbol{\xi}}~ \sum_{j=1}^{n} u_{j} + \eta\sum_{i=1}^{t} \xi_{i} ~~~
        \text{subject to } (\uv,\boldsymbol{\xi}) \in \Cnoisy \\  \text{ (or $(\uv,\boldsymbol{\xi}) \in \Cnoisyr$ if relaxed)}
    \label{lp_noisy_intro}
\end{multline}
where
\begin{equation}
    \begin{aligned}
        \Cnoisy = &\bigg\{ (\uv,\boldsymbol{\xi}) \in \{0,1\}^n \times \{0,1\}^t \,:\, \nonumber \\ &\quad \sum_{j=1}^{n} X_{ij}u_{j} = \xi_{i} \text{ when } Y_{i}=0 \\ 
        &\quad \sum_{j=1}^{n} X_{ij}u_{j} + \xi_{i} \geq 1 \text{ when } Y_{i}=1 \bigg\},
    \end{aligned} \label{eq:C_noisy}
\end{equation}
and $\Cnoisyr$ is defined analogously with $[0,1]^n \times [0,1]^t$ in place of $\{0,1\}^n \times \{0,1\}^t$.  Here the parameter $\eta$ balances the trade-off between having a sparse solution and having few flipped tests.  In both the noiseless as noisy settings, the non-relaxed variant is an Integer Linear Program (ILP), whereas the relaxed version is a Linear Program (LP).  There is numerical evidence that the LP solution is typically very close to the ILP solution in practice, while being cheaper to compute \cite{aldridge2019group,ciampiconi2020maxsat, Zabeti2021}.

% \olive{We notice for the noiseless constraints in (\ref{lp_noisy_intro}), which we denote as $\Cc_{noisy}$ for the sake of brevity, and the noisy constraints in (\ref{lp_noiseless_intro}), denoted as $\Cc_{noiseless}$, $u_{j}$ and $\xi_{i}$ are either $0$ or $1$ are discrete constraints. Problems of this form are called Integer Linear Programs (ILP), which generally make them NP-Hard and not easy to solve. However, we can relax the problem by simply requiring $u_{j} (\xi_{i}) \in [0,1]$ instead of $\{0,1\}$, and the transform the problem into standard LP (see Appendix \ref{lp_def} for a general definition). We call the relaxed constraints sets as $\Cc_{noisy\_relaxed}$ and $\Cc_{noiseless\_relaxed}$.}  

It is shown in \cite[Sec. 5.3]{ciampiconi2020maxsat} that the non-relaxed variants of \eqref{lp_noiseless_intro} and \eqref{lp_noisy_intro} can be interpreted as the \emph{maximum a posteriori} decoding rules under an i.i.d.~prior, and i.i.d.~symmetric noise in the noisy case.  Specifically, for this interpretation in the noisy case, one should set $\eta = \frac{\log\frac{1-\rho}{\rho}}{\log\frac{1-q}{q}}$, where $q$ is the defectivity probability and $\rho$ is the noise parameter. 

% In the later sections of the paper, we will denote the constraint sets in \eqref{lp_noiseless_intro}--\eqref{lp_noisy_intro} by $\Cless$ and $\Cnoisy$ respectively, and the relaxed versions by $\Clessr$ and $\Cnoisyr$ respectively.

\subsubsection{Group testing with prior information} \label{sec:related_prior}

Here we briefly outline some existing group testing literature where non-uniform models on the defective set have been considered.  

A common choice of non-uniform prior is the {\em i.-non-i.d.} prior, in which each item is randomly defective with independence across items, but the associated probabilities may differ \cite{li2014group,kealy2014capacity}.  This is an important special case of prior knowledge in group testing, but it is limited due to the absence of interactions/correlations between items.

Most related to our theoretical part is the concurrent work of \cite{gonen2022group}, who considered essentially the same model that we will consider: The defective set is known to lie in some set $\Sc$, with each $S \in \Sc$ being a size-$k$ (or size at most $k$) subset of $\{1,\dotsc,n\}$.  However, apart from the model, their work is very different from ours.  In particular, they focus on \emph{zero-error exact recovery}, which is a much more stringent goal than small-error approximate recovery, and accordingly, considerably more tests are required for non-adaptive strategies.\footnote{Specifically, the upper bound for non-adaptive testing in \cite{gonen2022group} is $O(k \log |\Sc|)$, whereas our less stringent criterion will give $O(\log |\Sc|)$ scaling.}  In addition, the focus in \cite{gonen2022group} is on scaling laws on the number of tests, whereas we are interested in how the underlying constants change even when the scaling laws match.

% \cite{gonen2022group} considers the following generalization of combinatorial/zero-error group  testing\footnote{\magenta{In combinatorial/zero-error group testing, the defective set is required to be recovered with probability 1, i.e., $\mathbb{P}[\text{err}] = 0$, instead of approaching 0 asymptotically.}}:   Instead of taking the space of defective sets $\Sc = \Sc^{\textrm{default}}$ to be the set of \textit{all} subsets of $V$ of size at most~$k$ or exactly~$k$, they take the space of defective sets $\Sc$ to be an \textit{arbitrary} set of subsets of $V$ of size at most~$k$ or exactly~$k$. That is,  $\Sc^{\textrm{default}} =  \{ \mathcal{L} \subseteq V : |\mathcal{L}| \le k \}$ or $\Sc^{\textrm{default}} = \{ \mathcal{L} \subseteq V : |\mathcal{L}| = k \}$; while $\Sc \subseteq \{ \mathcal{L} \subseteq V : |\mathcal{L}| \le k \}$ or $\Sc \subseteq \{ \mathcal{L} \subseteq V : |\mathcal{L}| = k \}$.  In particular, they provide lower and upper bounds on the (non-asymptotic) number of tests $t$ in terms of $|\Sc|$.

There are several additional works that are related to the more practical part of our work, and to our paper more generally.  We briefly outline them here, and discuss the main differences and advantages/disadvantages in Section \ref{sec:ising_discussion}.

A recent line of works has studied {\em community-aware group testing} \cite{nikolopoulos2021group,ahn2021adaptive,nikolopoulos2021overlap}, in which the items are arranged into known clusters, and those in the a common cluster are (relatively) highly correlated.  Various algorithms were proposed based on first identifying highly infected communities, and then using high-prevalence testing strategies (possibly even one-by-one testing) on the individuals therein.  This is another important special case of prior information, and is practically well-motivated, e.g., in testing contagious diseases where certain individuals share a house, office, etc., but in this paper we are also interested in broader kinds of structure beyond communities/clusters alone.

Two algorithms in \cite{goenka2021contact} are proposed for exploiting information from contact tracing with two different test models.  In particular, in the case of a binary-valued testing model (as we consider), they propose a Generalized Approximate Message Passing (GAMP) algorithm based on a dynamic infection model with known infection times and physical proximity levels.  Further graph-based models are given in \cite{arasli2021group} and \cite{nikpey2022group}.  In \cite{arasli2021group} the graph is modeled as being random, and after its realization is produced, anyone with a path to a ``patient zero'' becomes infected.  Along similar lines, in \cite{nikpey2022group}, given a known graph, each edge is dropped with some fixed probability, and then each connected component is independently (fully) defective with some fixed probability.

We note that various theoretical guarantees appear in \cite{nikolopoulos2021group,ahn2021adaptive,nikolopoulos2021overlap,arasli2021group,nikpey2022group} (as well as \cite{gonen2022group} discussed above), but they are largely incomparable to ours due to some combination of the following: (i) focusing on more specific defectivity models; (ii) giving an analysis that does not attempt to optimize constants (or sometimes even logarithmic factors), at least in the non-adaptive case that we focus on; and (iii) providing bounds (e.g., on the average number of errors) without attempting to characterize their asymptotic behavior, which is our focus.

\subsection{Contributions}
Our main contributions are as follows:
\begin{itemize}
    \item \emph{Theoretical bounds for model-based priors}: Focusing primarily on the noiseless setting, we characterize the information-theoretic limits for approximate recovery when the space of defective sets, $\mathcal{S}$, is a strict subset of all size-$k$ sets, and $|\mathcal{S}| \ll \binom{n}{k}$.  We specialize our general result to several specific examples, and illustrate the gains offered by this prior information, both in terms of $|\mathcal{S}|$ and other useful combinatorial properties.
    \item \emph{Ising model priors and relaxation-based decoders}:  We introduce the Ising model as a natural and flexible model for capturing dependencies in group testing, and show that it naturally leads to a Quadratic Programming (QP) based decoder, which can also be further relaxed to a more computationally efficient Linear Program (LP).  This is in contrast with previous approaches exploiting prior information, which often lead to Belief Propagation (BP) based decoders \cite{goenka2021contact,nikolopoulos2021group}.  We provide synthetic numerical experiments demonstrating the improvement of our decoders over standard baselines that do not exploit prior information.
\end{itemize}

\section{Theoretical Bounds for Model-Based Priors} \label{sec:model_based}
%!TEX root = main.tex

In this section, we study the information-theoretic limit of group testing with prior information, focusing on the approximate recovery criterion given in \eqref{eq:pe_dmax}.  Specifically, we generalize the standard combinatorial prior as follows: The number of defective items $k$ is still fixed, but the defective set $S$ is now uniform over some \emph{subset} $\Sc$, where each element of $\Sc$ is a size-$k$ subset of $[n]$.  In the case that $|\Sc| \ll \binom{n}{k}$, this amounts to having significant prior knowledge compared to the standard combinatorial prior.  Accordingly, we focus primarily on the following scaling regime:
\begin{equation}
\label{eq:size_of_def_spac}
    |\Sc| = 2^{ (\beta \,  k \log_2 \frac{n}{k})(1+o(1)) } \quad 
     \text{for some } \beta \in (0, 1).
\end{equation}
We observe that $\beta$ close to one corresponds to having little prior information, whereas $\beta$ close to zero amounts to having substantial prior information.  Varying $\beta$ between 0 and 1 covers a wide range of distinct scaling regimes, analogous to the consideration of $k = O(n^{\theta}) = n^{\theta(1+o(1))}$ with $\theta \in (0,1)$.

Since the preceding setup is a direct counterpart to that of model-based compressive sensing \cite{baraniuk2010model}, we adopt the terminology \emph{model-based priors}.

\subsection{General Achievability and Converse Bounds}

In this subsection, we provide general achievability and converse bounds for model-based priors. These bounds are typically not easy to evaluate directly, but they serve as the starting point for applying to specific cases and simplifying.

Throughout this section, we focus on the noiseless case for concreteness and relative ease of analysis, but sometimes also discuss noisy settings.

\subsubsection{Achievability}

The analysis in \cite{scarlett2016phase}, which takes $\Sc$
to be the entire space of $k$-sized subsets of $[n]$, contains the combinatorial terms
\begin{equation}
\label{eqn: combi term}
    \Ntau^{\text{default}} = \binom{k}{\tau} \binom{n-k}{\tau} 
    \quad \text{for } \tau = 0,1, \dotsc, k,
\end{equation}
counting the number of ways that the correct defective set $S$ can have $\tau$ of its items removed, and a different $\tau$ items added (from $[n] \setminus S$), to produce some incorrect 
defective set $S'$ of cardinality $k$. 
Note that the value $\Ntau^{\text{default}}$
is the same for each realization of the defective set $S$ by symmetry. 
Hence, for $\tau = 0,1, \dotsc, k$, we can rewrite \eqref{eqn: combi term} as
\begin{equation}
\label{eq:Ntau_default_maxform}
    \Ntau^{\text{default}} = 
    \max_{ S\,:\, |S| = k }
    \big| S' :   |S'| = k  \quad \text{and} \quad d(S,S') = \tau \big|,
\end{equation}
where $d(S,S') = \max\big\{ |S \setminus \Shat|, |\Shat \setminus S| \big\}$ is the distance measure associated with our approximate recovery criterion.

In our analysis, we generalize \eqref{eq:Ntau_default_maxform} for the setting of model-based priors, defining
\begin{equation}
\label{eq:Ntau}
    \Ntau = \max_{S \in \Sc}
    \big| S' \in \Sc : d(S,S') = \tau \big|.
\end{equation}
Coupling this with a suitably-modified information-theoretic decoder, we analyze how the final bound on the number of tests $t$ changes, leading to the following analog of \cite[Equation (3.47)]{scarlett2016phase}.

\begin{theorem}
\label{thm:card_bound_achiev}
    Let $n$ be the number of items, $k = \Theta\left(n^{\theta}\right)$ for some $\theta \in (0, 1)$ be the defective set size, $\Sc$ an arbitrary subset of the entire space of $k$-sized subsets of $[n]$, and $S$ drawn uniformly from $\Sc$. 
    Fix $\alpha^{\ast} \in (0, 1)$, and let $\dmax = \lfloor \alpha^{\ast} k \rfloor$.
    % For $\tau = \dmax +1, \dotsc, k$, let $\Ntau$ be as defined in \eqref{eq:Ntau}.
    % be the maximum number of ways that a
    % correct defective set $S \in \Sc$ can have $\tau$ of its $k$ items removed, and different $\tau$ items added, to produce     some incorrect defective set $S' \in \Sc$.
    Let $\mathsf{X} \in \{0,1\}^{t \times n}$ be a random binary matrix with i.i.d Bernoulli$\left(\frac{\nu}{k}\right)$ entries for some $\nu \in [0, 1]$. 
    Under the noiseless group testing model,
    % using Bernoulli testing with $p = \frac{\nu}{k}$, 
    the error probability $\pe$  of the information-theoretic threshold decoder 
    in Appendix~\hyperref[sect:inf_theoretic_decoder]{A-II} vanishes asymptotically as $n \to \infty$, 
    provided that
    \begin{equation}
    \label{eq:bound_T_Ntau}
        t \ge \max_{\alpha \in [\alpha^{\ast}, 1]}  
        \frac{\log_2 N_{\lceil \alpha k \rceil} + 2 \log_2 k }
        { e^{-(1-\alpha)\nu} H_2\big(e^{-\alpha \nu} \big)} \left(1 + o(1) \right), 
    \end{equation}
    where $H_2(\cdot)$ is measured in bits, and $N_{\tau}$ is given in \eqref{eq:Ntau}.
\end{theorem}
\begin{proof}
    See Appendix~\ref{app:general_achievability}.
\end{proof}

This expression can be simplified by substituting the standard choice $\nu = \ln 2$, which makes the denominator in \eqref{eq:bound_T_Ntau} equal to $2^{-(1-\alpha)}H_2(2^{-\alpha})$, in particular simplifying to $1$ when $\alpha=1$.  However, as we will see shortly, this choice of $\nu$ is not always optimal.

While we have focused on the noiseless setting in Theorem \ref{thm:card_bound_achiev}, symmetric noise is also straightforward to handle in the same way as \cite{scarlett2016phase}.  Non-symmetric noise can also be considered (e.g., see \cite[App.~A]{scarlett2018noisy}), but in such cases, the final expressions tend to be more complicated (e.g., not closed form, more free parameters, etc.).

\subsubsection{Converse}

Next, we provide an information-theoretic converse, i.e., a hardness result that holds for any non-adaptive test design and decoder.  To seek an ideal balance between the convenience of analysis vs.~generality, we consider two separate cases depending on whether the decoder is constrained to output an element of $\Sc$, and accordingly define
\begin{equation}
    \Sc_{\rm dec} = 
    \begin{cases}
        \Sc & \text{decoder always outputs $\Shat \in \Sc$} \\
        \{S \,:\, |S| = k \} & \text{unconstrained decoder output}.
    \end{cases} \label{eq:Sdec}
\end{equation}
In the second case, the reason we may still constrain $|S| = k$ is that outputting a size-$k$ set is without loss of optimality for the approximate recovery criterion that we consider \cite{reeves2013approximate,Scarlett2017LimitsSupportRecovery}.

We then introduce the following quantity:
\begin{equation}
\label{eq:Ntilde}
    \Ntil_{d_{\max}} = \max_{S \in \Sc_{\rm dec}} \big| S' \in \Sc : d(S,S') \le d_{\max} \big|,
\end{equation}
which is closely related to $\sum_{\tau=0}^{d_{\max}} N_\tau$ (see \eqref{eq:Ntau}), but is slightly different due to the order of sum/max and the possibility that $\Sc_{\rm dec}$ differs from $\Sc$.

% \magenta{\bf Comment: The analysis in my converse bounds seems to require a slightly different definition of $\Ntil_{\dmax}$, where constraint on the max is $S \in \Sc$. I am not sure if my analysis still holds under this definition. Should we change the constraint and hence potentially weaken our converse bounds, or perhaps there is a way to argue that WLOG we can always output $S \in \Sc$?}

\begin{theorem}
\label{thm:converse}
    Let $n$ be the number of items, $k = \Theta\left(n^{\theta}\right)$ for some $\theta \in (0, 1)$ be the defective set size, $\Sc$ an arbitrary subset of the entire space of $k$-sized subsets of $[n]$, and $S$ drawn uniformly from $\Sc$. 
    Fix $\alpha^{\ast} \in (0, 1)$, and let $\dmax = \lfloor \alpha^{\ast} k \rfloor$.
    Let $\mathsf{X} \in \{0,1\}^{t \times n}$ be an arbitrary non-adaptive test design.  Then, under the noiseless group testing model with approximate recovery, for any decoder to attain $\pe \not\to 1$ as $n \to \infty$, it must hold that
    \begin{equation}
    \label{eq:converse_bound_T_Ntau}
        t \ge 
        \Big( \log_2 |\Sc| - \log_2 \Ntil_{\dmax} \Big) \cdot \left(1 + o(1) \right), % = \Big( \beta k \log_2 \frac{n}{k} - \log_2 \Ntil_{\dmax} \Big) \cdot \left(1 + o(1) \right),
    \end{equation}
    \end{theorem}
    where $\Ntil_{\dmax}$ is given in \eqref{eq:Ntilde}.
\begin{proof}
    See Appendix \ref{app:pf_converse}.
\end{proof}

To simplify the analysis, we will sometimes apply this result to cases where the decoder must output an element of $\Sc$, i.e., the first case in \eqref{eq:Sdec}.  However, by a basic triangle inequality argument, e.g., as used in \cite[Sec.~4.2.2]{scarlett2019introductory}, a converse for such decoder translates to a converse for general decoders with $\dmax$ replaced by $2\dmax$.\footnote{The idea is that we may assume that the general decoder outputs $\Shat \in \{S \,:\, |S| = k \}$ within $\dmax$ of some $\Shat'\in \Sc$, since otherwise an error is guaranteed.  Then, for any $S \in \Sc$, the triangle inequality gives $d(S,\Shat') \le d(S,\Shat) + d(\Shat,\Shat') \le 2\dmax$.}  In particular, the resulting converse bounds will always coincide in the limit $\alpha^* \to 0$, which we believe to be the regime of primary interest.

For noisy group testing models in which the noiseless outcome is passed through a binary channel, we can obtain a similar result to Theorem \ref{thm:converse} with $\pe \to 0$ instead of $\pe \not\to 1$, and with the right-hand side of \eqref{eq:converse_bound_T_Ntau} being divided by the channel capacity.  Specifically, the analysis based on Fano's inequality in \cite[Sec.~4.1.2]{scarlett2019introductory} readily applies, and  $\Ntil_{\dmax}$ directly appears therein in the same way as \eqref{eq:converse_bound_T_Ntau}.  We also expect that the stronger statement $\pe \not\to 1$ can be obtained following the ideas of \cite[Thm.~1]{scarlett2016converse}, but we omit this direction since it is not our main focus.

\subsection{Cardinality-Bounded Structure}
\label{sect: trivial bound}

In this case, we specialize our general bounds to the case that only $|\Sc|$ is known, without more fine-grained bounds on each value of $\Ntau$.  This is motivated by the fact that given knowledge of $\Sc$, one would expect that $|\Sc|$ is typically easy to bound, whereas the individual $\Ntau$ may be much more complex.

For the achievability part, we characterize \eqref{eq:bound_T_Ntau} using the following trivial bound:
% In this section, we compute the achievability and limit if we replace the combinatorial
% term~\eqref{eqn: combi term} by a trivial bound
\begin{equation}
\label{eq:trivial_bound}
     \Ntau \le \min\left\{\Ntau^{\text{default}}, |\Sc|  \right\},
\end{equation}
and show that this leads to the following corollary.

\begin{corollary}
\label{cor:card_bound_achiev}
    Under the setup of Theorem~\ref{thm:card_bound_achiev}, if
    $|\Sc| = 2^{\left(\beta \,  k \log_2 \frac{n}{k}\right)(1+o(1))}$ for some $\beta \in (0, 1)$,
    % Let $n$ be the number of items, $k = \Theta\left(n^{\theta}\right)$ for some $\theta \in (0, 1)$ be the size of an unknown defective set $S$. Let  $\Sc$ be an arbitrary subset of the entire space of $k$-sized subsets of $[n]$ of size where $|\Sc| = 2^{\left(\beta \,  k \log_2 \frac{n}{k}\right)}$ for some $\beta \in (0, 1)$. 
    % Let $\alpha^{\ast} \in (0, 1)$ be arbitrary and $\dmax = \lfloor \alpha^{\ast} k \rfloor$.
    % % For $\tau = \dmax +1, \dotsc, k$, let $\Ntau$ be the number of ways that the correct defective set
    % % $S$ can have $\tau$ of its $k$ items removed, and different $\tau$ items added, to produce
    % % some incorrect defective set $S' \in \Sc$.
    % Let $\mathsf{X} \in \{0,1\}^{T \times n}$ be a random binary matrix with i.i.d Bernoulli$\left(\frac{\nu}{k}\right)$ entries for some $\nu \in [0, 1]$. 
    % Under the non-adaptive, noiseless, and approximate recovery setting  group testing model,
    % % using Bernoulli testing with $p = \frac{\nu}{k}$, 
    % as $n \to \infty$, 
    then the error probability $\pe$  of the information-theoretic threshold decoder 
    (Appendix~\hyperref[sect:inf_theoretic_decoder]{A-II}) vanishes asymptotically provided that
    \begin{equation}
    \label{eqn: simplified coeff}
    t \ge
        \left(
            \beta \, 
            \frac{ \mathrm{exp}\left(\nu \left(1 - \max \left\{ \alpha^{\ast}, \beta \right\} \right)\right)}
            {H_2\left(\mathrm{exp}(-\nu  \max \left\{ \alpha^{\ast}, \beta \right\} )\right)} 
        \right)
        \left( k \log_2 \frac{n}{k} \right)
        \left(1 + o(1) \right).
    \end{equation}
\end{corollary}
\begin{proof}
See Appendix~\ref{app:cardinality_bound}.
\end{proof}

\begin{figure}[t!]
    \centerline{\includegraphics[width = 7cm]{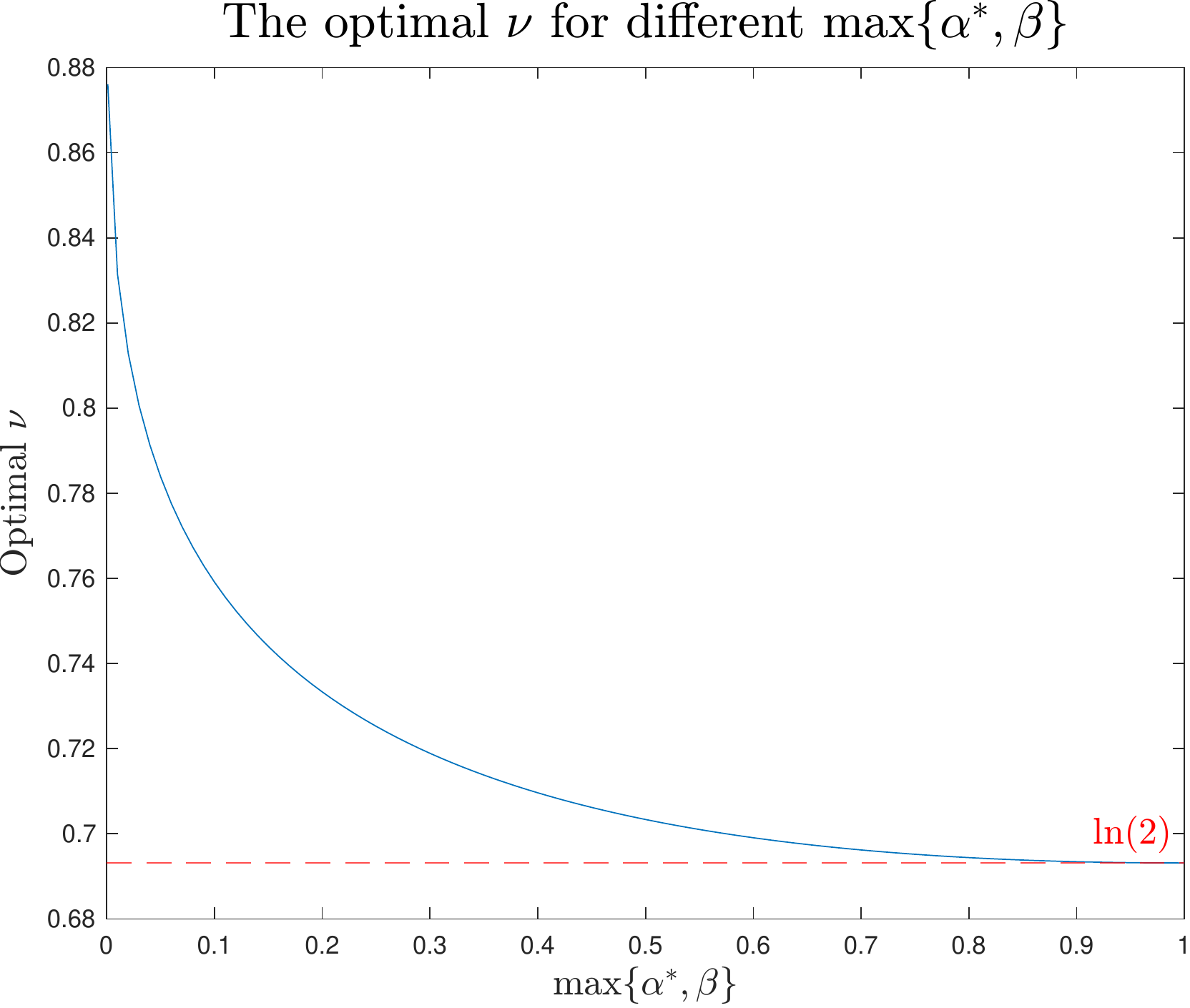}}
\caption{Numerical optimization of $\nu$.}
\label{img: optimal nu}
\end{figure}

Based on \cite[Thm. 3]{scarlett2016phase}, we expect the minimizing $\nu$ to approach $\ln 2$ when $\max \left\{ \alpha^{\ast}, \beta \right\} \to 1$. 
% the minimizing $\nu$ 
For fixed $\max \left\{ \alpha^{\ast}, \beta \right\}$, we can show by differentiation
% be solved analytically by simple differentiation. 
that the minimizing $\nu$  satisfies
\begin{multline}
\label{eq:card_bound_optimal_nv}
    H_2\left(\mathrm{exp}\left(-\nu  \max \left\{ \alpha^{\ast}, \beta \right\} \right) \right) \\ = 
    % H_2\left(e^{-(1-\gamma) \nu}\right) =
    - \max \left\{ \alpha^{\ast}, \beta \right\} 
    % (\gamma - 1) 
    \cdot
    \log_2 \left(1- \mathrm{exp}\left(- \nu \max \left\{ \alpha^{\ast}, \beta \right\} \right) \right),
\end{multline}
which appears to have no closed-form solution 
% for $\nu$ 
for general $\max \left\{ \alpha^{\ast}, \beta \right\}$. 
However, this can be solved numerically; from Figure \ref{img: optimal nu}, we see that the optimal $\nu$ starts higher and gradually decreases to $\ln 2$.

% \subsection{limits}
% For $\alpha^{\ast} \in (0, 1)$ and $\beta \in (0, 1]$, we are interested in comparing the achievability bound in Corollary~\ref{cor:card_bound_achiev} to (i) the achievability bound if we there is no prior knowledge (Lemma \ref{lem:standard_bounds}), and (ii) the benchmark $\log_2 |\Sc|$.

In the group testing literature, it is common to measure the performance using the following limiting ratio, or ``rate'':
\begin{equation}
    \label{eqn: limit n c k}
    \lim_{n \to \infty} \frac{\log_2\binom{n}{k}}{ t },
\end{equation}
where $k$ and $t$ implicitly depend on $n$.  This is motivated by the fact that the prior uncertainty without prior information is $\log_2\binom{n}{k}$.  Accordingly, in our setting, a natural counterpart is the following:
\begin{equation}
    \label{eqn: limit S}
    \lim_{n \to \infty} \frac{\log_2 |\Sc|}{t}.
    % = \frac{1}{\log_2 2}\frac{\log_2 |S|}{n(\alpha^{\ast}, \beta)}
\end{equation}
We will consider both of these limiting ratios, since \eqref{eqn: limit n c k} is also useful for the purpose of comparing to what one would attain in the absence of prior information (beyond knowledge of $k$).  When $|\Sc| = 2^{ (\beta \,  k \log_2 \frac{n}{k})(1+o(1)) }$ and $k = o(n)$ (as we consider), the two limiting ratios are simply related via $\lim_{n \to \infty} \frac{\log_2\binom{n}{k}}{ t } = \frac{1}{\beta} \lim_{n \to \infty} \frac{\log_2 |\Sc|}{t}$.

\begin{figure}[t!]
    \centering
    \includegraphics[width = 7cm]{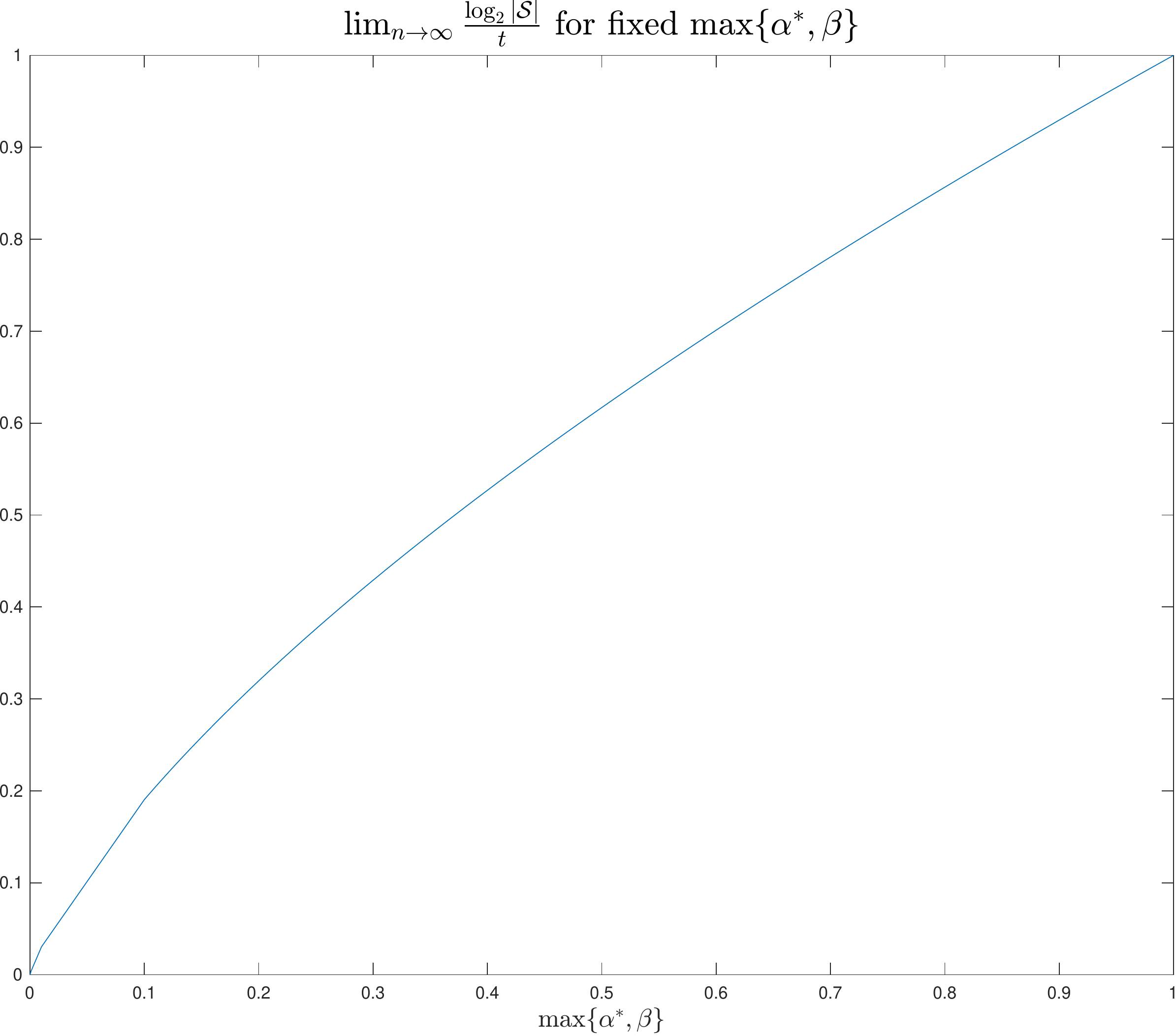}
        
     \includegraphics[width = 7cm]{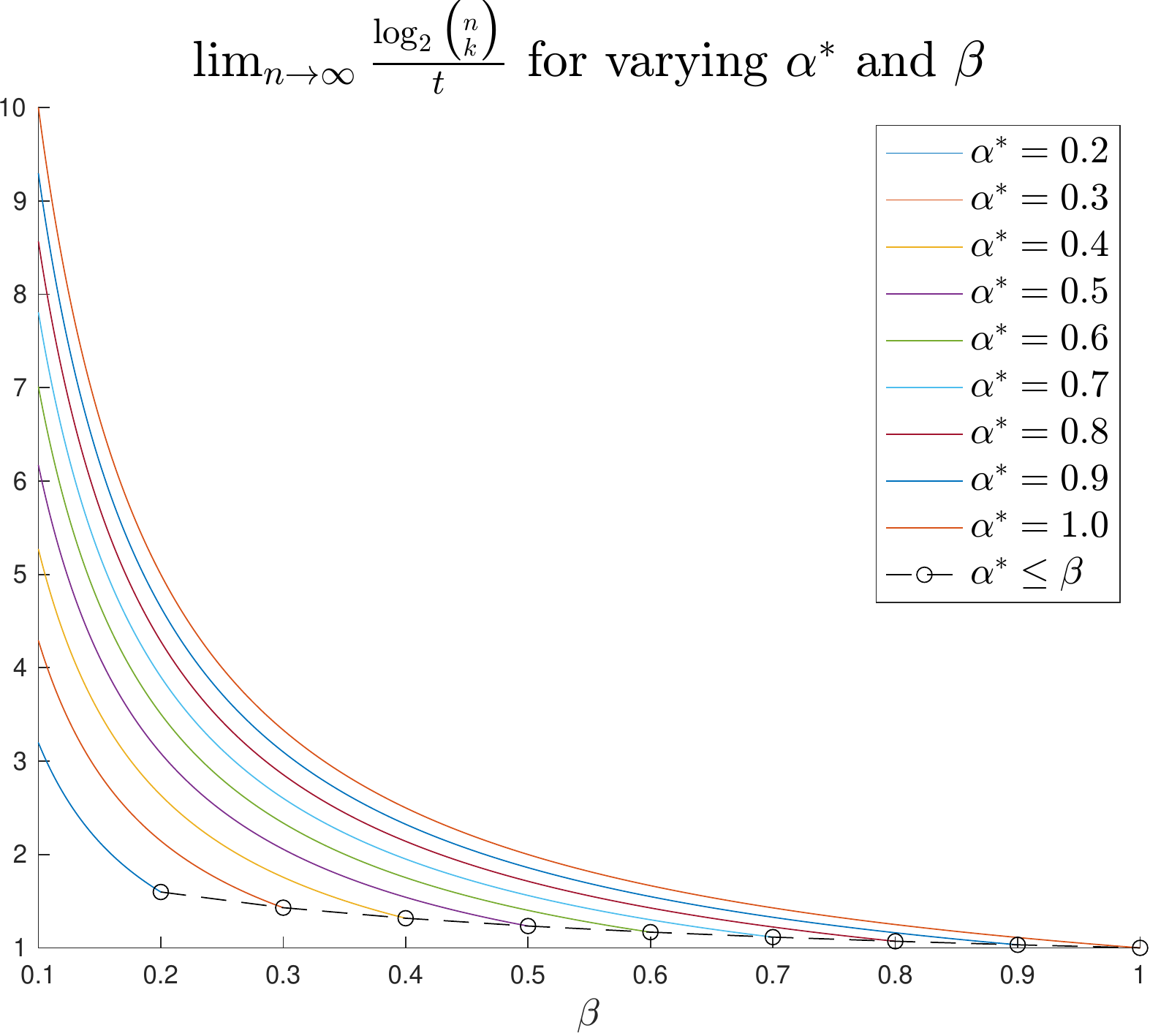}
     
\caption{(Top) Limiting ratio of $\frac{\log_2|\Sc|}{t}$ for various values of $\max \{\alpha^{\ast}, \beta \}$.  (Bottom) Limiting ratio of $\frac{\log_2 \binom{n}{k}}{t}$ for various values of $\alpha^{\ast}$ and $\beta$.  The circled markers correspond to $\beta = \alpha^*$.}
\label{img: limit card}
\end{figure}

Substituting \eqref{eqn: simplified coeff} into \eqref{eqn: limit S}, we obtain
\begin{align}
    \lim_{n \to \infty}
    \frac{\log_2 |\Sc|}{ t } &=
    % \frac{1}{\log_2 2} \frac{\log_2 |S|}{n(\alpha^{\ast}, \beta)} =
    \lim_{n \to \infty}
     \frac{\beta \, k \log_2 \frac{n}{k}}{ t } \nonumber \\ &=
     \frac{H_2\left(\mathrm{exp}(-\nu  \max \left\{ \alpha^{\ast}, \beta \right\} )\right)}{\mathrm{exp}\left(\nu \left(1 - \max \left\{ \alpha^{\ast}, \beta \right\} \right)\right)}, \label{eq:card_lim}
\end{align}
where $\nu$ takes the value that minimizes \eqref{eqn: simplified coeff}.
It follows that for a fixed value of $\max \{\alpha^{\ast}, \beta \}$, decreasing $\min \{\alpha^{\ast}$, $\beta\}$
does not affect the limit. The limit is shown in Figure~\ref{img: limit card} (Top).
% We can apply similar manipulations to the limit defined in \eqref{eqn: limit n c k} to obtain
% \begin{equation}
%     \lim_{n \to \infty}
%     \frac{\log_2\binom{n}{k}}{ t } =
%     % \frac{1}{\log_2 2} \frac{\log |S|}{n(\alpha^{\ast}, \beta)} =
%     \lim_{n \to \infty}
%     \frac{k \log_2 \frac{n}{k}}{ t } (1+o(1)) =
%     \frac{1}{\beta} 
%     \frac{H_2\left(\mathrm{exp}(-\nu  \max \left\{ \alpha^{\ast}, \beta \right\} )\right)}
%     {\mathrm{exp}\left(\nu \left(1 - \max \left\{ \alpha^{\ast}, \beta \right\} \right)\right)},
% \end{equation}
% where $\nu$ takes the value that minimizes \eqref{eqn: simplified coeff}.
As discussed above, the limit of $\lim_{n \to \infty} \frac{\log_2\binom{n}{k}}{ t }$ is simply a $\frac{1}{\beta}$ factor higher than \eqref{eq:card_lim}.  
For fixed $\max \{\alpha^{\ast}, \beta \}$, 
if we have $\alpha^{\ast} \le \beta$, then decreasing~$\alpha^{\ast}$ does not affect the limit;
however, if we have $\beta \le \alpha^{\ast}$, then decreasing~$\beta$ still increases the limit.  In particular, as $\beta \to 0$ with fixed $\alpha^*$, the resulting value becomes unbounded, indicating an arbitrarily large reduction in the number of tests compared to no prior information.  The limits for both cases are shown in Figure~\ref{img: limit card} (Bottom).

As seen in Figure~\ref{img: limit card} (Bottom), for all $\beta < 1$, we need 
strictly fewer tests than $\log_2 \binom{n}{k}$ (corresponding to $1$ on the y-axis), i.e., the achievability bound in the absence of prior information (Lemma~\ref{lem:standard_bounds}). Furthermore, smaller $\beta$ naturally leads to smaller $t$.  
On the other hand, we observe in Figure~\ref{img: limit card} (Bottom) that a smaller
value $\max\{\alpha^*, \beta\}$ actually leads to smaller $\lim_{n \to \infty}
\frac{\log_2 |\Sc|}{ t }$.  This may be in part due to the crude upper bound in \eqref{eq:trivial_bound} becoming looser for small $\beta$ (it is tight for $\beta = 1$).

% \begin{figure}%
%     \centering
%     \subfloat[\centering label 1]{{\includegraphics[width = 7.5cm]{limit_1.png} }}%
%     \
%     \subfloat[\centering label 2]{{\includegraphics[width = 7.5cm]{limit_2.png}}}%
%     \caption{2 Figures side by side}%
%     \label{fig:example}%
% \end{figure}

For the converse, if all that we know about $\Sc$ is a cardinality bound, then we have the following generalization of \eqref{eq:test_conv}.

\begin{corollary}
\label{cor:converse_card}
    Consider the setup of Theorem \ref{thm:converse}, and suppose that
    \begin{equation} \label{eq:Sbound}
        |\Sc| = 2^{\left(\beta \,  k \log_2 \frac{n}{k}\right)(1+o(1))}
    \end{equation}
    for some $\beta \in (0,1)$.  Then, for any decoder to attain $\pe \not\to 1$ as $n \to \infty$, it must hold that
    \begin{equation}
    \label{eq:converse_bound_T_Ntau_cardinality}
        t \ge \max\{0,\beta - \alpha^*\} \Big( k \log_2\frac{n}{k} \Big) \cdot \left(1 + o(1) \right).
    \end{equation}
\end{corollary}
\begin{proof}
    See Appendix \ref{app:cardinality_bound}.
\end{proof}

A regime of particular interest is $\alpha^*$ becoming arbitrarily small, meaning that the recovery is required to be increasingly close to exact.  In this case, the lower bound approaches $\big(\beta k \log_2\frac{n}{k}\big) (1+o(1))$, which is simply $(\log_2 |\mathcal{S}|)(1+o(1))$ as one would expect. 

We observe that Corollary \ref{cor:converse_card} is non-vacuous only if $\beta > \alpha^*$.  While this may appear limiting, this condition is unavoidable: It is straightforward to show that a feasible choice of $\Sc$ (subject to the cardinality bound \eqref{eq:Sbound}) is the set of all $k$-sparse binary vectors within distance $\beta k (1+o(1))$ of $\{1,\dotsc,k\}$.  In this case, if $\alpha^* > \beta$, then simply returning $\{1,\dotsc,k\}$ will be considered non-erroneous with probability one, even when there are no tests.

We will shortly see that both Corollary \ref{cor:card_bound_achiev} and Corollary \ref{cor:converse_card} can be significantly improved when more specific knowledge of the structure of $\Sc$ (beyond $|\Sc|$ alone) is available.

\subsection{Block Structure}
%!TEX root = main.tex

We now look at an extreme scenario where we have exact knowledge of all 
$\{ \Ntau \}_{\tau = \dmax + 1}^k$ in \eqref{eq:Ntau},
and illustrate that the achievability can be further improved as compared to using only the cardinality bound \eqref{eq:trivial_bound}.
% \subsection{Block graphs with infinite edge strength}
Specifically, we consider the scenario where the $n$ items have the following structure: 
% a block graph with infinite edge strength, % \footnote{the higher the edge strength, the more likely the connected vertices are to infect each other}, i.e., 
The $n$ items are arranged in $n/q$ groups\footnote{We assume that $n$ and $k$ are multiples of $q$.} of size $q$, and each group is either all defective  or all nondefective, with $k/q$ defective groups. 

This model can be viewed as capturing (albeit in a highly simplified manner) the fact that infections or defects may occur in \emph{groups} in applications.  It is also a special case of community-aware models studied in \cite{nikolopoulos2021group,ahn2021adaptive}.  Unlike those works, we do not seek to identify new test designs to handle community structure, but instead explore the benefits of improved decoding alone (while maintaining i.i.d.~designs).

Then, $\Ntau = 0$ if~$\tau$ is not a multiple of $q$, while if $\tau$ is a multiple of $q$, then
\begin{equation}
\label{eqn: block graph Nl bound}
    \Ntau = \binom{k/q}{\tau/q} \binom{\frac{n-k}{q}}{\tau/q}.
    % \le \min\left\{\binom{k/q}{\tau/q} \binom{\frac{n-k}{q}}{\tau/q} , |\Sc| =  \binom{n/q}{k/q}  \right\} 
\end{equation}
% For these $\tau$, it follows that, when $n \to \infty$
% \begin{equation}
% \label{eqn: block graph Nl bound}
%     \log_2 \Ntau \le \min\left\{\log_2 \binom{k/q}{l/q} + \log_2 \binom{\frac{n-k}{q}}{l/q} , \left(\frac{k}{q} \, \log_2 \frac{n}{k}\right)\left(1 + o\left(1\right) \right)
%     \right\}.
% \end{equation}
% We now use similar steps and notations in Appendix~\ref{sect:appendix_B} to characterize
The exact knowledge of $\Ntau$ gives the following improved achievability result.
We note that while this result in itself is not ``surprising'',\footnote{By interpreting the problem as recovering $k/q$ defective groups out of the $n/q$ groups, Lemma~\ref{lem:standard_bounds} gives the same achievability bound.} it serves as a useful illustration of the benefit of characterizing each $\Ntau$ individually in Theorem \ref{thm:card_bound_achiev}, instead of only $|\Sc|$. 

\begin{corollary}
\label{cor: block_graph_achiev}
    Let $1 \le q \le k \le n$ be integers such that $n$ and $k$ are multiples of~$q$,  $k = \Theta\left(n^{\theta}\right)$ for some $\theta \in (0, 1)$, and $\dmax = \Theta(k)$.
    Consider $\Sc$ defined according to the above block-structured model. 
    % Let  $\Sc$ be an arbitrary subset of the entire space of $k$-sized subsets of $[n]$ of size where $|\Sc| = 2^{\left(\beta \,  k \log_2 \frac{n}{k}\right)}$ for some $\beta \in (0, 1)$. 
    % Let $\alpha^{\ast} \in (0, 1)$ be arbitrary and 
    % Let $\dmax = \Theta(k)$.
    % For $\tau = \dmax +1, \dotsc, k$, let $\Ntau$ be the number of ways that the correct defective set
    % $S$ can have $\tau$ of its $k$ items removed, and different $\tau$ items added, to produce
    % some incorrect defective set $S' \in \Sc$.
    Let $\mathsf{X} \in \{0,1\}^{t \times n}$ be a random binary matrix with i.i.d. Bernoulli$\left(\frac{\nu}{k}\right)$ entries for some $\nu \in [0, 1]$. 
    Under the non-adaptive, noiseless, and approximate recovery setting  group testing model,
    % using Bernoulli testing with $p = \frac{\nu}{k}$, 
    as the number of items $n \to \infty$, the error probability $\pe$  of the information-theoretic threshold decoder (see Appendix~\hyperref[sect:inf_theoretic_decoder]{A-II}) vanishes asymptotically provided that
    \begin{equation}
    % \label{eqn: simplified coeff}
    t \ge
    \label{eq:block_graph_achievability}
        \frac{1}{q} \frac{1}{H_2(e^{-\nu})} 
        \left( k \log_2 \frac{n}{k} \right)
        \left(1 + o(1) \right).
        % \frac{1}{q}
        % \left( k \log_2 \frac{n}{k} \right)
        % \left(1 + o(1) \right).
    \end{equation}
\end{corollary}

\begin{proof}
See Appendix~\ref{app:block_structure}.
\end{proof}

The minimizing $\nu$ in \eqref{eq:block_graph_achievability} is $\ln 2$, which gives the condition 
\begin{multline}
    t  \ge \frac{1}{q} \left( k \log_2 \frac{n}{k} \right) \left(1 +o(1) \right) \\
    = \left( \log_2  \binom{n/q}{k/q} \right)\left(1 +o(1) \right)
    = \left( \log_2 |\Sc| \right)\left(1 +o(1) \right).
\end{multline}
Observe that $\alpha^*$ is absent, as it only affects the higher-order asymptotics.

For fixed $q$, 
% since $\log_2 |\Sc| = \log_2 \binom{n/q}{k/q} = \frac{1}{q} \big( k \log_2 \frac{n}{k} \big) (1+o(1))$, 
we simply have $\lim_{n \to \infty} \frac{\log_2 |\Sc|}{ t } = 1$ and $\lim_{n \to \infty} \frac{\log_2\binom{n}{k}}{ t } = q$.  Hence, the required number of tests is a $1/q$ fraction of that without prior information (Lemma~\ref{lem:standard_bounds}).

For comparison, we now consider the achievability result obtain by only considering the naive cardinality bound \eqref{eq:trivial_bound} on~$\Ntau$, instead of \eqref{eqn: block graph Nl bound}. 
Note that the value of $\beta$ in \eqref{eq:size_of_def_spac} evaluates to $\beta = \frac{\log_2 |\Sc|}{ k \log_2 \frac{n}{k}} = \frac{1}{q} (1+o(1))$.
% $ |\Sc| =  \binom{n/q}{k/q}$ and so $\beta = 1/q$.
For fixed $\alpha^{\ast} = \dmax /k \in (0, 1)$, it follows from the discussion in Section~\ref{sect: trivial bound} that the coefficient of $\left( k \log_2 \frac{n}{k} \right)\left(1 + o(1) \right)$ is given by
\begin{equation}
    \frac{1}{q}
    \left(
    \inf_{\nu > 0}
    \frac{\mathrm{exp}\left(\nu \left(1 - \max \left\{ \alpha^{\ast}, 1/q\right\} \right)\right)}
    {H_2\left(\mathrm{exp}(-\nu  \max \left\{ \alpha^{\ast}, 1/q \right\} )\right)}
    \right) >
    \frac{1}{q}, \label{eq:card_comparison}
\end{equation}
% the maximizing~$\alpha$ is $\max\{\alpha^{*}, 1/q \}$, 
and is no longer independent of $\alpha^{*}$.
From Figure~\ref{img: limit card}, we see that if
 $\max\left\{\alpha^{\ast}, 1/q \right\} < 1$, then the values are lower than 1 (i.e., more tests are needed), and significantly so when $\max\left\{\alpha^{\ast}, 1/q \right\}$ is small (i.e., $d_{\max}$ is small and $q$ is large).  
%  Likewise, based on Figure \ref{img: limit n c k}, we see that with
%  $\max\left\{\alpha^{\ast}, 1/q \right\} < 1$, the limits are lower than $q = 1/\beta$, and hence lower than  the one in \eqref{eqn: limit q}.
This illustrates that having a good control over $\{ \Ntau \}_{\tau = \dmax + 1}^k$ can give a significantly tighter achievability bound.

The converse is also simple in this case, and could likely be obtained by slightly modifying the analyses in \cite{nikolopoulos2021group,ahn2021adaptive}, but we provide a self-contained proof for completeness.

\begin{corollary}
\label{cor:converse_block}
    Consider the setup of Corollary \ref{cor: block_graph_achiev}, but with an arbitrary non-adaptive test matrix $\mathsf{X}$, and suppose that $\dmax = \lfloor \alpha^* k \rfloor$ for some $\alpha^* \in (0,1)$.  Then, for any decoder that outputs $\Shat \in \Sc$ to attain $\pe \not\to 1$ as $n \to \infty$, it must hold that
    \begin{align}
    \label{eq:converse_bound_block}
        t &\ge  (1-\alpha^*)\left( \frac{k}{q} \log_2 \frac{n}{k} \right)\left(1+o(1) \right) \nonumber \\
        &= (1-\alpha^*) \left( \log_2 |\Sc| \right)\left(1+o(1) \right).
    \end{align}
\end{corollary}
\begin{proof}
    See Appendix~\ref{app:block_structure}.
\end{proof}

Similarly to Lemma \ref{lem:standard_bounds}, the achievability and converse match to within a factor of $1-\alpha^*$.  The converse is non-vacuous for all $\alpha^* \in (0,1)$, demonstrating that we can overcome the limitation of Corollary \ref{cor:converse_card} discussed above given more precise knowledge about $\Sc$ rather than only the cardinality.

\subsection{Imbalanced Structure}
%!TEX root = main.tex

We now turn to a more complex example that further demonstrates the possible gaps between the various bounds of interest.  We consider the following setup: The $n$ items consist of two disjoint blocks of size
$n_1 = o(n)$ and $n_2 = n - n_1 = n(1-o(1))$ items, and there are $k_1$ defectives in the first block and $k_2 = k - k_1$ in the second.  We let $\Sc$ be the set of all $k$-size subsets satisfying these conditions.  This model captures (in a simplified manner) scenarios where some items are known \emph{a priori} to be more likely to be defective, analogous to previous works on exact recovery such as \cite{li2014group,kealy2014capacity}.

Although we do not provide its details, the case $k_1 = k(1-o(1))$ and $k_2 = k - k_1 = o(k)$ (i.e., almost all of the~$k$ defective items are concentrated in the smaller block) turns out to be straightforward: Similarly to the block example, the achievability part gives $\lim_{n \to \infty} \frac{\log_2 |\Sc|}{ t } = 1$, and the converse matches this up to multiplication by $1-\alpha^*$.  We instead focus on the more challenging case in which $k_1 = \gamma k$ for some $\gamma \in (0, 1)$, i.e., both $k_1$ and $k_2$ are linear in~$k$, while taking $n_1 = \Theta(n^c)$ for some $c \in (\theta, 1)$ (with $k = \Theta(n^{\theta})$).

\begin{corollary}
\label{cor: imbal_achiev_linear_k2}
    Let $n$ be the number of items, and $k = \Theta\left(n^{\theta}\right)$ for some $\theta \in (0, 1)$ 
    be the size of an unknown defective set $S \subset [n]$.
    Fix $\alpha^{\ast} \in (0, 1)$, and let $\dmax = \lfloor \alpha^{\ast} k \rfloor$.
    Let $\Sc$ be defined as above with parameters $n_1, n_2, k_1, k_2$ satisfying
    $n_1 = \Theta(n^c)$ for some $c \in (\theta, 1)$ and $n_2 = n - n_1$,
    as well as $k_1 = \gamma k$ and $k_2 = (1-\gamma) k$ for some $\gamma \in (0, 1)$. %
    Let $\mathsf{X} \in \{0,1\}^{t \times n}$ be a random binary matrix with i.i.d. Bernoulli$\left(\frac{\nu}{k}\right)$ entries for some $\nu \in [0, 1]$. 
    Under the noiseless group testing model,
    % using Bernoulli testing with $p = \frac{\nu}{k}$, 
    the error probability $\pe$ of the information-theoretic threshold decoder 
    (see Appendix~\hyperref[sect:inf_theoretic_decoder]{A-II}) vanishes asymptotically as $n \to \infty$, provided that
    \begin{multline}
        \label{eq:imbal_achiev_linear_k_2}
        t \ge
        \max_{\alpha \in [\max \{\alpha^*,  1- \gamma\}, 1]}
        \frac{\left(1 - \theta\right) - \gamma\left(1-c\right)  - (1-\alpha)\left(c-\theta\right) }
        % \frac{\left(1 - \gamma\right)\left(1-c\right) + \alpha \left(c -\theta\right) }
        {e^{-(1-\alpha)\nu} H_2(e^{-\alpha \nu})} \\ \times 
        \left( k \log_2 n \right) 
        \left(1 +o(1) \right)
    \end{multline}
\end{corollary}
\begin{proof}
See Appendix~\ref{app:imbalanced_structure}.
\end{proof}

Based on various numerical calculations, we observed that the maximizing $\alpha$ is often at one of the endpoints, i.e., $\alpha = \max \{\alpha^*,  1- \gamma\}$ or $\alpha = 1$.  This is supported by the following.

\begin{corollary}
\label{cor: imbal_achiev_linear_k2_convexity}
Under the setup of Corollary~\ref{cor: imbal_achiev_linear_k2}, if $\alpha^* \le 1-\gamma$
and 
\begin{equation}
    \left(\frac{\nu}{c-\theta}\right)
    (1-\gamma)(1-c)
    > 
    % \frac{0.029}{\ln 2} \ge 
    0.029,  \label{eq:nu_cond}
\end{equation}
then the achievability bound \eqref{eq:imbal_achiev_linear_k_2} can be simplified to 
    \begin{multline}
    \label{eq:imbal_achiev_linear_k_2_twoends}
         t \ge  
        %     % \inf_{\nu > 0} \ 
            \max
            \bigg\{
            (1-\theta) \, 
            \frac{ (1-\gamma)  e^{ \gamma \nu} }
            {H_2(e^{-(1-\gamma) \nu})} k \log_2 n, \\
            % \left((1-\gamma)(1-c) + (c-\theta) \right)
            \frac{(1-\theta) - \gamma(1-c)}
            {H_2(e^{-\nu})} k \log_2 n 
            \bigg\} 
            % \left( k \log_2 n \right) 
            \left(1 +o(1) \right). 
            % \max
            % \left\{
            % \frac{ (1-\gamma)  \left(e^{ \gamma \nu} \right) }
            % {H_2(e^{-(1-\gamma) \nu})}, \ 
            % % \left((1-\gamma)(1-c) + (c-\theta) \right)
            % \frac{1}{H_2(e^{-\nu})}
            % \left(1 - \frac{\gamma(1-c)}{1-\theta}\right)
            % \right\} 
            % \left( k \log_2 \frac{n}{k} \right) 
            % \left(1 +o(1) \right). 
    \end{multline}
\end{corollary}
\begin{proof}
See Appendix~\ref{app:imbalanced_structure}.
\end{proof}

We note that the condition $\alpha^* \le 1-\gamma$ is mild since we typically want small $\alpha^*$, and the condition \eqref{eq:nu_cond} is also mild since it is only violated when $\gamma$ and/or $c$ are very close to one (we have $\frac{1}{c-\theta} > 1$, and we will discuss below how the optimal $\nu$ is at least $\ln 2$).

\begin{figure}[htbp!]
    \centerline{\includegraphics[width = 8cm]{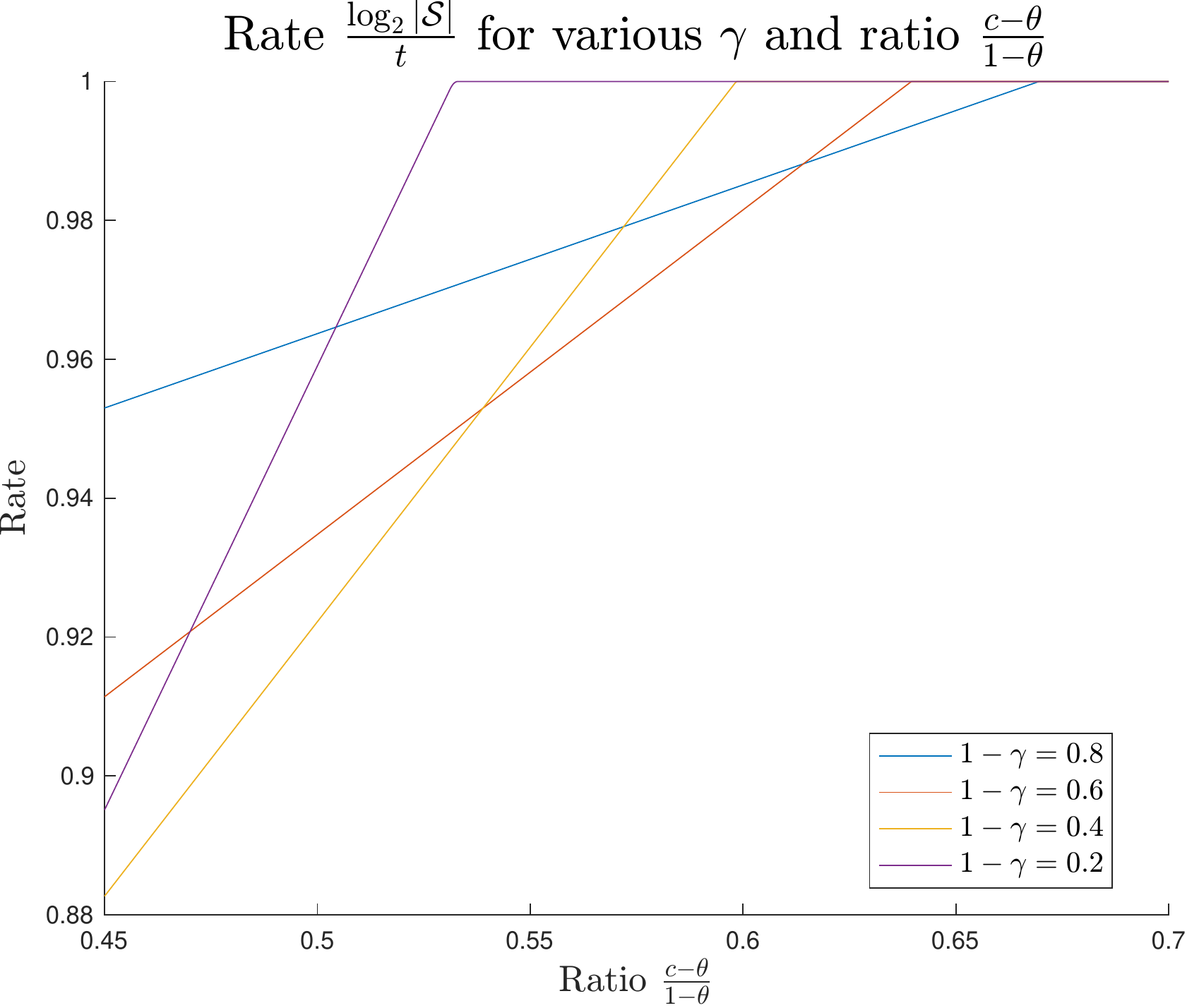}}
\caption{Limits as defined in \eqref{eqn: limit S} for various $\gamma$ and $\frac{c - \theta}{1-\theta}$; for all values shown, the conditions of 
Corollary~\ref{cor: imbal_achiev_linear_k2_convexity} hold.}
\label{fig:limits_imbalanced}
\end{figure}

We now consider minimizing \eqref{eq:imbal_achiev_linear_k_2_twoends} with respect to $\nu$. 
% which gives
% \begin{equation}
%     t \ge  
%             \inf_{\nu > 0} \ 
%             \max
%             \left\{
%             (1-\theta) \, 
%             \frac{ (1-\gamma) e^{ \gamma \nu} }
%             {H_2(e^{-(1-\gamma) \nu})} k \log_2 n, \ 
%             % \left((1-\gamma)(1-c) + (c-\theta) \right)
%             \frac{(1-\theta) - \gamma(1-c)}
%             {H_2(e^{-\nu})} k \log_2 n 
%             \right\} 
%             % \left( k \log_2 n \right) 
%             \left(1 +o(1) \right).  \label{eq:imb_inf_nu}
% \end{equation}
We show in \eqref{eq:imb_spars_log_2_S_asym_bound_with_k2} of Appendix~\ref{app:imbalanced_structure} that $\log_2|\Sc| = \left( \left(1 - \theta\right) - \gamma\left(1-c\right)  \right)\left( k \log_2 n \right) \left(1+o(1)\right)$, and combining this with \eqref{eq:imbal_achiev_linear_k_2_twoends} gives the following for the limit introduced in \eqref{eqn: limit S} upon optimizing $\nu$:
\begin{multline}
\label{eq:rate_imbal_linear_k2}    
    \lim_{n \to \infty}
        \frac{\log_2 |\Sc|}{t} = 
        \max_{\nu > 0} \,
        \min\Bigg\{
        H_2(e^{-\nu}), \\ \,
        \frac{H_2\left(e^{-(1-\gamma) \nu}\right)}
             {e^{ \gamma \nu} } 
             \left(1 + \frac{\gamma\left(c - \theta \right)}{(1-\gamma)(1-\theta)} \right)
        \Bigg\}.
\end{multline}
We denote the two minimands by
\begin{gather}
    f_1(\nu) \coloneqq H_2(e^{-\nu}), \\ 
    f_2(\nu) \coloneqq 
    \frac{H_2\left(e^{-(1-\gamma) \nu}\right)}
             {e^{ \gamma \nu} } 
             \left(1 + \frac{\gamma\left(c - \theta \right)}{(1-\gamma)(1-\theta)} \right),
\end{gather}
and analyze their maximizing $\nu$ separately:
\begin{itemize}
    \item The first minimand, $f_1$, is maximized at $\nu = \ln 2$, 
    % and denoting the corresponding value by $t_2$, 
    It follows that if $1 =  f_1\left(\ln\left(2\right)\right) 
        \le f_2\left(\ln\left(2\right)\right)$, 
    or equivalently,
    \begin{equation}
        \frac{c-\theta}{1-\theta} \ge
        \frac{1-\gamma}{\gamma}
        \left(\frac{2^{\gamma}}{H_2(2^{-(1-\gamma)})} - 1 \right),
    \end{equation}
    then \eqref{eq:rate_imbal_linear_k2} can be simplified to $1$.
    \item For the second minimand $f_2$, we can show by differentiation that,
    similar to \eqref{eq:card_bound_optimal_nv}, the maximizing~$\nu$  satisfies
    \begin{equation}
        H_2\big(e^{-(1-\gamma) \nu}\big) =
        (\gamma - 1) \log_2 \big(1- e^{-(1-\gamma) \nu}\big),
    \end{equation}
    which appears to have no closed-form solution in general.
    The numerically optimized $\nu$ gives similar behavior to that shown in Figure~\ref{img: optimal nu} (with $1-\gamma$ in place of $\max\{\alpha^*, \beta \}$).  Defining $\nu^* \coloneqq \argmax_{\nu > 0} f_2(\nu)$ accordingly, it follows that if $f_2\left(\nu^*\right) \le  f_1\left(\nu^*\right)$, 
    or equivalently,
    \begin{equation}
        \frac{c-\theta}{1-\theta} \le
        \frac{1-\gamma}{\gamma}
        \left(
        H_2\big(e^{-\nu^*}\big) 
        \frac{ e^{ \gamma \nu^*} }{ H_2(e^{-(1-\gamma) \nu^*}) } - 1
        \right),
    \end{equation}
    then \eqref{eq:rate_imbal_linear_k2} can be simplified to $f_2\left(\nu^*\right)$. 
\end{itemize}
% {\bf \color{magenta} [Suggestion: May be good to explicitly give the equivalent conditions on the $\frac{c-\theta}{1-\theta}$, e.g., in the first case above, we can lengthen it to ``if $1 =  f_1\left(\ln\left(2\right)\right) \le f_2\left(\ln\left(2\right)\right)$, or equivalently, $\frac{c-\theta}{1-\theta} > ...$, then (45) can...''.]}
While these cases on $\frac{c-\theta}{1-\theta}$ are not exhaustive, we found numerically that one of the two usually holds; see Figure~\ref{fig:limits_imbalanced}, where the horizontal parts correspond to $\nu = \ln 2$, the diagonal parts to $\nu = \nu^*$, and a seemingly imperceptible (but non-zero) curved region in between corresponds to other $\nu$ values.

To assess the tightness of our achievability results, we now turn to the converse, for which we have the following.
% {\bf \color{blue} [TODO (Jon) Go over proof.  Consider whether we should omit the second part because it already follows from Corollary \ref{cor:converse_card}.]}
% {\bf \color{magenta} [Remark (Ivan) Just for intuition/interpretation. We can get the same converse bound by doing standard group testing separately on two blocks, where the second block (size $n(1-o(1))$) is ``granted maximum error quota'', while the first block get error quota only if $\alpha^* > 1- \gamma$, i.e., the allowable error is bigger than number of defective in the second block]}

\begin{corollary}
\label{cor:converse_imbalanced_linear_k2}
    Consider the setup of Corollary \ref{cor: imbal_achiev_linear_k2}, but with an arbitrary non-adaptive test matrix $\mathsf{X}$.
    If $\alpha^* > 1-\gamma$, then for any decoder that outputs $\Shat \in \Sc$ to attain $\pe \not\to 1$ as $n \to \infty$, it must hold that
    \begin{equation}
    % \label{eq:converse_bound_block}
        t \ge  (1-\alpha^*)\left( \frac{c-\theta}{1-\theta} \right) \left( k \log_2 \frac{n}{k} \right)\left(1+o(1) \right).
    \end{equation}
    If $\alpha^* \le 1-\gamma$, then for any decoder to attain $\pe \not\to 1$ as $n \to \infty$, it must hold that
    \begin{align}
        t &\ge  
        \left(1 -\frac{\gamma(1-c)}{1-\theta} - \alpha^* \right) 
        \left( k \log_2 \frac{n}{k} \right)\left(1+o(1) \right) \nonumber
        \\ &= (\beta -\alpha^*) \left( k \log_2 \frac{n}{k} \right) \left(1+o(1) \right).
    \end{align}
\end{corollary}
\begin{proof}
   See Appendix~\hyperref[app:imbalanced_structure]{D-I}.
\end{proof}

The limit $\alpha^* \to 0$ is of particular interest, and gives a converse approaching $\beta k \log_2\frac{n}{k}$, which corresponds to a horizontal line with value $1$ in Figure \ref{fig:limits_imbalanced}.  Hence, we see that the achievability and converse match for many values of $\gamma$ and $\frac{c-\theta}{1-\theta}$, but do not always do so.  In fact, if we move beyond i.i.d.~designs, then we can adopt the block-structured design outlined in \cite[Sec.~5.6]{aldridge2019group} (essentially running group testing separately on the two groups of items); then, a minor modification of the analysis therein reveals that the resulting bound matches the converse for all values of $\gamma$ and $\frac{c-\theta}{1-\theta}$.  Hence, this example illustrates that sometimes adjusting the decoder alone is enough to maintain optimality in the presence of prior information, but sometimes it is necessary to also modify the test design.

\section{Ising Model Priors and Practical Decoding Rules}  \label{sec:ising}
%!TEX root = main.tex

The model-based setup considered in Section \ref{sec:model_based} is convenient for theoretical analysis, but has notable limitations that may prevent its use in practice.  Firstly, it assumes that $k$ is pre-specified, which is typically unrealistic when infections are random (e.g., i.i.d.~infections would lead to $k$ following a binomial distribution).  Secondly, in general, $\mathcal{S}$ may be a complicated set of subsets, and accordingly may be inconvenient to specify explicitly for use in the group testing algorithm, let alone clear how to effectively exploit once it is specified.

Motivated by these considerations, in this section, we introduce a natural generalization of the i.i.d.~prior that is versatile, admits a compact representation, and naturally leads to practical decoding algorithms for exploiting the availability of prior information.

\subsection{Ising Model}

We incorporate prior knowledge in the form of an Ising model.  Since the $\{-1,1\}$-valued Ising model is more standard than the $\{0,1\}$-valued counterpart, we work with $\tilde{\uv} \in \{-1,1\}^n$ with $+1$ for defectives and $-1$ for non-defectives.  Hence, $\uv \in \{0,1\}^n$ and $\tilde{\uv} \in \{-1,1\}^n$ are related via $\tilde{\uv} = 2\uv - \mathbf{1}$, where $\mathbf{1}$ is a vector of $1$s.

Considering the items $V = [n]$ as vertices, the Ising model is defined as 
\begin{equation}
    P(\tilde{\uv}) = 
    \frac{1}{Z} 
    \exp
    \left(
    \sum_{(j,j') \in E} \lambda_{jj'}  \tilde{u}_{j}\tilde{u}_{j'} - \sum_{j \in V} \phi_{j} \tilde{u}_{j}
    \right), \label{eq:Ising}
\end{equation}
where 
\begin{itemize}
    % \item $\tilde{\uv} \in \{-1,1\}^n$ is a group of lattice sites;
    \item $G = (V, E)$ is a graph with vertices $V = [n]$ and edges $E \subseteq [n] \times [n]$, where edges represent interactions among vertices.
    \item For each edge $(j,j') \in E$, $\lambda_{jj'}$ is the edge strength.  We focus mainly on the case of a common non-negative value $\lambda_{jj'}=\lambda$ (higher means that connected vertices are more likely to have the same status), but differing values and/or negative values are also straightforward to handle.
    \item For each vertex $j \in V$, $\phi_{j}$ is a parameter controlling the sparsity; again, we focus primarily on the case of a common non-negative value $\phi_{j}=\phi$ ($\phi = 0$ means around half the items are defective ($+1$), whereas $\phi \gg 0$ means that very few are);
    \item $Z$ is a normalizing constant.
\end{itemize}
We believe that this model serves as a natural way to incorporate prior information, with edges corresponding to interactions (e.g., making infections more likely). 

For the more standard case of a 0-1 valued defectivity vector, we can substitute $\tilde{\uv} = 2\uv - \mathbf{1}$ into \eqref{eq:Ising} to obtain
\begin{multline}
    P(\uv) = \frac{1}{Z} 
    \exp 
    \bigg(
    \lambda \sum_{(j,j') \in E} \left(2u_{j}-1 \right)\left(2u_{j'}-1\right) \\ -
    \phi\sum_{j \in V}(2u_{j}-1) 
    \bigg). \label{eq:Ising01}
\end{multline}
% or in the matrix form as 
% \begin{equation}
%     P_{U}(\uv) = \frac{1}{Z} \exp{(\lambda (2\uv - \mathbf{1})^{T}\mathsf{Q}(2\uv - \mathbf{1}) - \phi \mathbf{1}^{T}(2\uv - \mathbf{1}))},
%     \label{ising_matrix}
% \end{equation}
% where $\mathsf{Q}$ is the corresponding adjacency matrix for the graph $G$. 
% Incorporating this type of prior information is reasonable since connected vertices in $V$ tend to be more likely to share the similar defectiveness, and this should reduce the necessary number of tests.

In the case that $E = \emptyset$, this model reduces to the standard i.i.d.~prior, with a one-to-one correspondence between our parameter $\phi$ and the defectivity probability $q$.  Recall that in this case, previous works have considered the (Integer) Linear Programming decoding methods stated in \eqref{lp_noiseless_intro} (noiseless) and \eqref{lp_noisy_intro} (noisy), and the integer versions can be interpreted as maximizing $P(\uv|\mathsf{X},\Yv)$ \cite{ciampiconi2020maxsat}.  We will derive the analogs for the Ising model, re-using the definitions of $\Cless$ and $\Cnoisy$ in \eqref{eq:C_noiseless} and \eqref{eq:C_noisy} (and their relaxations $\Clessr$ and $\Cnoisyr$).

\subsection{Advantages, Limitations, and Comparisons} \label{sec:ising_discussion}

An advantage of adopting the Ising model is that it is a versatile and well-studied model, with existing applications including statistical physics \cite{Gla63}, image processing \cite{Gem84}, computational biology \cite{Dur98}, natural language processing \cite{Man99}, and social network analysis \cite{Was94}.  As a result, an extensive range of graph learning algorithms are already available (e.g., see \cite{lokhov2018optimal} and the references therein) and could potentially be used provided that suitable historical data is available.  Further advantages are highlighted at the end of this subsection when comparing with previous approaches.

On the other hand, it is also important to note the limitations of the model.  Notably, accurate learning of the graph (and/or the associated edge strengths $\{\lambda_{jj'}\}_{(j,j') \in E}$) can require significant amounts of data that may not always be available.  Inaccurate graph learning may degrade the performance, though we provide some experimental evidence of robustness in Section \ref{sec:experiments}.  Finally, as we discuss further below, our model is static, and thus does not model dynamic behavior that may arise in applications such as medical testing (e.g., via the well-known SIR model \cite{kermack1927contribution}).

Finally, we give some discussion on this model compared to previous models (see Section \ref{sec:related_prior} for their summary): 
\begin{itemize}
    \item In \cite{nikolopoulos2021group,ahn2021adaptive,nikolopoulos2021overlap}, the items are arranged into known clusters with high correlations within clusters.  The Ising model can also capture such structure, e.g., by including a large number of edges (or even all edges) within each cluster, but also has the added versatility of allowing arbitrary cross-community edges.  On the other hand, unlike the preceding works, we do not attempt to give theoretical guarantees on the number of tests for our graph-based model.
    \item The works \cite{goenka2021contact,arasli2021group,nikpey2022group} also allow general graphs, and can thus model structure beyond clustering alone.  In all of these works, adding an edge to the graph amounts to increasing the chances of the two nodes simultaneously being defective.  However, the graphs in these works and ours are all used in a different manner, making a direct comparison difficult.  A potential limitation of \cite{arasli2021group,nikpey2022group} is the ``all or nothing'' nature of defectivity in each connected component (after possible edge removals), though this may be reasonable in many cases of interest.  The model in \cite{goenka2021contact} appears to be especially powerful when precise knowledge infection times, dynamics, and proximity are known.  In contrast, ours is a static model suited to scenarios where interactions are known to have occurred, but not necessarily the finer details (though such details could potentially be incorporated into varying $\{\lambda_{jj'}\}_{(j,j') \in E}$).  Overall, our model is not intended to ``replace'' any other, but we believe that it is a useful addition with a good balance between modeling power and simplicity.  The following two aspects appear to be unique to our model compared to the existing ones:
    \begin{itemize}
        \item By considering negative values of $\lambda_{jj'}$, we can easily model \emph{negative} associations between items (i.e., if one is defective, the other is less likely to be).  While this is unlikely to be of interest in medical testing scenarios, it may be useful more generally.
        \item Our model lends itself naturally to {\em relaxation-based decoding methods} based on Quadratic Programming and Linear Programming, in contrast to existing techniques that instead lead to message passing algorithms \cite{nikolopoulos2021group,goenka2021contact}.
    \end{itemize}
\end{itemize}
Overall, in view of the preceding advantages and limitations, we believe that the Ising model is primarily suited to settings that (i) are static in nature, (ii) have historical data or prior knowledge available for (at least approximately) inferring a graph and an edge strength $\lambda$ (or varying edge strengths $\{\lambda_{jj'}\}_{(j,j') \in E}$), and (iii) benefit from the use of simple and efficient decoding algorithms.

\subsection{Quadratic Programming Decoder}

In the following, we provide a simple theorem showing that maximum a posteriori (MAP) decoding rule (i.e., choosing $\uv$ to maximize $P(\uv|\mathsf{X},\Yv)$) naturally leads to an Integer Quadratic Program.

\begin{theorem}
\label{thm:qp_map}
    In the noiseless setting, under an Ising model prior with parameter $(\lambda,\phi)$, the MAP decoder has the same solution as the following minimization problem:
    \begin{equation}
    \begin{aligned}
        &{\rm minimize}_{\uv} - \Big(\lambda \sum_{(j,j') \in E} (2u_{j}-1)(2u_{j'}-1) \\
            &\hspace*{5cm} - \phi\sum_{j \in V}(2u_{j}-1) \Big) \\
        &\text{subject to } \uv \in \Cless.
    \end{aligned}
    \label{qp_noiseless}
\end{equation}
Moreover, in the case of i.i.d.~symmetric noise with parameter $\rho \in \big(0,\frac{1}{2}\big)$, the MAP decoder has the same solution as the following minimization problem (upon keeping $\uv$ and discarding $\boldsymbol{\xi}$):
\begin{equation}
    \begin{aligned}
        &\min_{\uv,\boldsymbol{\xi}} - \Big(\lambda \sum_{(j,j') \in E} (2u_{j}-1)(2u_{j'}-1) - \phi\sum_{j \in V}(2u_{j}-1) \Big) \\
        &\hspace*{6cm} - \eta\sum_{i=1}^{t} \xi_{i} \\
        &\text{subject to } (\uv,\boldsymbol{\xi}) \in \Cnoisy,
    \end{aligned}
    \label{qp}
\end{equation}
where $\eta = \log\frac{\rho}{1-\rho}$.
\end{theorem}
\begin{proof}
    See Appendix \ref{sec:map_proof}.
\end{proof}

As with the LP formulations in \eqref{lp_noiseless_intro} and \eqref{lp_noisy_intro}, we can relax the constraints in \eqref{qp_noiseless} and \eqref{qp} by replacing $\Cless$ and $\Cnoisy$ by $\Clessr$ and $\Cnoisyr$.  This removes the seemingly difficult integer-valued nature of the problems, though as we discuss further below, it does not immediately imply that they can be solved efficiently.  We also note that the objective function \eqref{qp} can be rewritten in a matrix-vector form that is more typical of QP problems (and similarly for \eqref{qp_noiseless}):
\begin{equation}
    - (\lambda (2\uv - \mathbf{1})^{T}\mathsf{Q}(2\uv - \mathbf{1}) - \phi \mathbf{1}^{T}(2\uv - \mathbf{1})) - \eta \mathbf{1}^{T} \boldsymbol{\xi},
    \label{ising_matrix}
\end{equation}
where $\mathsf{Q}$ is the ``upper'' adjacency matrix for the graph $G$, i.e., it is upper triangular since we only count each edge between $u_j$ and $u_j'$ once, and all entries on the main diagonal are zeros because there are no self-edges.

\subsection{Linearized Quadratic Programming Decoder}
\label{linearizedqp}
Since the adjacency matrix $\mathsf{Q}$ in \eqref{ising_matrix} is not positive semidefinite,\footnote{Even if we symmetrize $\mathsf{Q}$, the fact that the diagonals are zero (but the off-diagonals contain non-zeros) prevents the matrix from being PSD.  For instance, if there are two nodes connected by a single edge, the symmetrized matrix becomes $\begin{bmatrix} 0 & 1 \\ 1 & 0 \end{bmatrix}$, which has eigenvalues $-1$ and $1$.} the optimization problem is non-convex.  This is somewhat unfortunate, because even with real-valued linear constraints, non-convex QPs are NP hard to solve in general \cite{karp1972reducibility,cook1971complexity}.

To overcome this potential difficulty, we apply an idea introduced in \cite{10.1287/opre.22.1.180} and surveyed in \cite{burer2012non}, which ``linearizes'' the QP at the expense of adding further constraints.  In general there can be up to $O(n^2)$ such additional constraints, but in our setting there are only $O(|E|)$, which is much smaller than $O(n^2)$ for typical (sparse) graphs.  We call this approach Linearized QP.

In both the noiseless and noisy cases, the only non-linear terms in the objective are the $u_{j}u_{j'}$ terms.  We convert the Integer QP into an Integer LP by replacing each such term $u_{j}u_{j'}$ by a new binary variable $u_{jj'} \in \{0,1\}$, and these variables $\{u_{jj'}\}_{(j,j') \in E}$ are included as binary optimization variables.  We need to ensure that the property $u_{jj'} = u_{j}u_{j'}$ is maintained, and to do so, it suffices to constrain $u_{j} \geq u_{jj'}$, $u_{j'} \geq u_{jj'}$, and $u_j + u_{j'} - u_{jj'} \le 1$ \cite{burer2012non}.  In the noiseless case, this leads to the following: $\hat{\uv}$ is equal to
% a quadratic function of $n$ binary variables by adding $o(n^2)$ additional variables and constraints. We call this approach as Linearized QP in this paper. By incorporating the Ising model, we obtain the non-linear terms $u_{j}u_{j'}$ and hence are motivated to linearize them. Specifically, we follow the standard formulation for transforming 0-1 polynomial programming problem into 0-1 linear problem introduced in \cite{10.1287/opre.22.1.180}: Replace any cross product $u_{j}u_{j'}$ with a new binary variable $u_{jj'}$ such that $u_{jj'} \in \{0,1\}$, and set the constraints $u_{j}(u_{j'})\in \{0,1\}$, $u_{j} + u_{j'} - u_{jj'} \leq 1$, and $- u_{j} - u_{j'} + 2u_{jj'} \leq 0$. In the noiseless setting, we have
%\red{We also let the set $E_{j} \subseteq \{1, \dots, n\}$ represents the indices of all neighboring vertices that interacted with $u_{j}$ through the edges in $E$, such that we define the degree of $u_j$ as $d_{j} = \sum_{j' \in E_j} u_{j'}$.}
{\fontsize{9.5}{10} \selectfont
\begin{align}
    &\argmin_{\uv \in \Cless} - \Big(\lambda \sum_{(j,j') \in E} (2u_{j}-1)(2u_{j'}-1) - \phi\sum_{j \in V}(2u_{j}-1) \Big) \\
    = &\argmin_{\uv \in \Cless} - \Big(\lambda \sum_{(j,j') \in E} (4u_{j}u_{j'} -2u_{j} -2u_{j'}) - \phi\sum_{j \in V}2u_{j} \Big) \\
    = &\argmin_{\uv \in \Cless} - \Big( 2\lambda \sum_{(j,j') \in E} u_{jj'} - \lambda \sum_{(j,j') \in E} (u_j + u_j') - \phi\sum_{j \in V} u_{j} \Big).
    \label{qp_linear_argmin}    
\end{align}}
Hence, we are left with the following optimization problem:
\begin{equation}
    \begin{aligned}
        &\min_{\uv, \{u_{jj'}\}_{(j,j') \in E}} - \Big( 2\lambda \sum_{(j,j') \in E} u_{jj'} - \lambda \sum_{(j,j') \in E} (u_j + u_j') \\
        & \hspace*{6cm} - \phi\sum_{j \in V} u_{j} \Big) \\ 
        &\text{subject to } \uv \in \Cless,  \text{ (or $\Clessr$ if relaxed)} \\ 
        % &u_{jj'} \in \{0,1\} \text{\ or\ } u_{j} \in [0,1] \\ 
        % & - u_{j} - u_{j'} + 2u_{jj'} \leq 0  \text{\ (or\ } u_{j} \geq u_{jj'}, u_{j'} \geq u_{jj'}) \\
        & \qquad\qquad~~\, u_{jj'} \in \{0,1\}, \text{ (or $[0,1]$ if relaxed)} \\
        & \qquad\qquad~~\, u_{j} \geq u_{jj'} \\
        & \qquad\qquad~~\, u_{j'} \geq u_{jj'} \\
        & \qquad\qquad~~\, u_{j} + u_{j'} - u_{jj'} \leq 1.
    \end{aligned}
    \label{atl_noiseless}
\end{equation}
Similarly for the noisy case, the objective function is given as
\begin{equation}
    \begin{aligned}
        &\min_{\uv,\boldsymbol{\xi},\{u_{jj'}\}_{(j,j') \in E}} - \bigg( 2\lambda \sum_{(j,j') \in E} u_{jj'} -\lambda \sum_{(j,j') \in E} (u_j + u_j') \\
        &\hspace*{5cm} - \phi\sum_{j \in V} u_{j} \bigg) - \eta\sum_{i=1}^{t} \xi_{i} \\
        &\text{subject to } (\uv,\boldsymbol{\xi}) \in \Cnoisy \text{ (or $\Cnoisyr$ if relaxed)} \\ 
        & \qquad\qquad~~\, u_{jj'} \in \{0,1\}, \text{ (or $[0,1]$ if relaxed)} \\
        & \qquad\qquad~~\, u_{j} \geq u_{jj'} \\
        & \qquad\qquad~~\, u_{j'} \geq u_{jj'} \\
        & \qquad\qquad~~\, u_{j} + u_{j'} - u_{jj'} \leq 1,
        % &u_{jj'} \in \{0,1\} \text{\ or\ } u_{j} \in [0,1] \\ 
        % & - u_{j} - u_{j'} + 2u_{jj'} \leq 0 \text{ (or\ } u_{j} \geq u_{jj'}, u_{j'} \geq u_{jj'} \text{ if\ relaxed)} \\
        % & u_{j} + u_{j'} - u_{jj'} \leq 1. \\
    \end{aligned}
    \label{qp_linear}
\end{equation}
where $\eta = \log(\frac{\rho}{1-\rho})$.  We observe that the relaxed versions are indeed Linear Programs, for which numerous efficient solvers exist both in theory and in practice.  While the integer versions remain equivalent to those given in Theorem \ref{thm:qp_map}, this equivalence may be lost upon relaxing the problems.  Despite this, we will see in the following section that the relaxed versions still give strong empirical performance, analogous to the findings for standard group testing in \cite{6288622}.

\section{Numerical Experiments} \label{sec:experiments}
%!TEX root = main.tex

In this section, we provide some simple proof-of-concept experiments for the QP and LP based decoding rules introduced in the previous section, without attempting to be comprehensive.\footnote{For reproducibility, the code for our experiments is available at \url{https://github.com/ethangela/priors_group_testing}.}

\subsection{Ising Model Details}

% We consider the Ising model with two distinct graphs. The first is a $28\times28$ square grid graph where each vertex is connected with its (up to $4$) neighboring vertices (up, down, left and right). The second graph, with the same size $28\times28$, is divided into $16$ blocks, and the size of each block is $7\times7$. Within each block all vertices are still connected in the form of grid, but there are no connections among different blocks. Two visual examples of these types of graphs are shown in the top row of Figure \ref{fig:samples}.  We manually selected the parameters $\lambda$ and $\phi$ to be $0.5$ and $0.006$ for the grid graph, and $0.6$ and $0.035$ for the block graph, to obtain a suitable level of defectivity and correlation. 

\begin{figure}[!t]

\begin{minipage}[b]{.4\linewidth}
  \centering
  \centerline{\includegraphics[width=3cm]{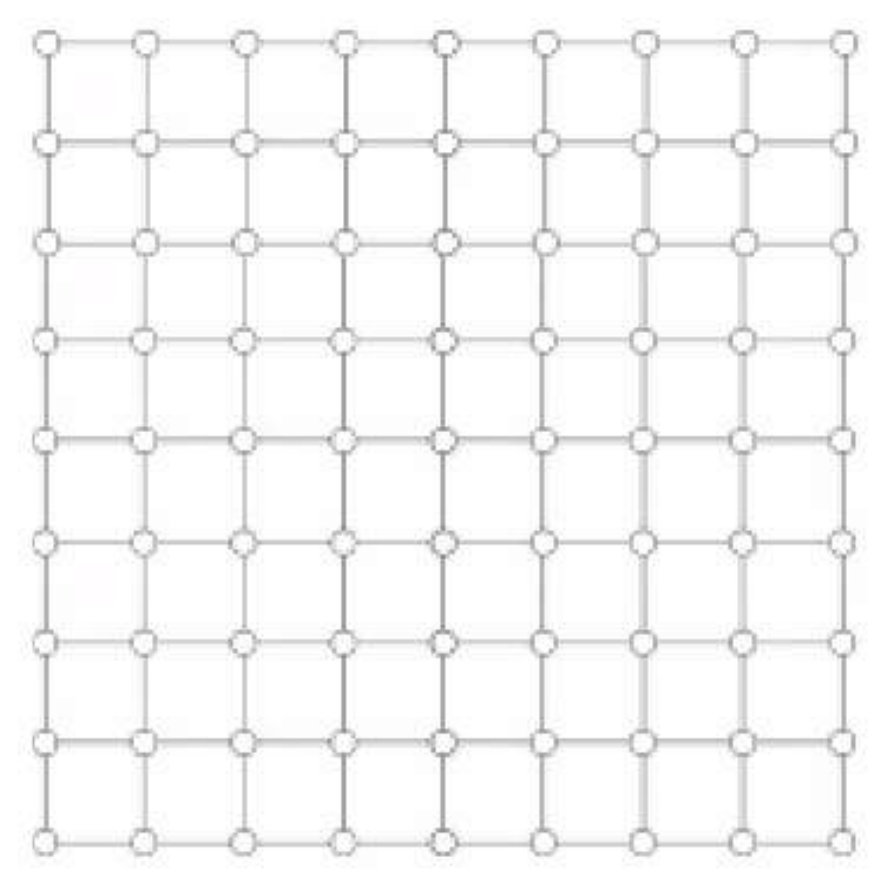}}
   \centerline{Grid graph example}\medskip
\end{minipage}
\hfill
\begin{minipage}[b]{0.4\linewidth}
  \centering
  \centerline{\includegraphics[width=3cm]{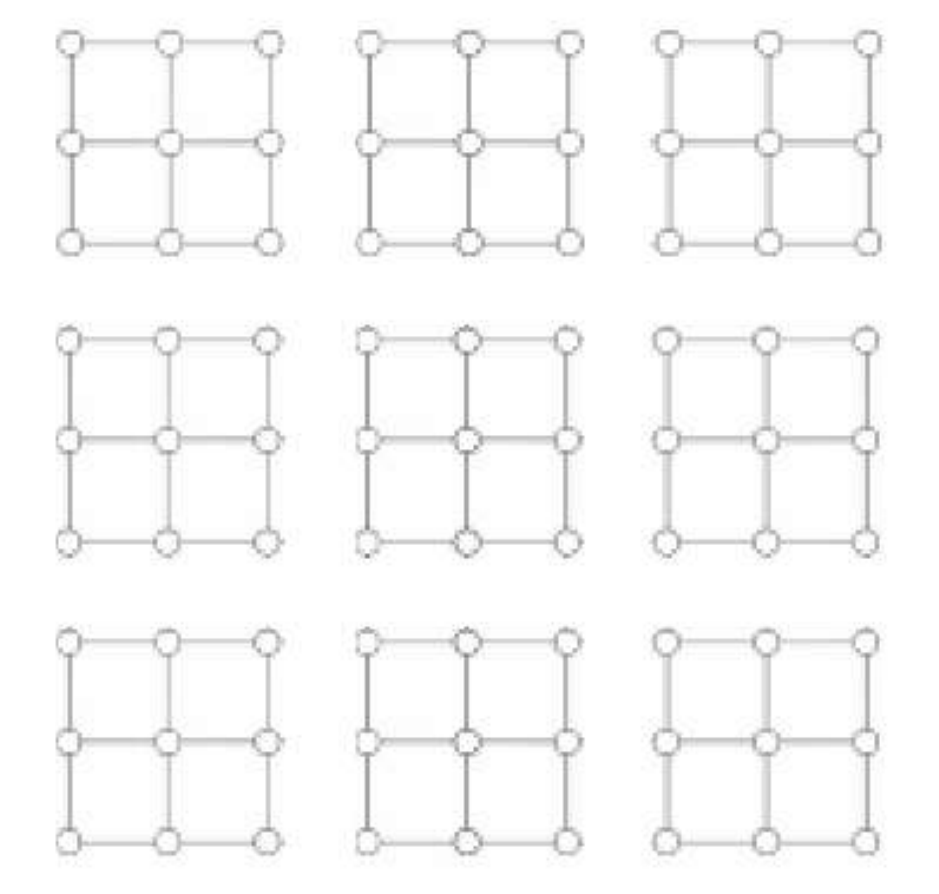}}
   \centerline{Block graph example}\medskip
\end{minipage}

\begin{minipage}[b]{.4\linewidth}
  \centering
  \centerline{\includegraphics[width=4cm]{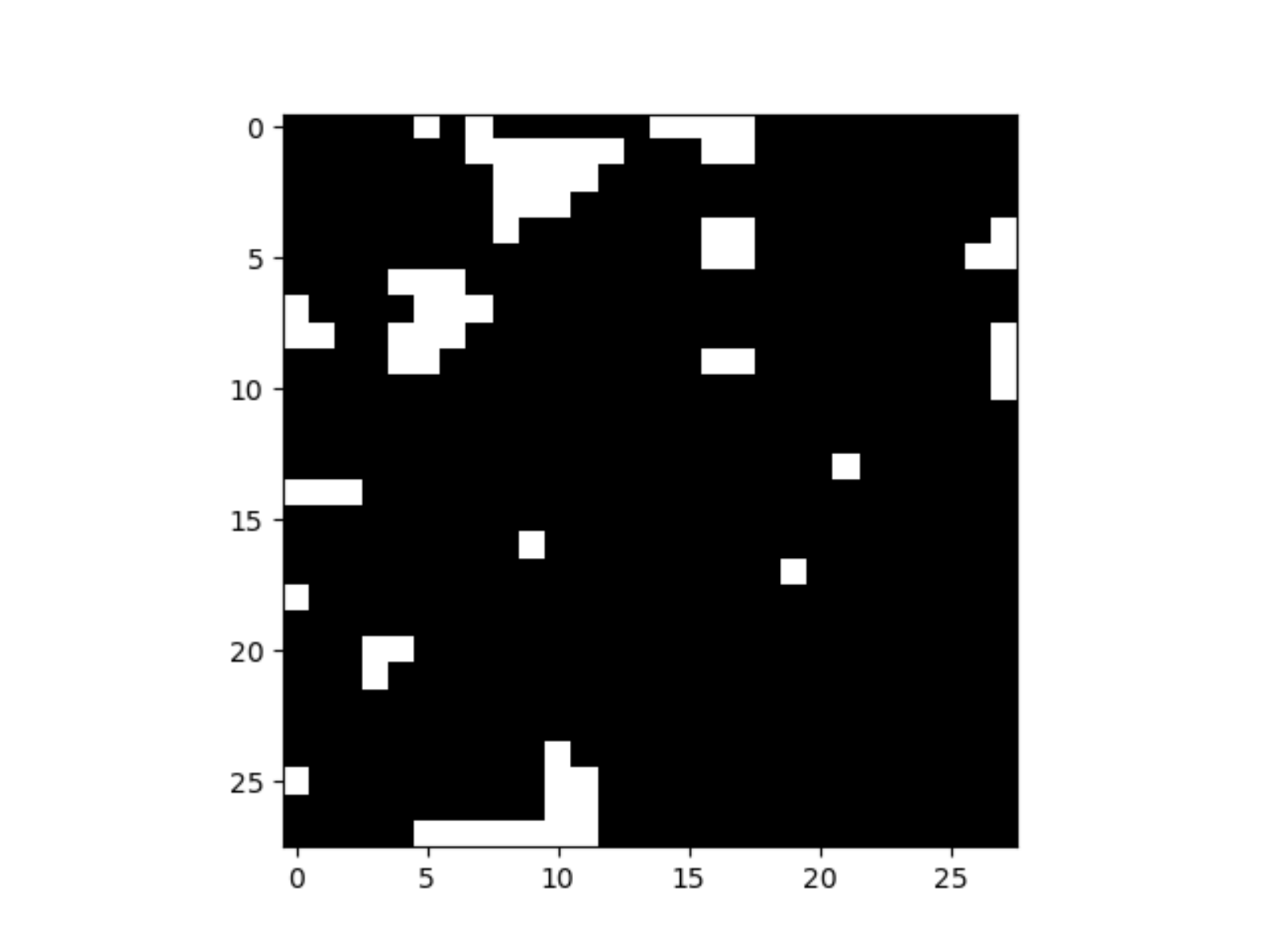}}
  \centerline{Grid graph sample}
  \centerline{($28 \times 28$)}\medskip
\end{minipage}
\hfill
\begin{minipage}[b]{0.4\linewidth}
  \centering
  \centerline{\includegraphics[width=4cm]{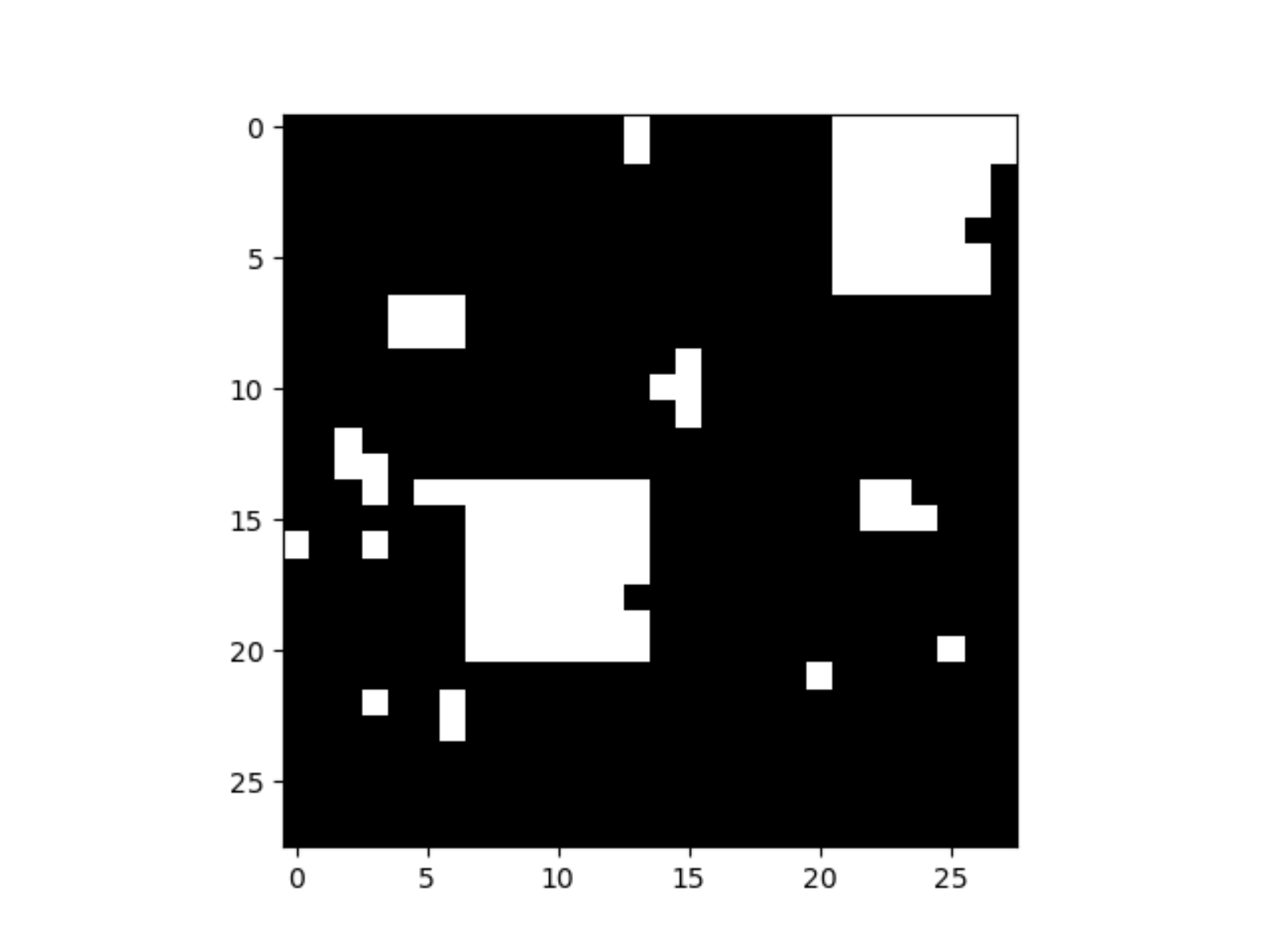}}
  \centerline{Block graph sample}
  \centerline{($4 \times 4$ blocks, $7 \times 7$ each)}\medskip
\end{minipage}
\caption{Ising model graphs and samples.}
\label{fig:samples}
\end{figure}

We consider the Ising model, focusing initially on two simple choices of the graph. The first is a square grid graph where each vertex is connected with its (up to $4$) neighboring vertices (up, down, left and right). The second graph, with the same size as the first one, is further divided into several blocks of equal size. Within each block, all vertices are still connected in the form of grid, but there are no connections among different blocks. Two examples of these types of graphs are shown in the top row of Figure \ref{fig:samples}.  Roughly, the grid graph can be viewed as capturing a simple form of proximity-based correlation, and the block graph can be viewed as capturing a simple form of grouped/clustered structure.

We set the graph size as $28\times28$ and the block size as $7\times7$ ($16$ blocks in total). We manually set the parameters $\lambda$ and $\phi$ to be $0.5$ and $0.006$ for the grid graph, and $0.6$ and $0.035$ for the block graph, which we found to be suitable for producing defectivity patterns that are sparse (as desired for group testing) while also exhibiting visible correlations (as desired for understanding the gains of prior information).  We apply Gibbs sampling \cite{4767596} to (approximately) generate samples from these models, using $1000$ iterations.  The generated samples are shown in the bottom row of Figure \ref{fig:samples}.

%Gibbs Sampling is a Monte Carlo Markov Chain method that iteratively draws an instance from the distribution of each variable, conditional on the current values of the other variables, in order to approximately sample from complex joint distributions. 

\subsection{Implementation}
For each generated sample with size $28\times28$, we flatten it as a vector $\uv' \in \{-1,1\}^n$ with $n=784$, and form $\uv \in \{0,1\}^n$ by replacing $-1$s by $0$s.  The test matrix is randomly drawn according to a Bernoulli design, where each item $u_j$ is included in each test independently with probability $\frac{\ln 2}{k}$.

For both graphs, we keep $\uv$ fixed to the samples indicated in Figure \ref{fig:samples}, and for each trial of our experiment, we only change the test matrix $\mathsf{X}$ applied on them.  We generate $50$ trials to get the averaged $\mathrm{FP}/k$, $\mathrm{FN}/k$, and time for each algorithm. For each number of tests, the same $50$ test matrices are used for three different algorithms and for two different noise levels, where $\rho \in \{0, 0.01\}$.

We use Gurobi \cite{GUROBI} as the mathematical optimization solver to implement LP, QP, and Linearized QP, and compare these three decoding algorithms' results using three metrics: the normalized number of false positives ($\mathrm{FP}/k$), the normalized number of false negatives ($\mathrm{FN}/k$), and the computation time.  Note that by ``LP'' we mean the decoder in \eqref{lp_noiseless_intro} (noiseless) or \eqref{lp_noisy_intro} (noisy), which is only based on sparsity and does not use the graph information.  For all continuous relaxations, we form the $\{0,1\}$-valued estimate by rounding to the nearer value.

\subsection{Results}

\begin{figure*}[htbp!]

\begin{minipage}[b]{0.3\linewidth}
  \centering
  \centerline{\includegraphics[width=6.0cm]{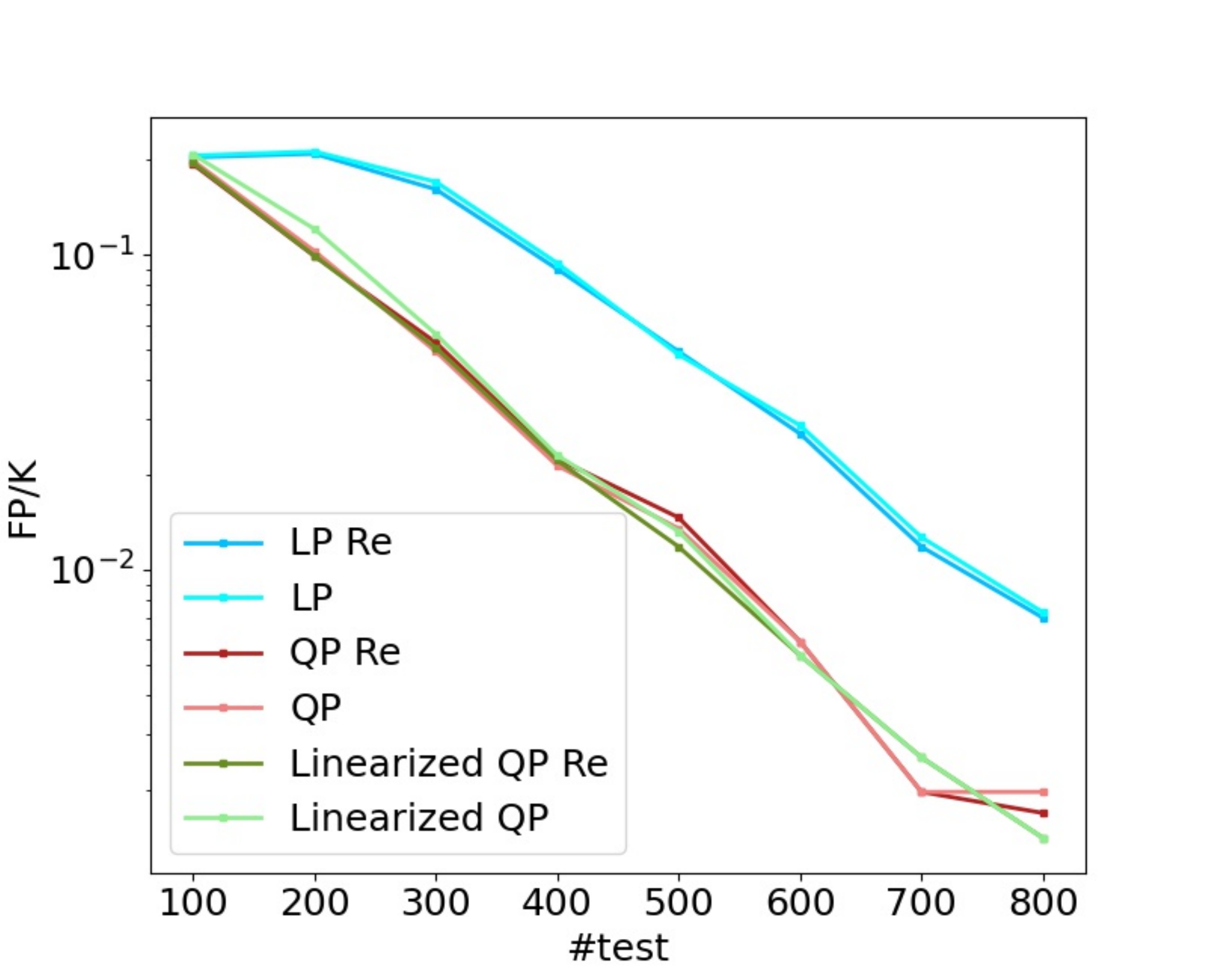}}
\end{minipage}
\hfill
\begin{minipage}[b]{0.3\linewidth}
  \centering
  \centerline{\includegraphics[width=6.0cm]{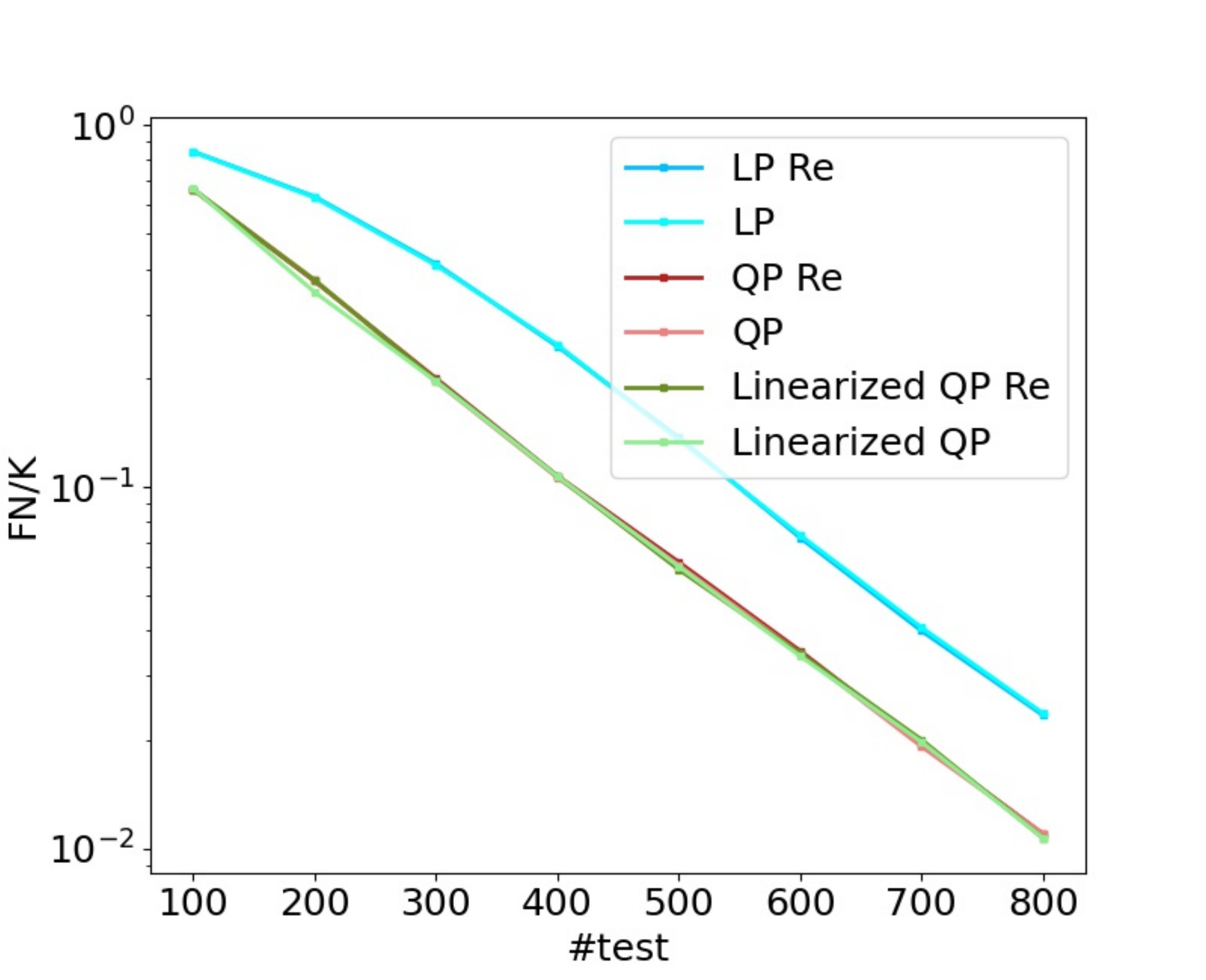}}
\end{minipage}
\hfill
\begin{minipage}[b]{0.3\linewidth}
  \centering
  \centerline{\includegraphics[width=6.0cm]{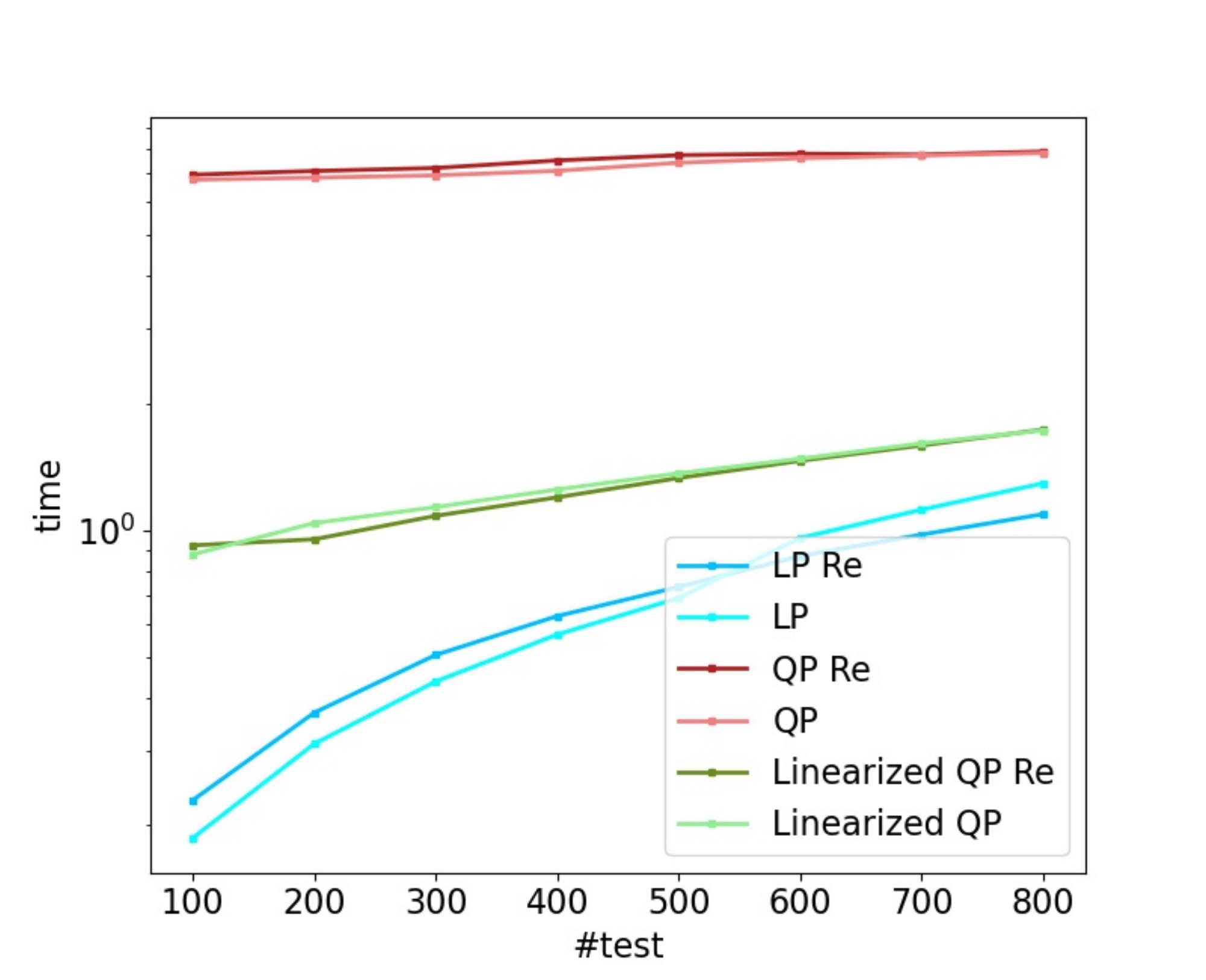}}
\end{minipage}

\begin{minipage}[b]{0.3\linewidth}
  \centering
  \centerline{\includegraphics[width=6.0cm]{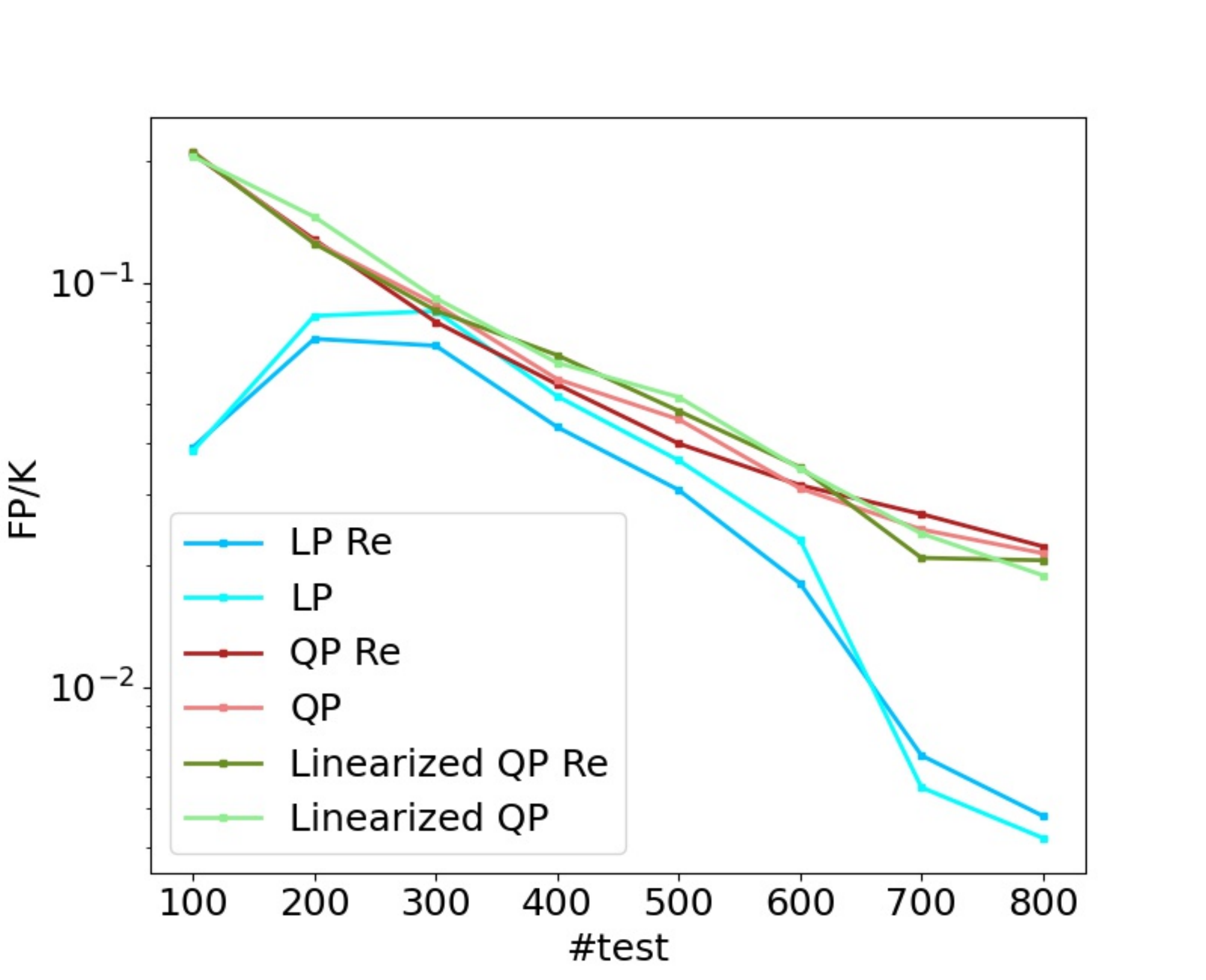}}
  \subfloat{$\mathrm{FP}/k$ (false positive rate)}
\end{minipage}
\hfill
\begin{minipage}[b]{0.3\linewidth}
  \centering
  \centerline{\includegraphics[width=6.0cm]{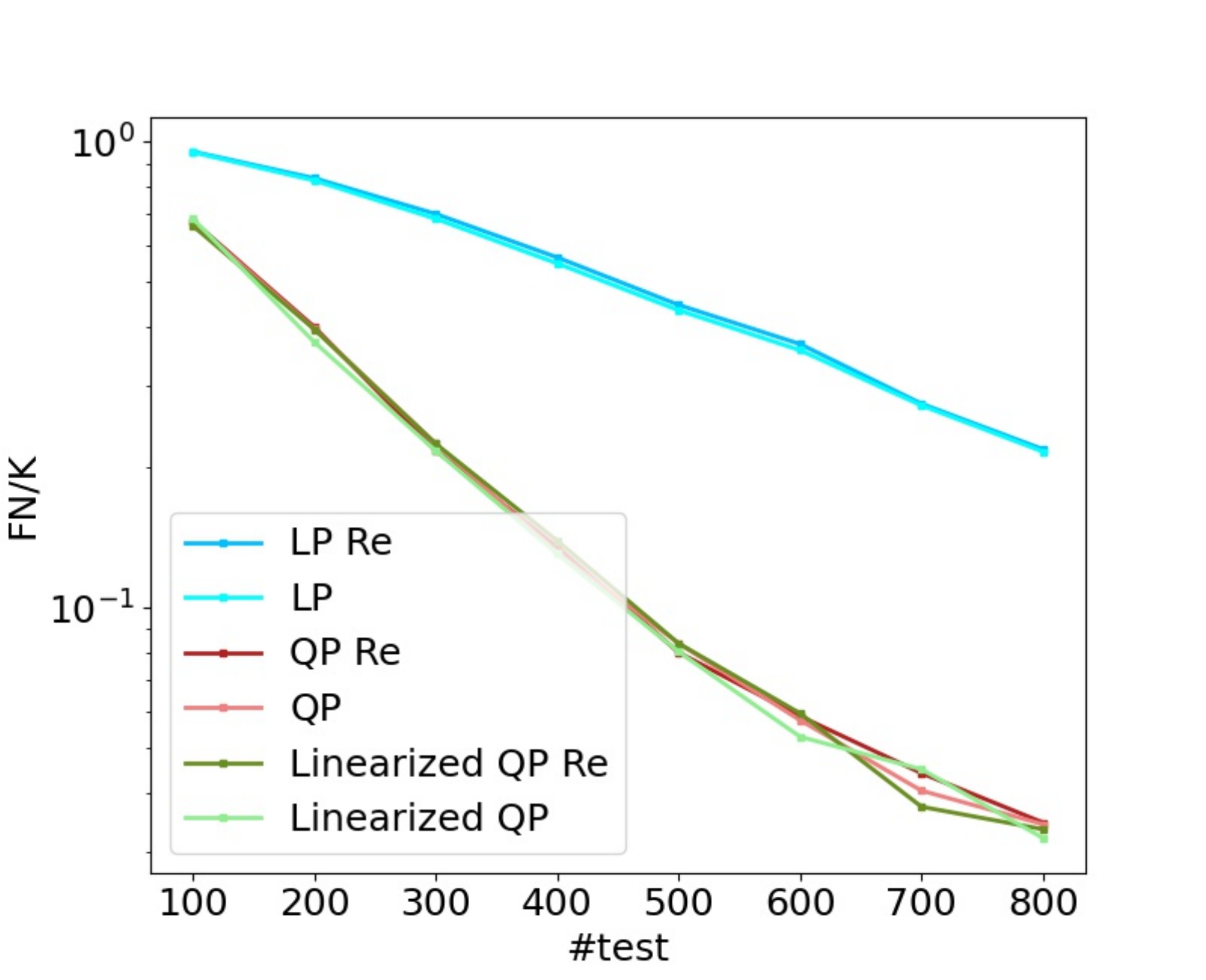}}
  \subfloat{$\mathrm{FN}/k$ (false negative rate)}
\end{minipage}
\hfill
\begin{minipage}[b]{0.3\linewidth}
  \centering
  \centerline{\includegraphics[width=6.0cm]{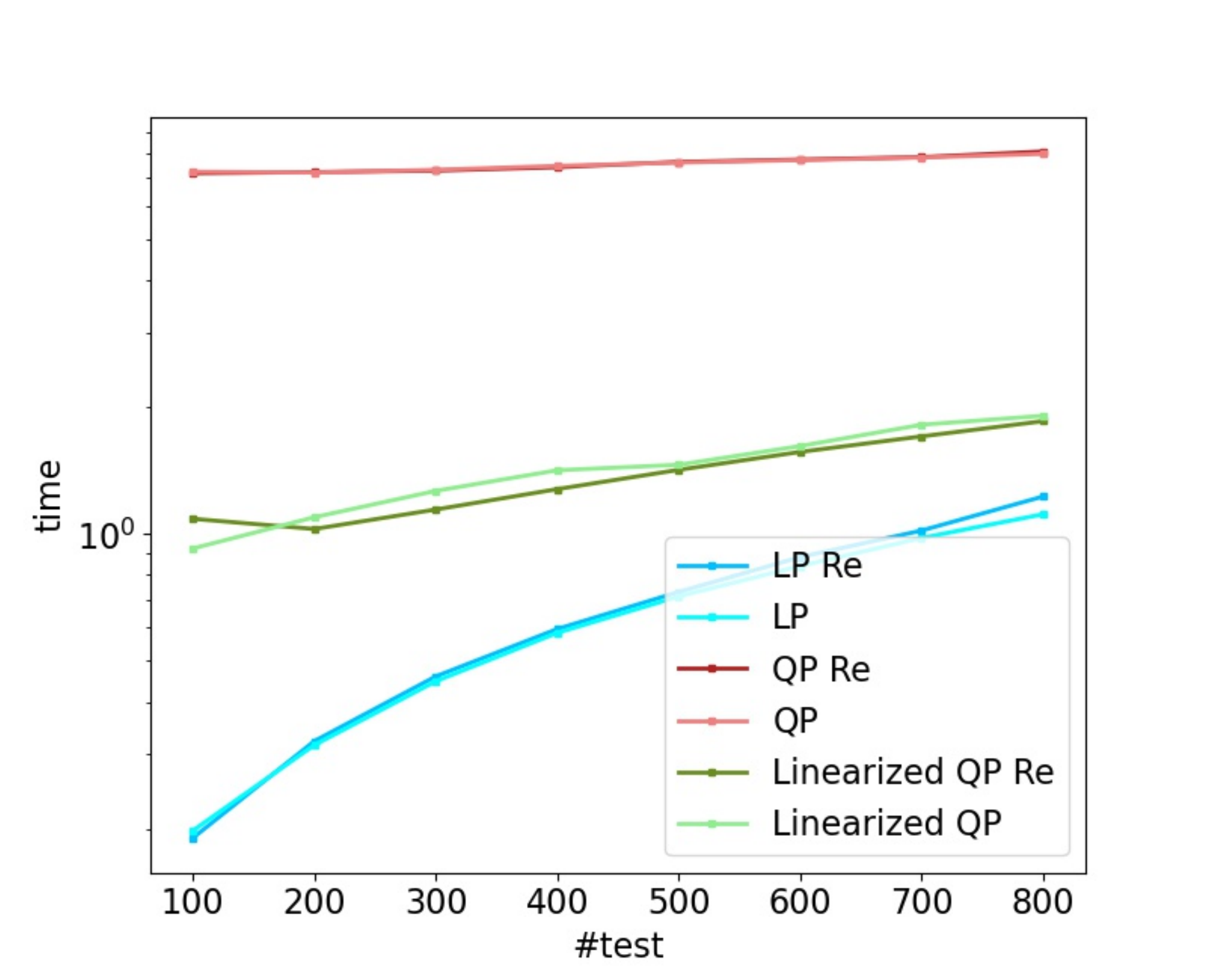}}
  \subfloat{computation time}
\end{minipage}

\caption{Results on grid graph with $\rho=0$ (top) and $\rho=0.01$ (bottom).}
\label{relaxed_grid}
\end{figure*}

\begin{figure*}[htbp!]

\begin{minipage}[b]{0.3\linewidth}
  \centering
  \centerline{\includegraphics[width=6.0cm]{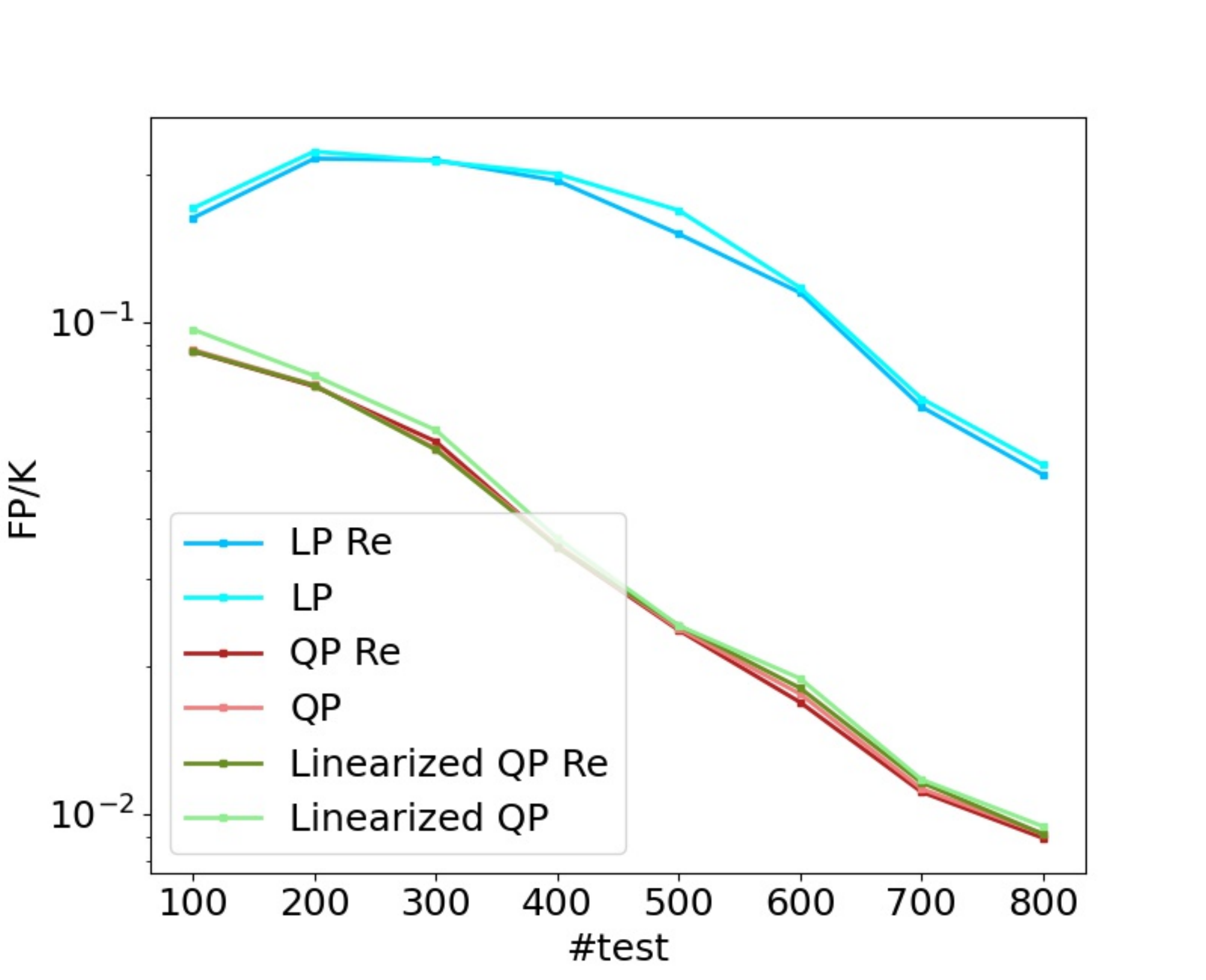}}
\end{minipage}
\hfill
\begin{minipage}[b]{0.3\linewidth}
  \centering
  \centerline{\includegraphics[width=6.0cm]{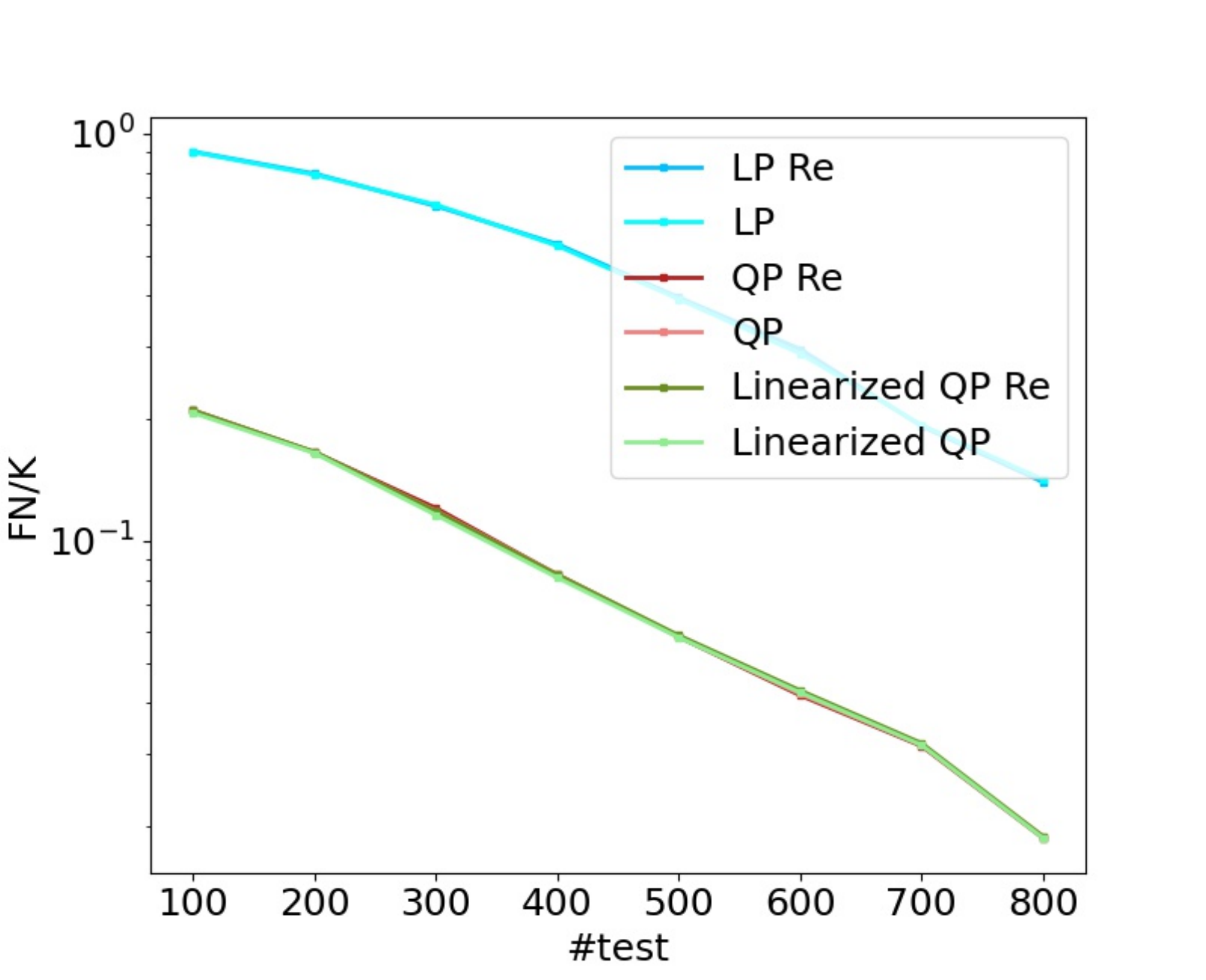}}
\end{minipage}
\hfill
\begin{minipage}[b]{0.3\linewidth}
  \centering
  \centerline{\includegraphics[width=6.0cm]{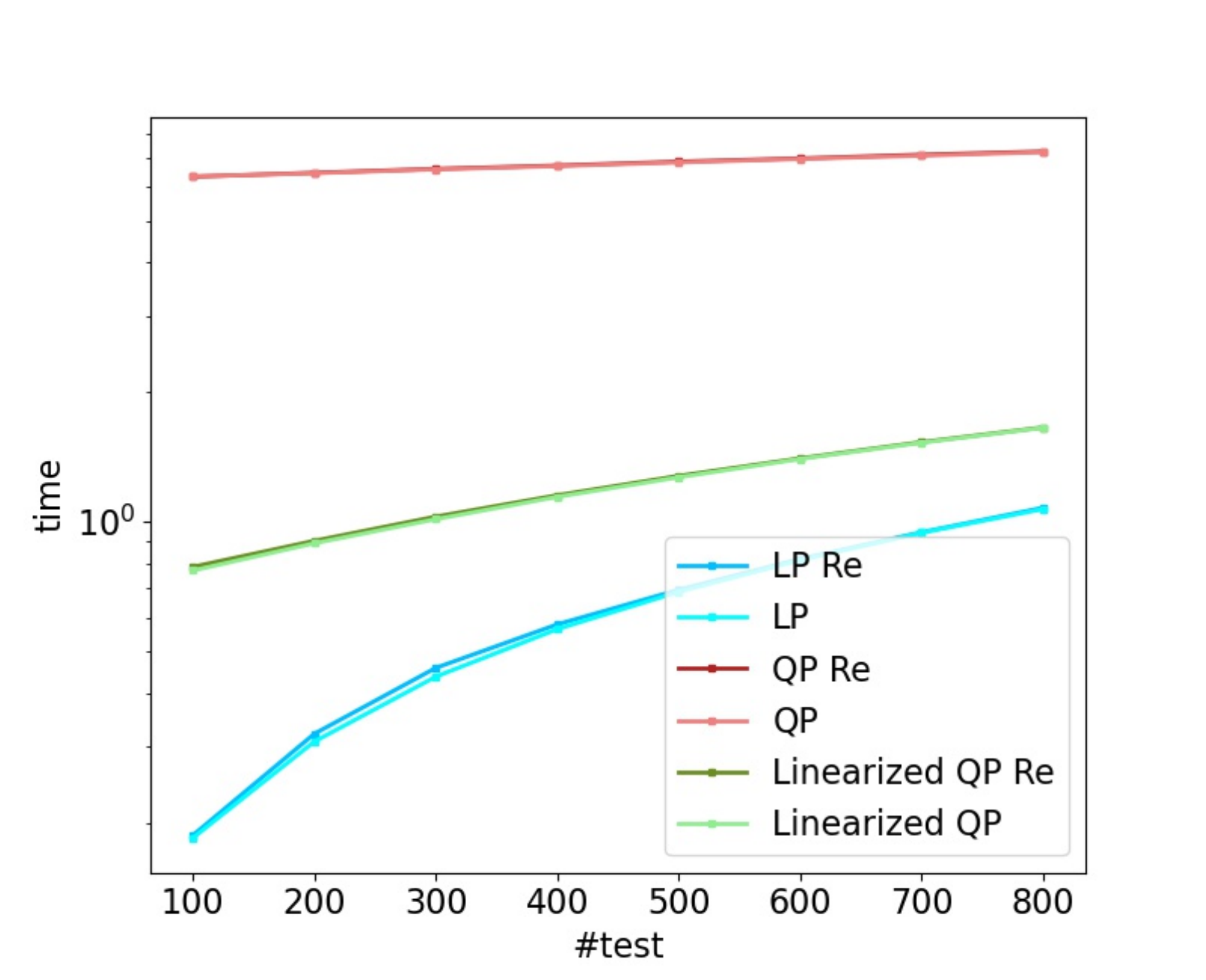}}
\end{minipage}

\begin{minipage}[b]{0.3\linewidth}
  \centering
  \centerline{\includegraphics[width=6.0cm]{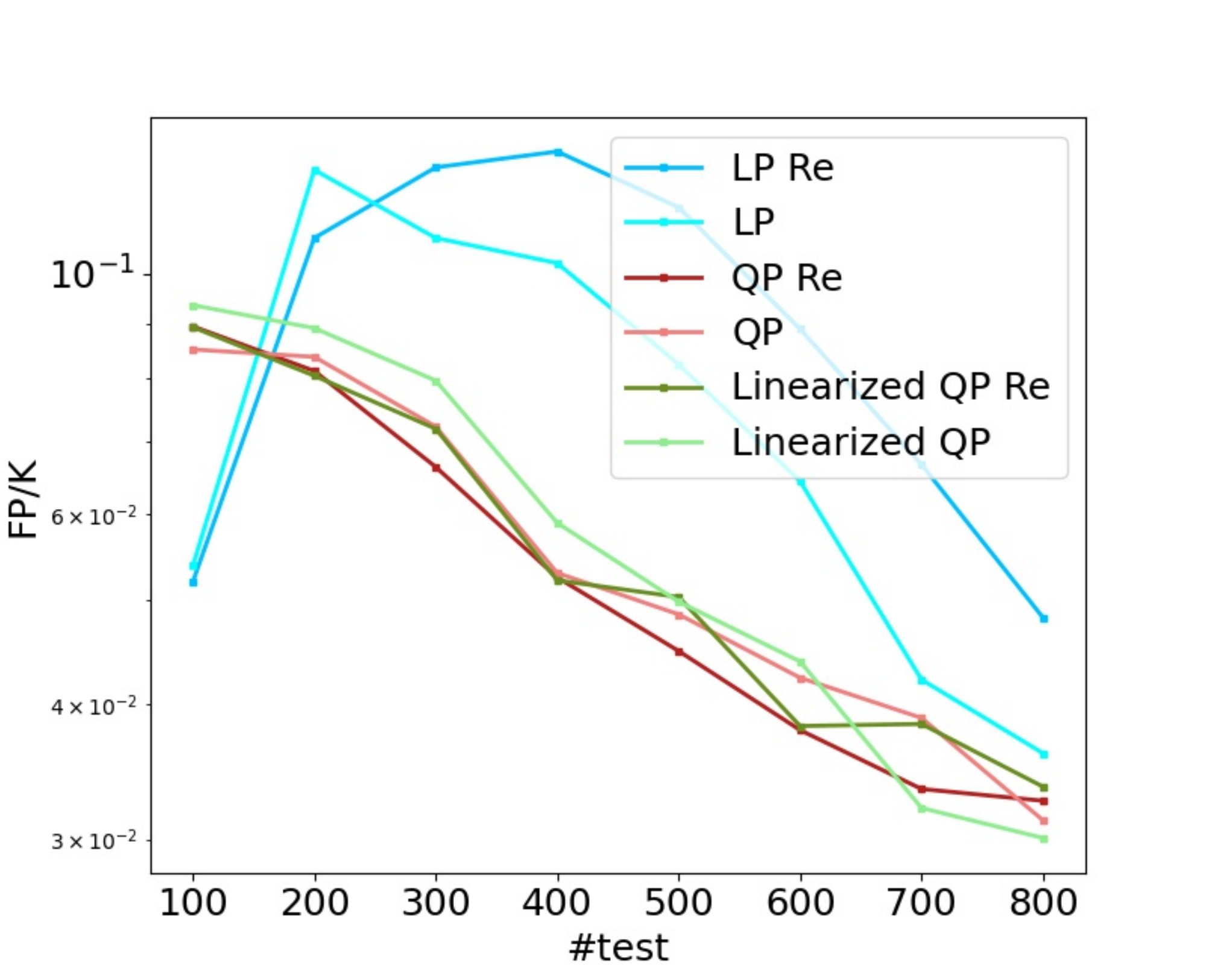}}
  \subfloat{$\mathrm{FP}/k$ (false positive rate)}
\end{minipage}
\hfill
\begin{minipage}[b]{0.3\linewidth}
  \centering
  \centerline{\includegraphics[width=6.0cm]{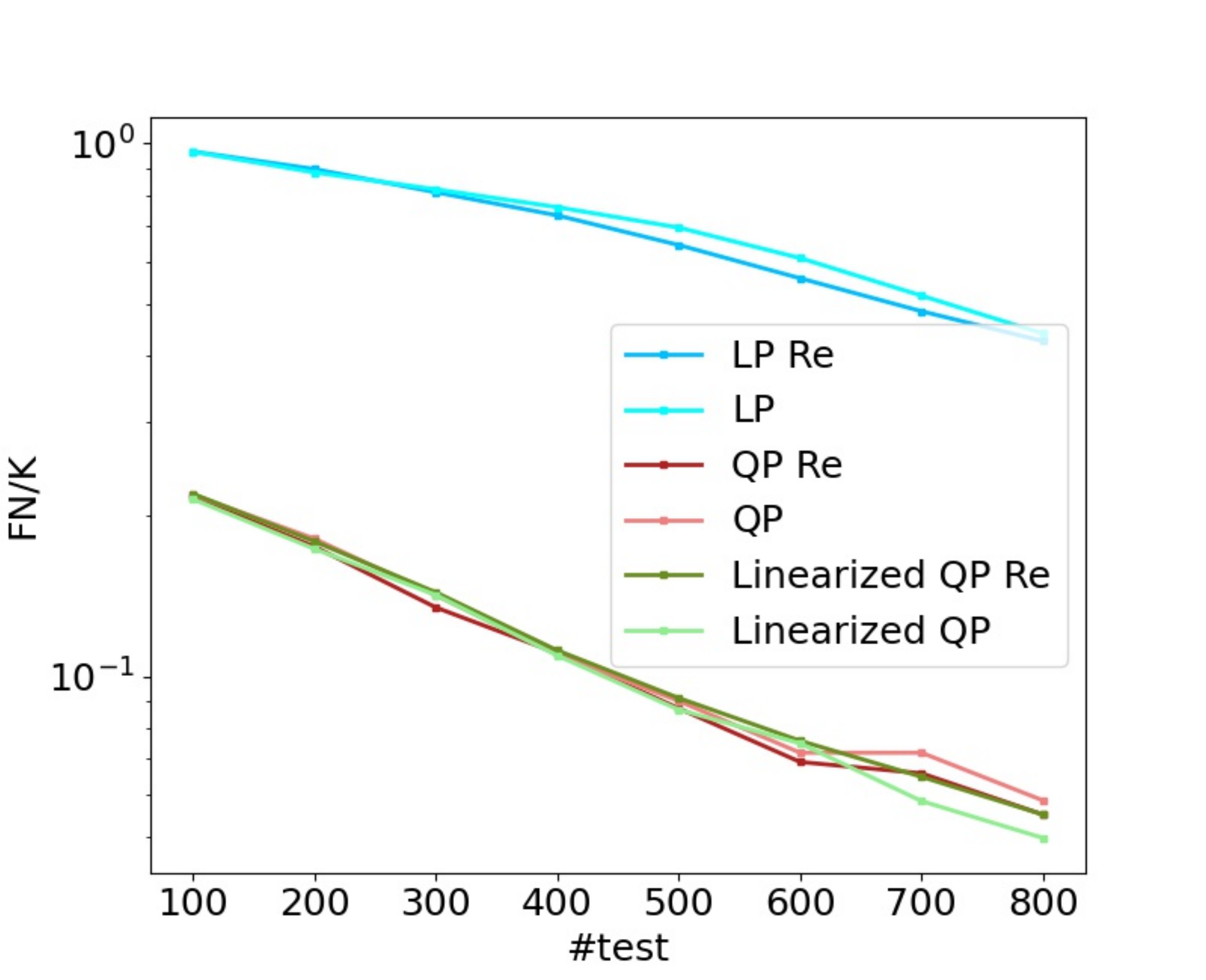}}
  \subfloat{$\mathrm{FN}/k$ (false negative rate)}
\end{minipage}
\hfill
\begin{minipage}[b]{0.3\linewidth}
  \centering
  \centerline{\includegraphics[width=6.0cm]{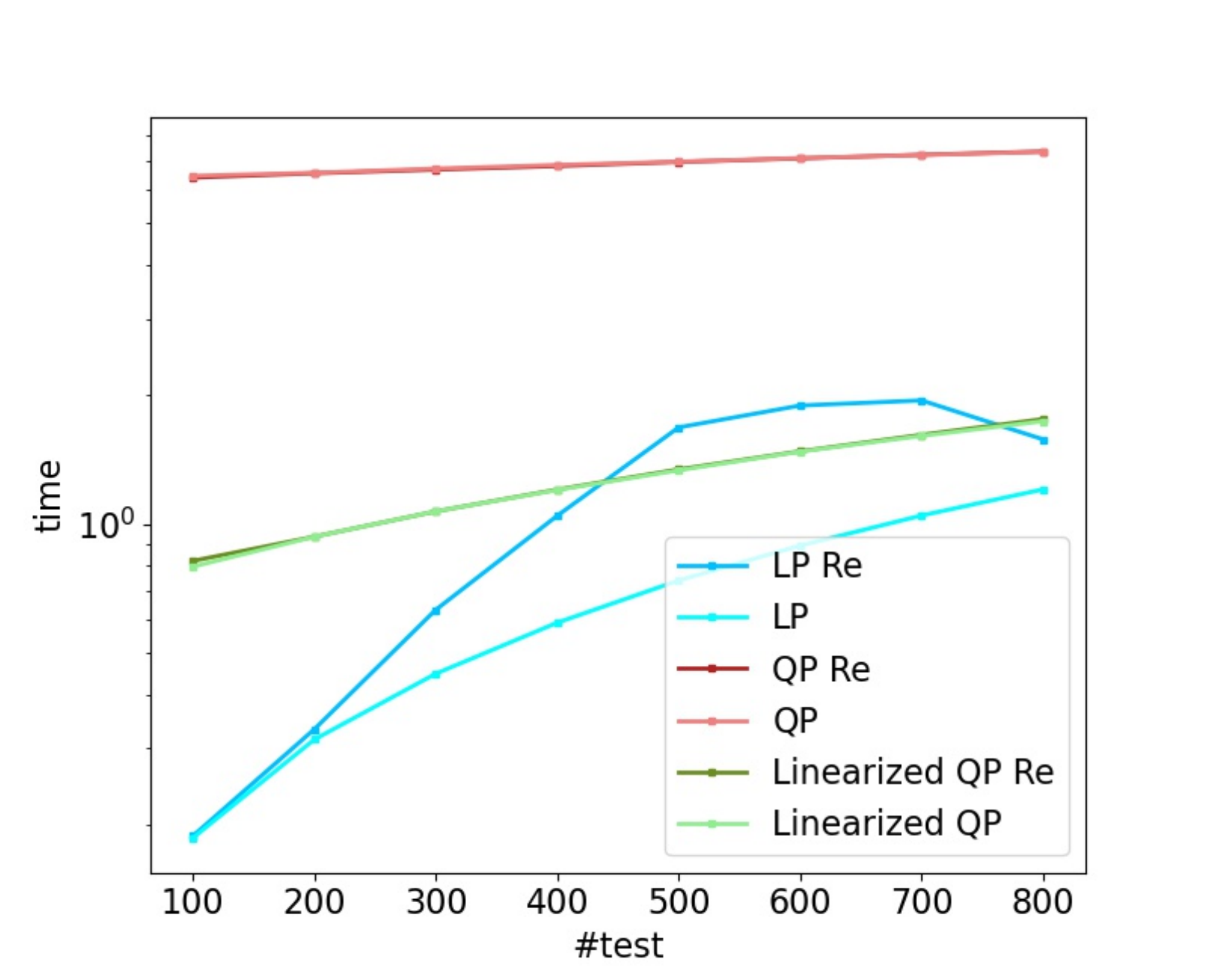}}
  \subfloat{computation time}
\end{minipage}

\caption{Results on block graph with $\rho=0$ (top) and $\rho=0.01$ (bottom).}
\label{relaxed_block}
\end{figure*}

We present the results in Figures \ref{relaxed_grid} (grid) and \ref{relaxed_block} (block).  We show the error rates $\mathrm{FP}/k$ (left column) and $\mathrm{FN}/k$ (middle column), and the computation time (right column), both for the non-relaxed and relaxed (``Re'') variants, and in both noiseless and noisy settings.

% {\bf \color{blue} [TODO (Yang): Can have one or two plots demonstrating how binary vs.~relaxed behave similarly in terms of FP and FN.]}
% {\bf \color{red} [To discuss]} 
% \olive{TODO: Add new computation time graphs only for more complicated graph (e.g., grid/block size as 72/18)}

% {\bf \color{blue} [TODO (Yang): Make the plots more compact, using a $2 \times 3$ structure (top row for noiseless, bottom row for noisy)]} \olive{Done.}

We find that for both graphs and for both noise levels, the recovery performance of QP improve on that of LP, and often significantly so (note that we use a logarithmic y-scale).  The only exception is that when $\rho=0.01$, LP tends to generate fewer FPs, possibly due to there being too few tests to reliably mark many items as positive (hence the very high FN). These results support the idea that by incorporating the Ising model as prior information, we are able to attain more accurate recovery with fewer tests.  Moreover, although QP is naturally slower than LP, we still manage to reduce its computation time to a large extent by utilizing Linearized QP, which shows almost identical recovery performance as that of QP.  Therefore, we believe that Linearized QP provides a good solution for practical use.      

Perhaps surprisingly, in Figures \ref{relaxed_grid} and \ref{relaxed_block}, the computation times are similar for the non-relaxed and relaxed variants, except for LP in the final sub-figure of Figure \ref{relaxed_block}, and for certain other cases with few tests (e.g., 100).  However, as we exemplify using larger $72 \times 72$ graphs in Figure \ref{fig:comp_large} in the supplementary material, the gap can be significant for Linearized QP as well.  Figure \ref{fig:comp_large} also highlights the fact that both relaxed and non-relaxed QPs can be slow to solve (as discussed previously), whereas the relaxed Linearized QP can be much faster than the non-relaxed variant. 

Overall, it is difficult to pinpoint the precise factors that make these methods faster or slower, particularly when the time may not even be monotonically increasing with respect to the number of tests (this was also observed in \cite{ciampiconi2020maxsat}, with the intuition being that more tests can help to rule out suboptimal parts of the search space faster).  Nevertheless, our results suggest that both the non-relaxed and relaxed variants can be practically feasible, with the latter typically being preferable when computation is a bottleneck, and with (relaxed) Linearized QP being preferable to QP for large problem sizes.

\subsection{Sensitivity to Model Mismatch}

In most practical scenarios, one cannot know the graph $G$ and edge strength $\lambda$ perfectly (e.g., they may only be approximately learned from historical data).  Here we consider the effect of mismatch with respect to each of these, albeit without seeking to be comprehensive.

\textbf{Graph mismatch:}
In this experiment, we consider varying degrees of mismatch in the graph estimate.  To do so, starting with the true graph, we remove a certain fraction of its edges uniformly at random, and simultaneously add the same number of edges uniformly chosen from the original non-edges.  Clearly, the higher the fraction, the `more mismatched' the graph estimate is.

We consider two separate scenarios: (i) noiseless tests, the grid structure, and 300 tests, and (ii) noisy tests, the block structure, and 500 tests.  In both cases, we vary the fraction of removed (and re-added) edges from $0$ to $\frac{1}{2}$.  All other settings are the same as before.

\begin{figure*}[htbp!]
\centering
\begin{minipage}[b]{0.4\linewidth}
  \centering
  \centerline{\includegraphics[width=6.5cm]{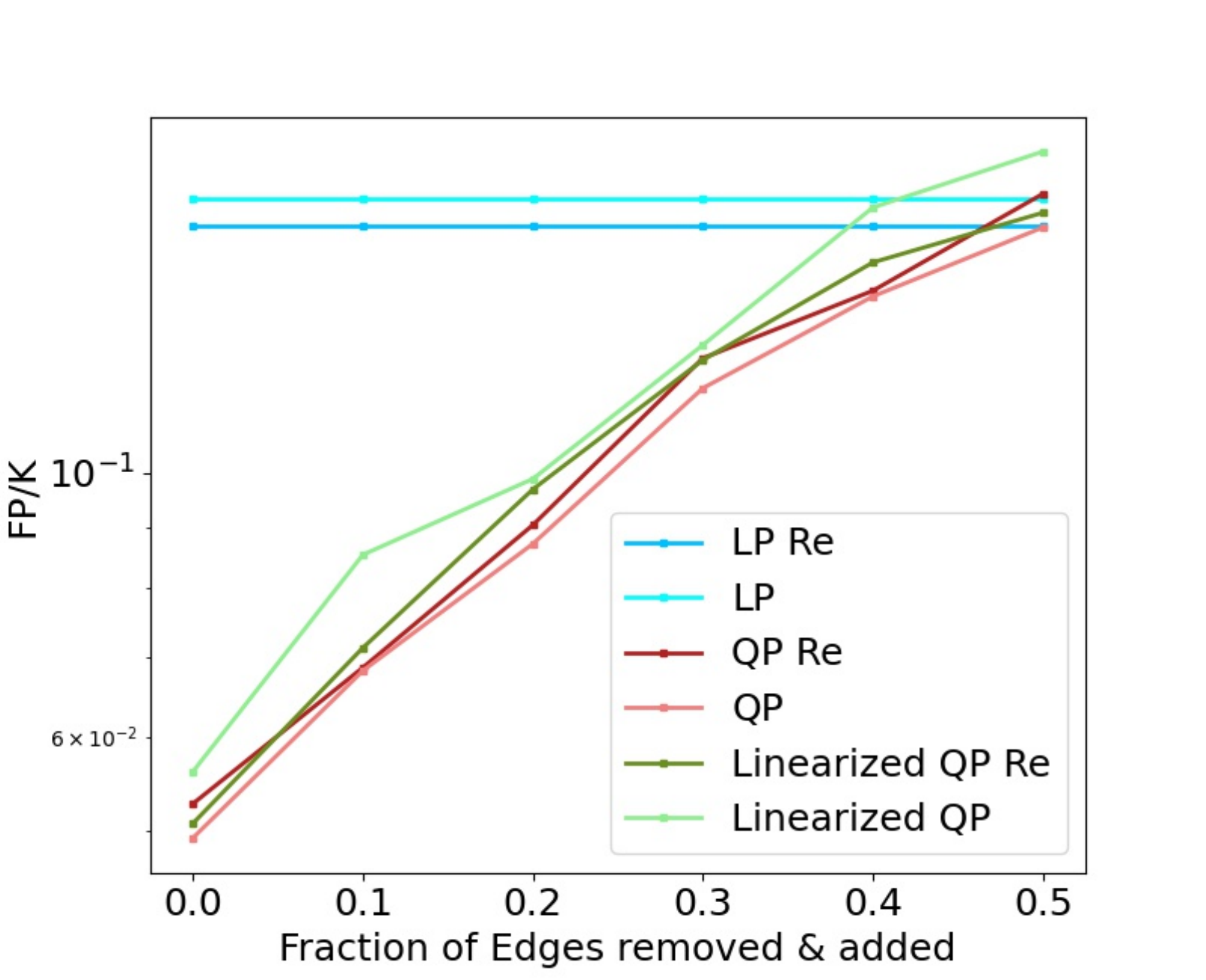}}
\end{minipage}
\quad
\begin{minipage}[b]{0.4\linewidth}
  \centering
  \centerline{\includegraphics[width=6.5cm]{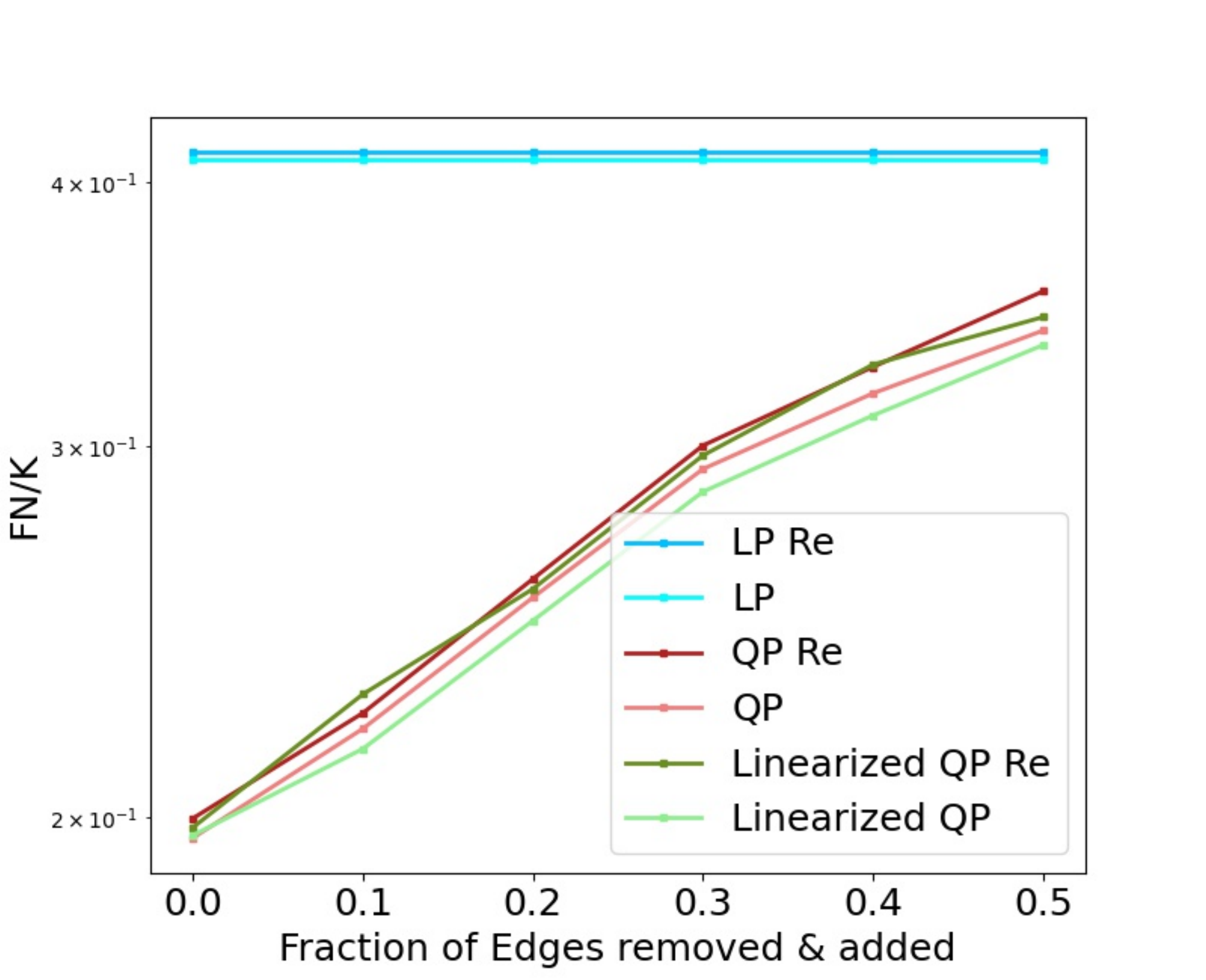}}
\end{minipage}

\begin{minipage}[b]{0.4\linewidth}
  \centering
  \centerline{\includegraphics[width=6.5cm]{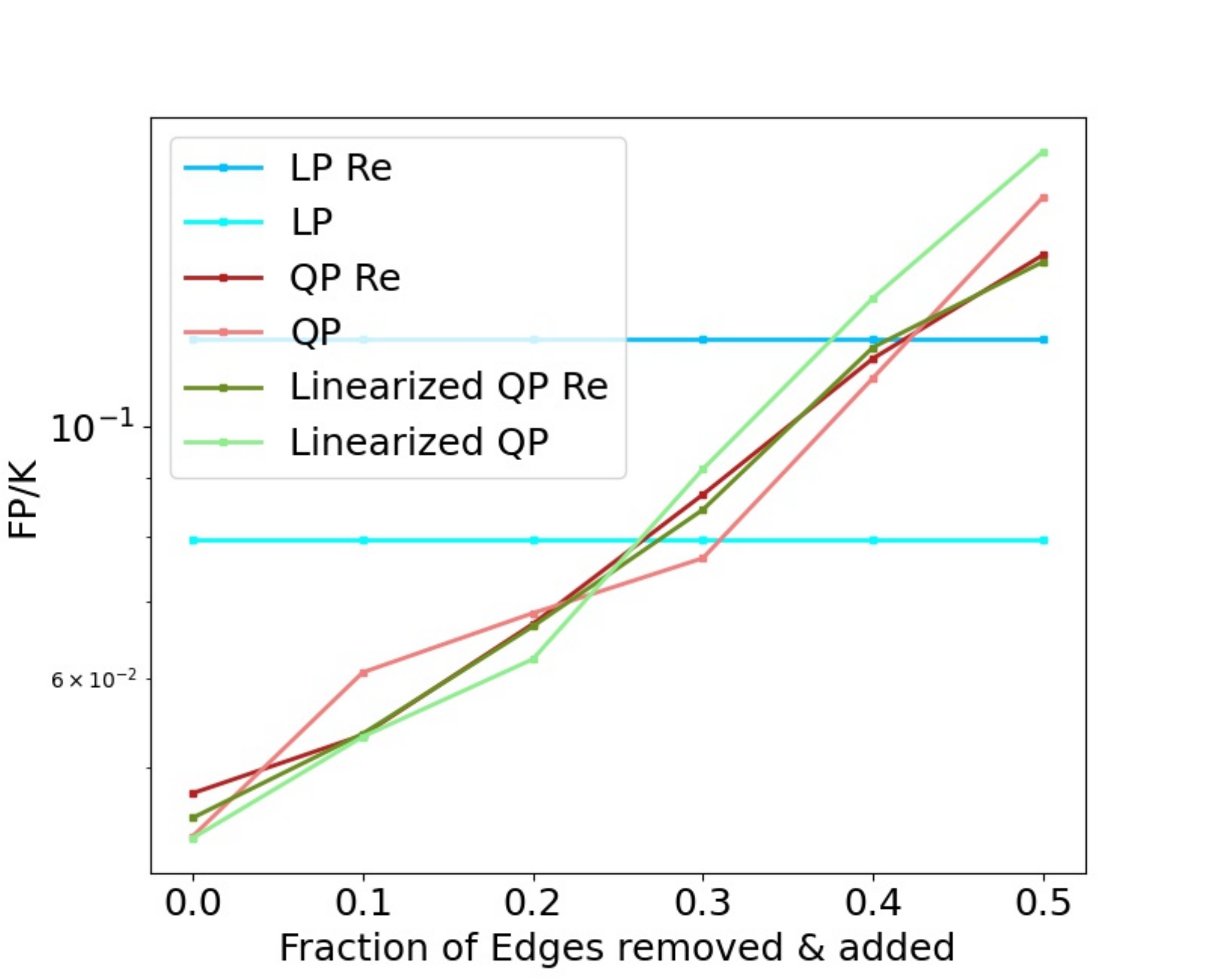}}
  \subfloat{$\mathrm{FP}/k$ (false positive rate)}
\end{minipage}
\quad
\begin{minipage}[b]{0.4\linewidth}
  \centering
  \centerline{\includegraphics[width=6.5cm]{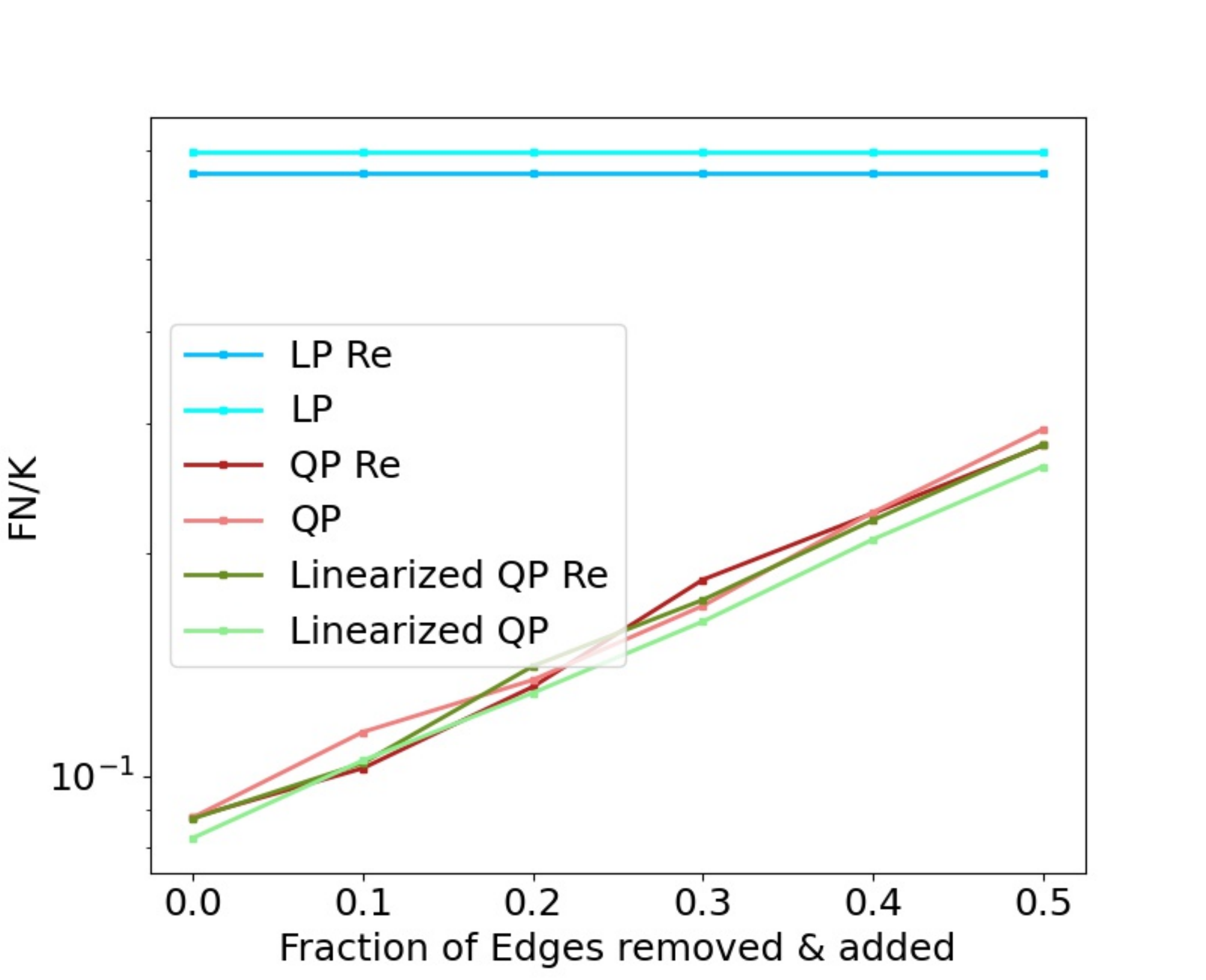}}
  \subfloat{$\mathrm{FN}/k$ (false negative rate)}
\end{minipage}

\caption{Results under graph mismatch: (Top) noiseless, grid graph, 300 tests, (Bottom) noisy, block graph, 500 tests.}
\label{edge}
\end{figure*}

The results are presented in Figure \ref{edge}.  Naturally, for all algorithms that use the prior information, the performance worsens as the degree of mismatch increases.  Nevertheless, we observe graceful degradation rather than a sharp drop in performance, with the FN curves lying entirely below those of the LP approach, and the FP curves doing the same until the degree mismatch becomes considerable.

\textbf{Parameter mismatch:}
Here we consider the effect of varying the edge strength $\lambda$ that the algorithm assumes, which may now differ from the true value.  We consider two scenarios: (i) noiseless, block graph, 300 tests, and (ii) noisy, grid graph, 500 tests.  We vary the algorithm's value of $\lambda$ from $0.01$ to $2.0$, when the true value is $0.5$ (grid) or $0.6$ (block).

\begin{figure*}[htbp!]
\centering
\begin{minipage}[b]{0.4\linewidth}
  \centering
  \centerline{\includegraphics[width=6.5cm]{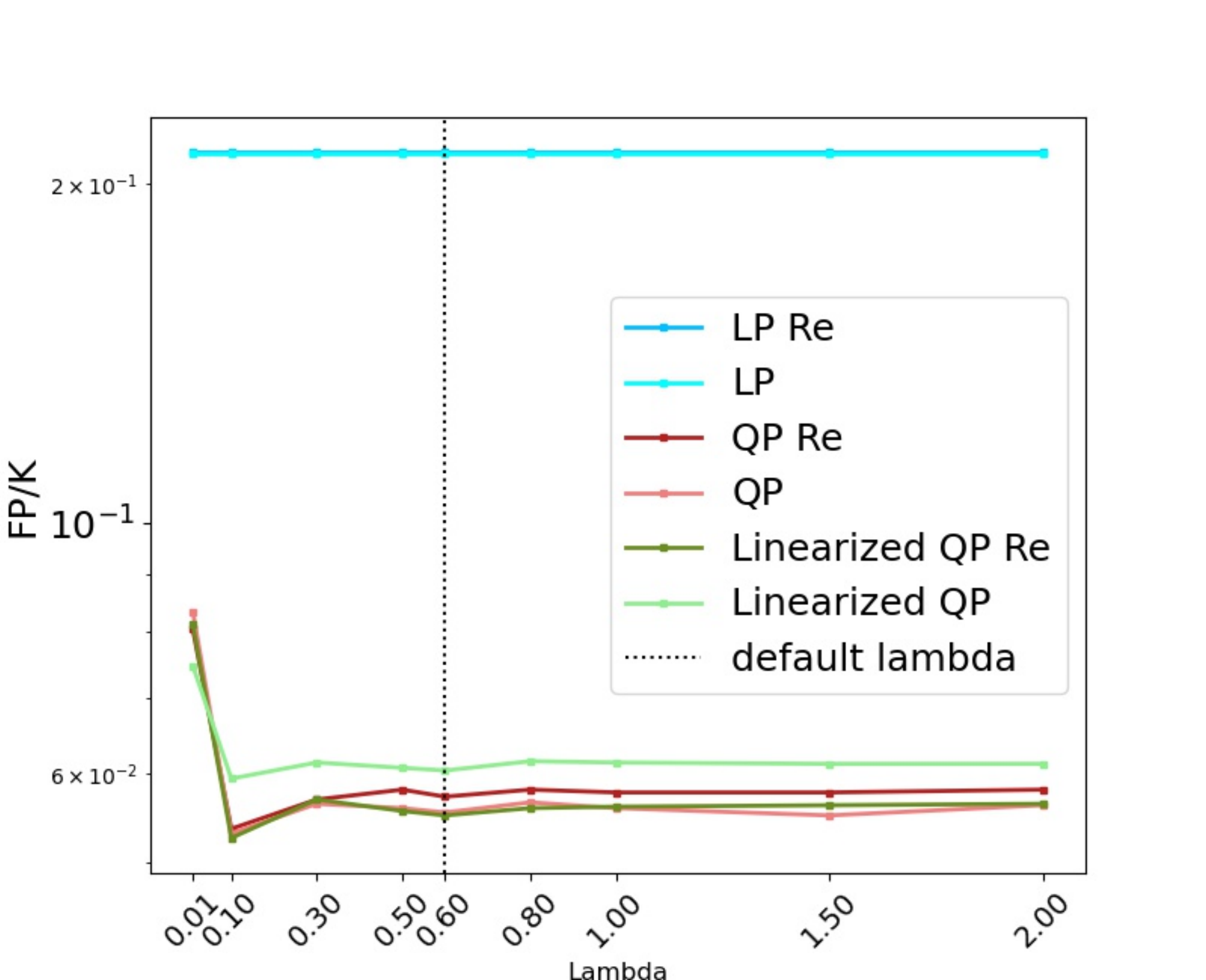}}
\end{minipage}
\quad
\begin{minipage}[b]{0.4\linewidth}
  \centering
  \centerline{\includegraphics[width=6.5cm]{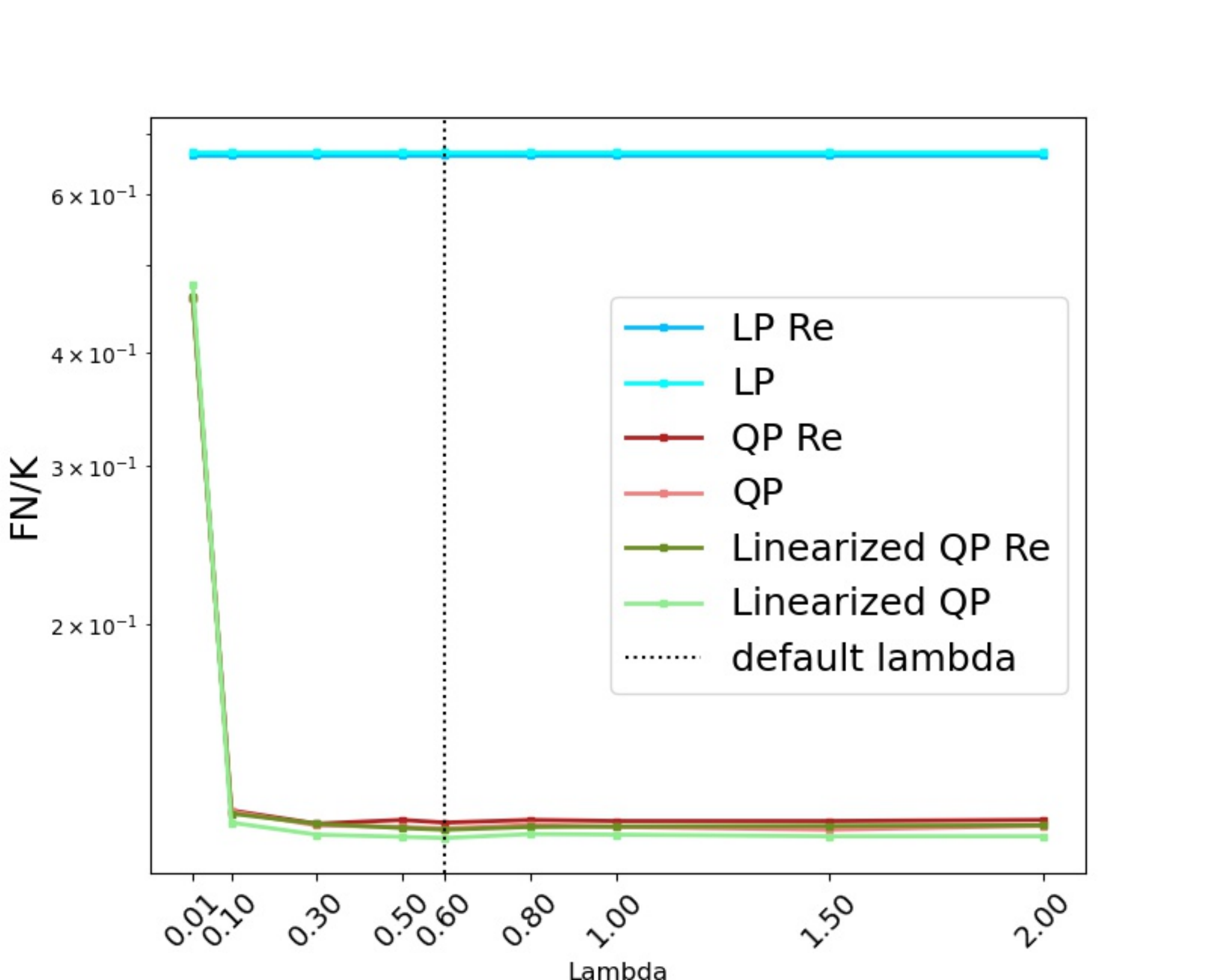}}
\end{minipage}

\begin{minipage}[b]{0.4\linewidth}
  \centering
  \centerline{\includegraphics[width=6.5cm]{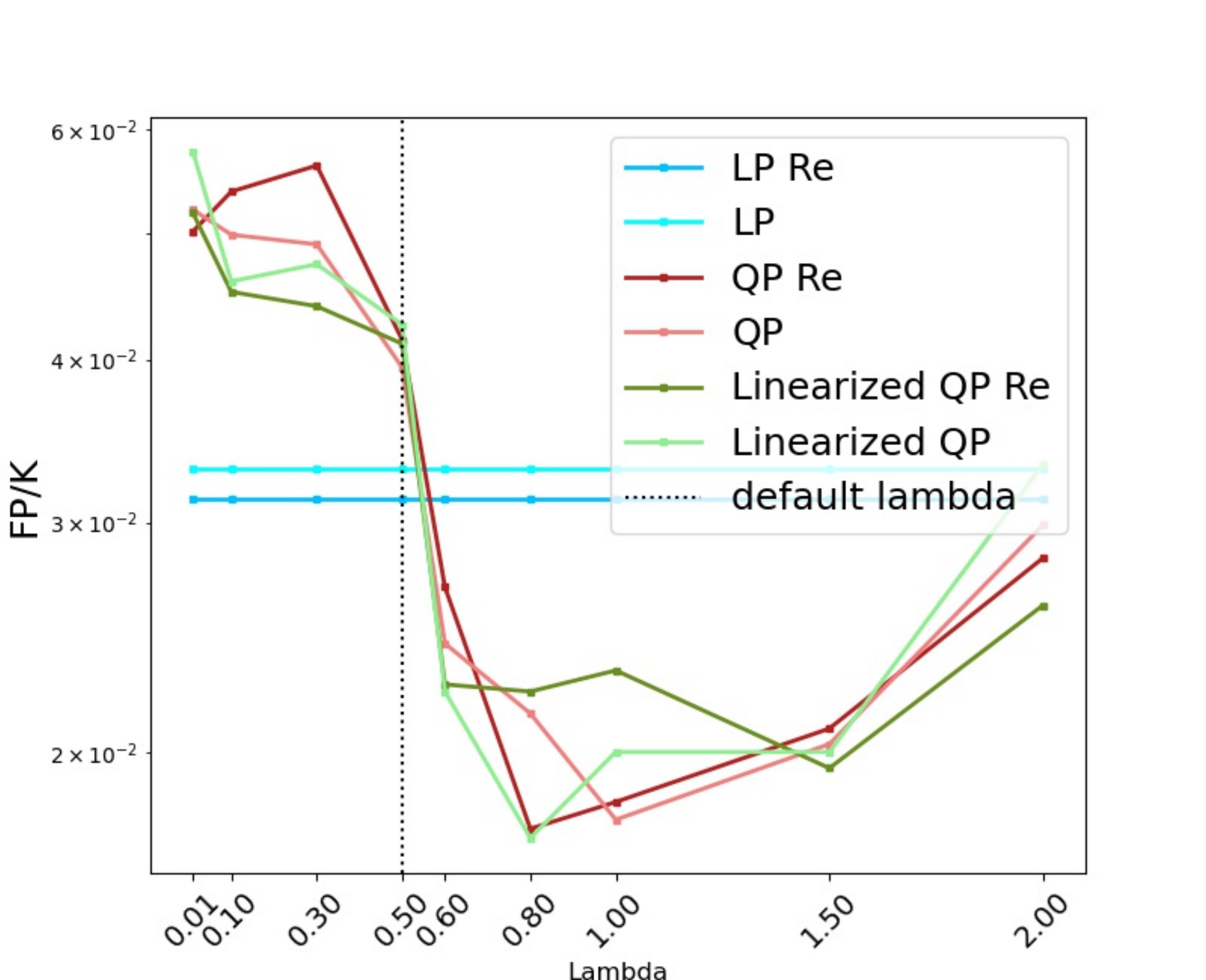}}
  \subfloat{$\mathrm{FP}/k$ (false positive rate)}
\end{minipage}
\quad
\begin{minipage}[b]{0.4\linewidth}
  \centering
  \centerline{\includegraphics[width=6.5cm]{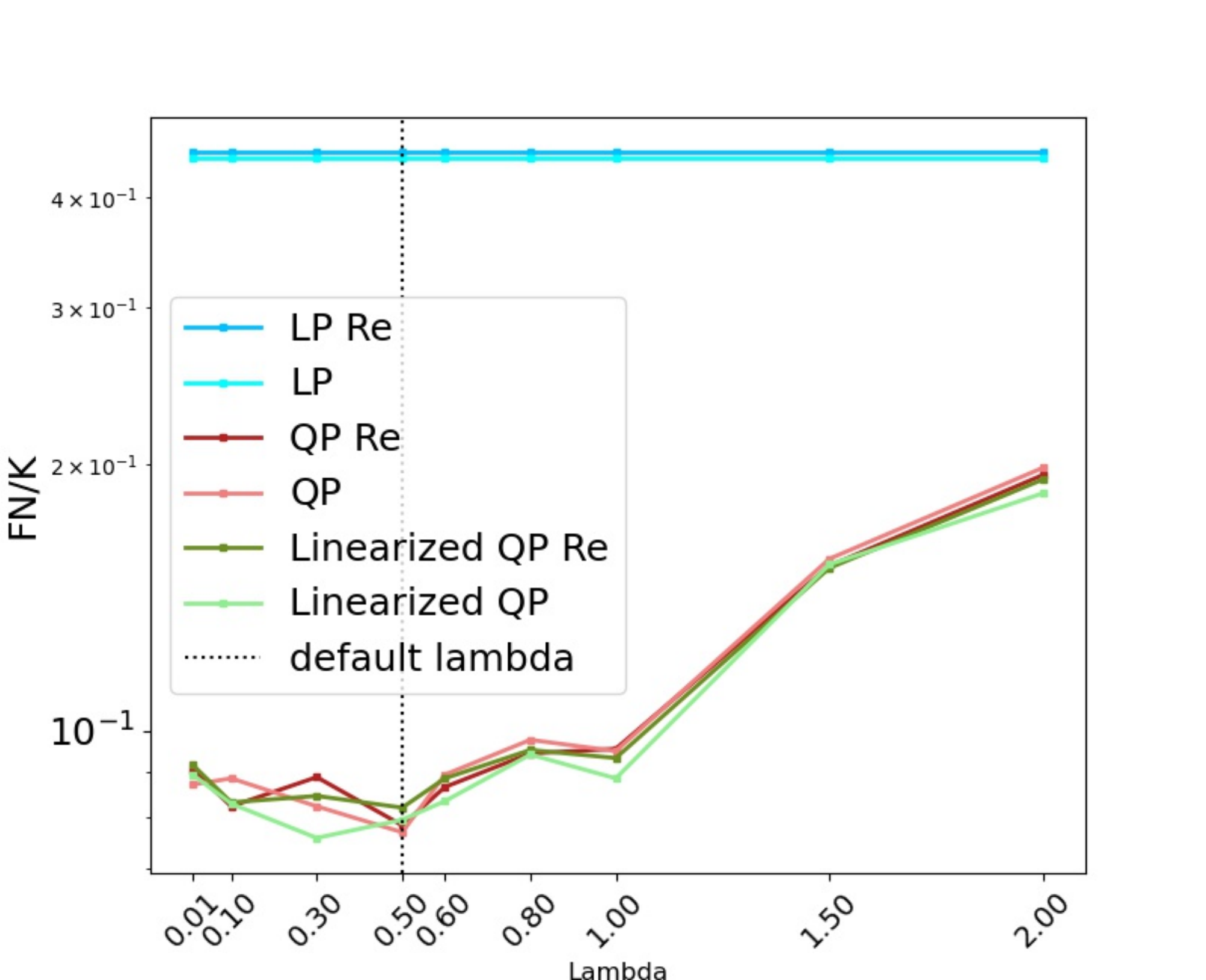}}
  \subfloat{$\mathrm{FN}/k$ (false negative rate)}
\end{minipage}

\caption{Results under $\lambda$ mismatch: (Top) noiseless, block graph, 300 tests, (Bottom) noisy, grid graph, 500 tests.}
\label{lambda}
\end{figure*}

The results are presented in Figure \ref{lambda}.  In the noiseless case, we observe a remarkably high degree of robustness, with the performance only degrading significantly when $\lambda$ is severely underestimated.  In the noisy case, the degradation is much more obvious, but the behavior is still generally robust (note that the $y$-axis is for ${\rm FP}/k$ is from around $0.02$ to $0.06$).  The FN increases slowly as the algorithm's $\lambda$ increases, whereas the FP exhibits a distinctive ``U''-shape; having $\lambda$ slightly above its true value can even reduce the FP (though slightly increasing the FN).

\subsection{Example on Real-World Data}

While the grid and block graphs exhibits certain properties of interest (e.g., clustering structure), there are naturally limitations to what they can capture.  To address this limitation, here we consider a graph created from real-world data.  Specifically, we consider an `ego-Facebook' dataset \cite{leskovec2012learning} collected from $4039$ survey Facebook users (nodes) consisting of $88234$ `social interactions' (edges), and we reduce its size by randomly selecting 500 nodes and extracting the corresponding sub-network.

\begin{figure*}
\centering
\begin{minipage}[b]{0.4\linewidth}
  \centering
  \centerline{\includegraphics[width=6cm]{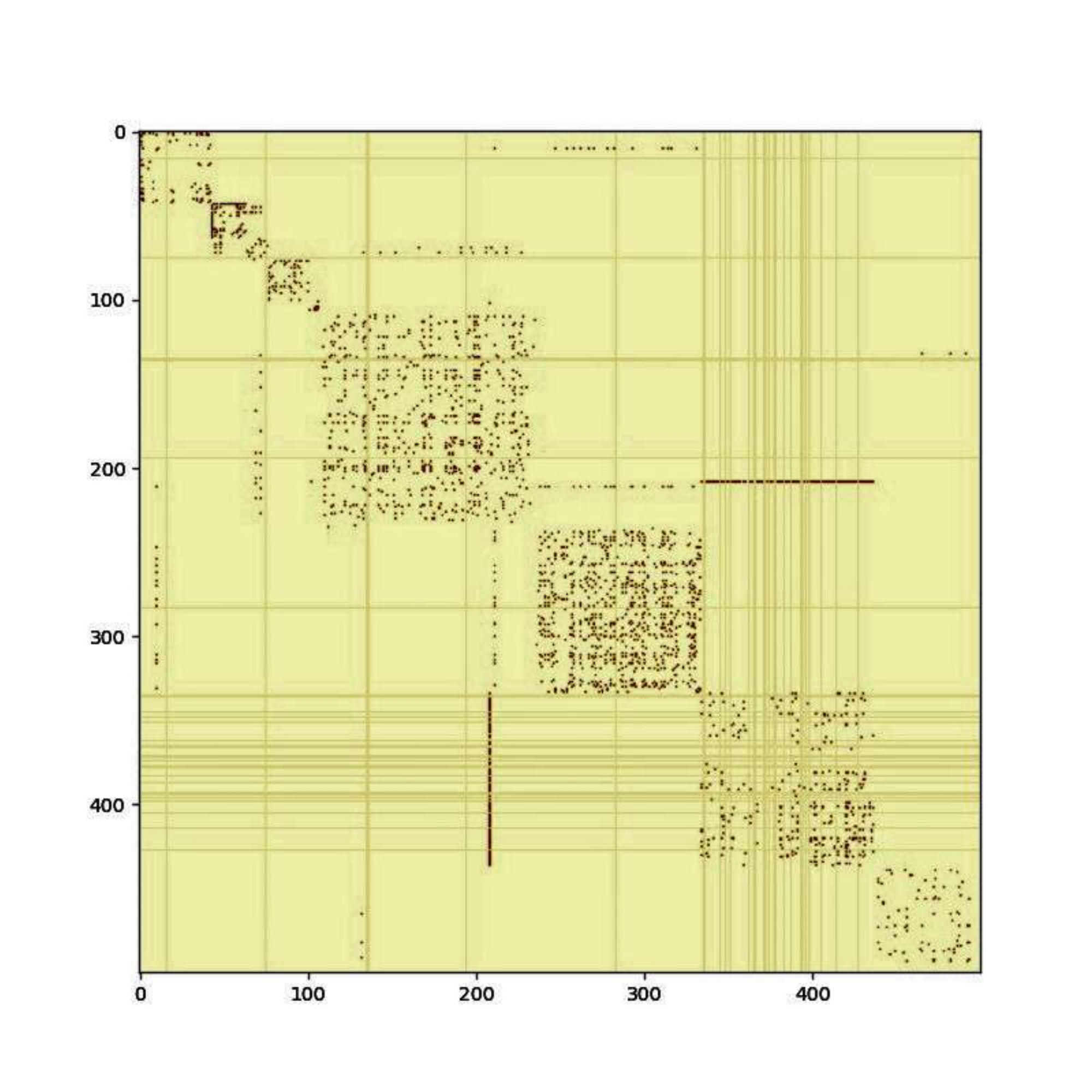}}
  \centerline{Edges (dark dots) and infections (straight lines)}
%   \centerline{($500 \times 500$)}\medskip
\end{minipage}
\quad
\begin{minipage}[b]{0.4\linewidth}
  \centering
  \centerline{\includegraphics[width=7.2cm]{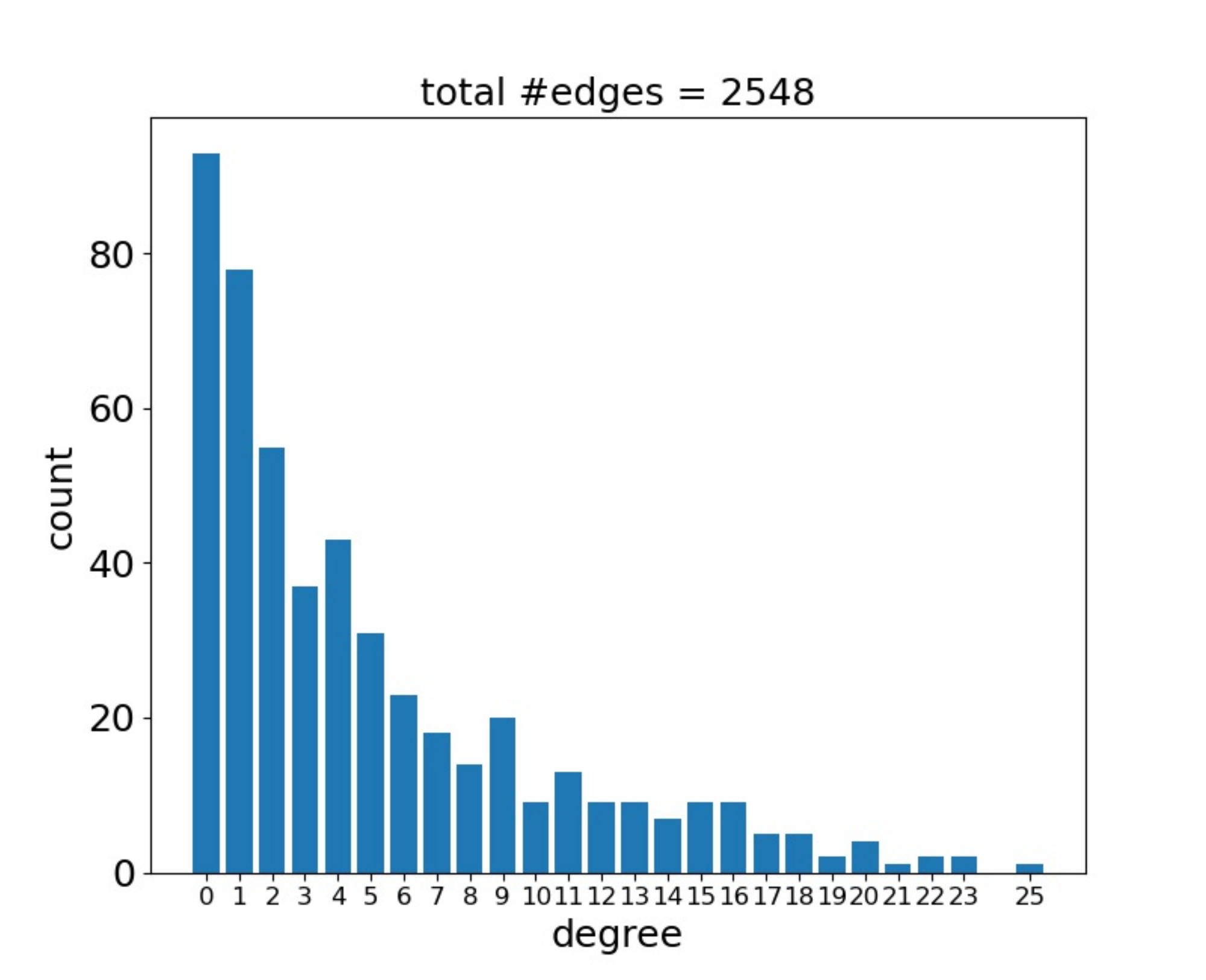}}
  \centerline{Degree distribution of the nodes}
%   \centerline{($4 \times 4$ blocks of size $18 \times 18$ each)}\medskip
\end{minipage}
\caption{Visualizations of the graph created from Facebook data.  In the histogram on the right, one outlier node is omitted (degree=$102$, count=$1$) for compactness.} \label{fig:facebook}
\end{figure*}

In Figure \ref{fig:facebook}, we visualize the resulting adjacency matrix (Left) and degree distribution (Right).  While there is still some block structure, we observe that this graph is significantly more complex than those above, with several ``clusters'' of varying sizes, and a number of ``heavily connected'' nodes connected to multiple clusters. 
For the Ising model, we choose $\lambda$ and $\phi$ as $1.5$ and $1.0$, respectively.  A resulting defective set from Gibbs sampling is also depicted in Figure \ref{fig:facebook} (Left) using vertical and horizontal lines.

% Apart from the grid and block structure, we are also interested in the decoding performance in more complex networks such as social graphs. As social network from Facebook app is a great real-world example, , and perform some data pre-processing steps as follows: 1) Randomly select a subset (size=$500$) of the nodes and extract the sub-network that only contains these picked nodes to form a `Facebook graph'; 2) Plot the degree distribution of the obtained Facebook graph as shown in Figure \ref{fig:facebook} right (where we don't show an outlier point (degree=$102$, count=$1$) due to the limited space); 3) Visualize the symmetrical adjacency matrix of the edges (black dots) of the Facebook graph and the defectives samples (yellow lines) generated by Gibbs sampling under the Ising model (where parameters $\lambda$ and $\phi$ are manually tuned as $1.5$ and $1.0$, respectively) in Figure \ref{fig:facebook} left.  

\begin{figure*}[htbp!]

\begin{minipage}[b]{0.3\linewidth}
  \centering
  \centerline{\includegraphics[width=6.0cm]{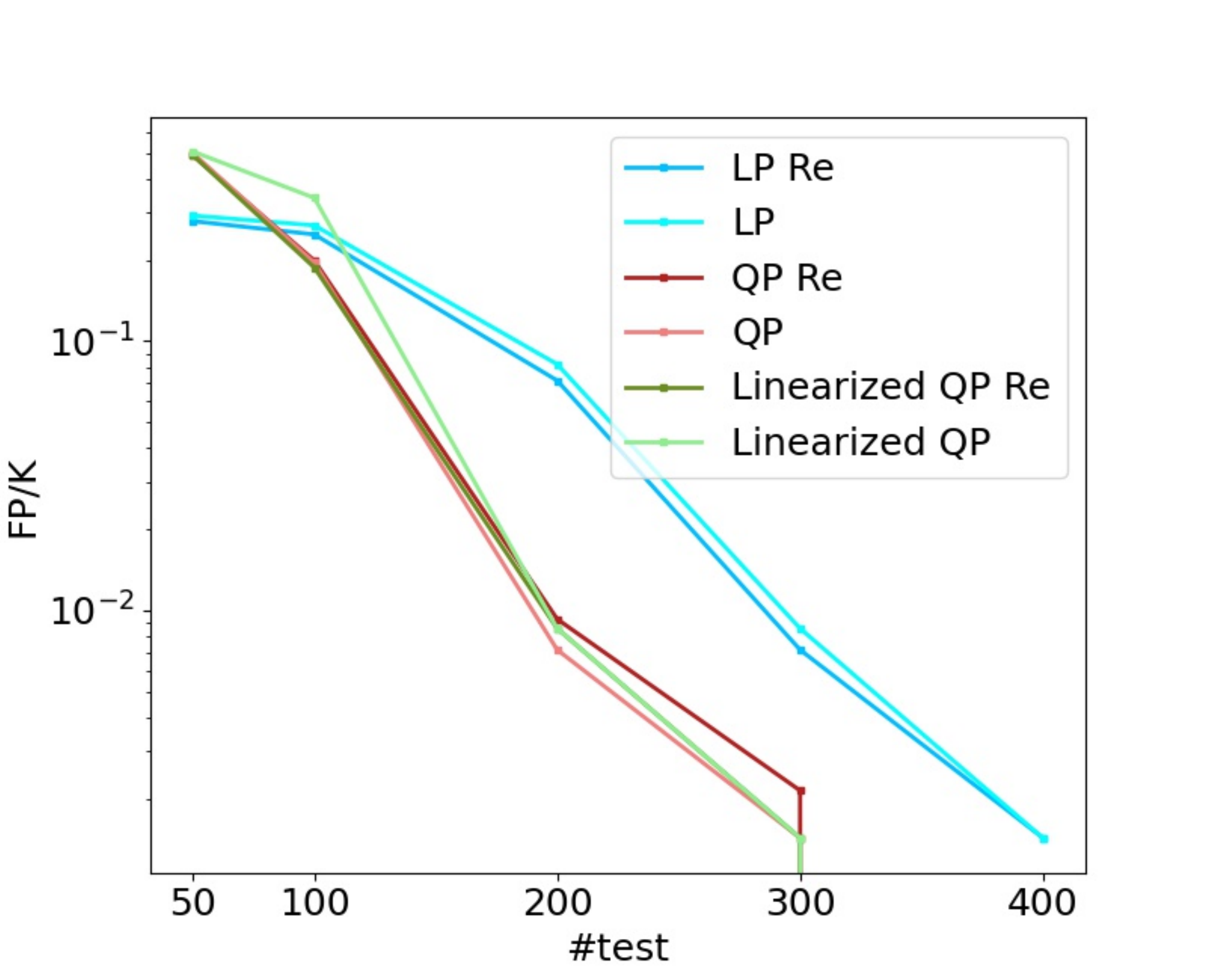}}
\end{minipage}
\hfill
\begin{minipage}[b]{0.3\linewidth}
  \centering
  \centerline{\includegraphics[width=6.0cm]{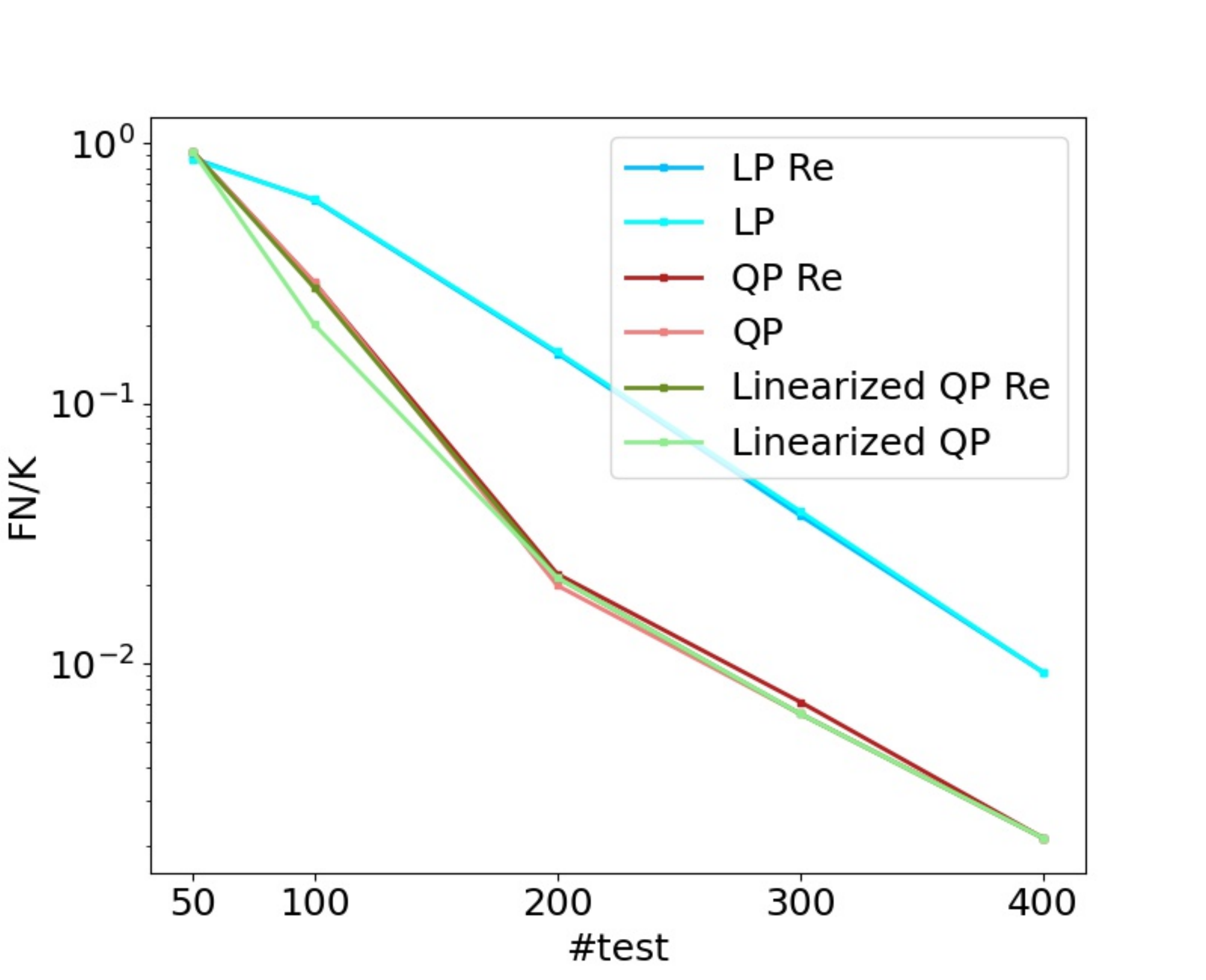}}
\end{minipage}
\hfill %quad
\begin{minipage}[b]{0.3\linewidth}
  \centering
  \centerline{\includegraphics[width=6.0cm]{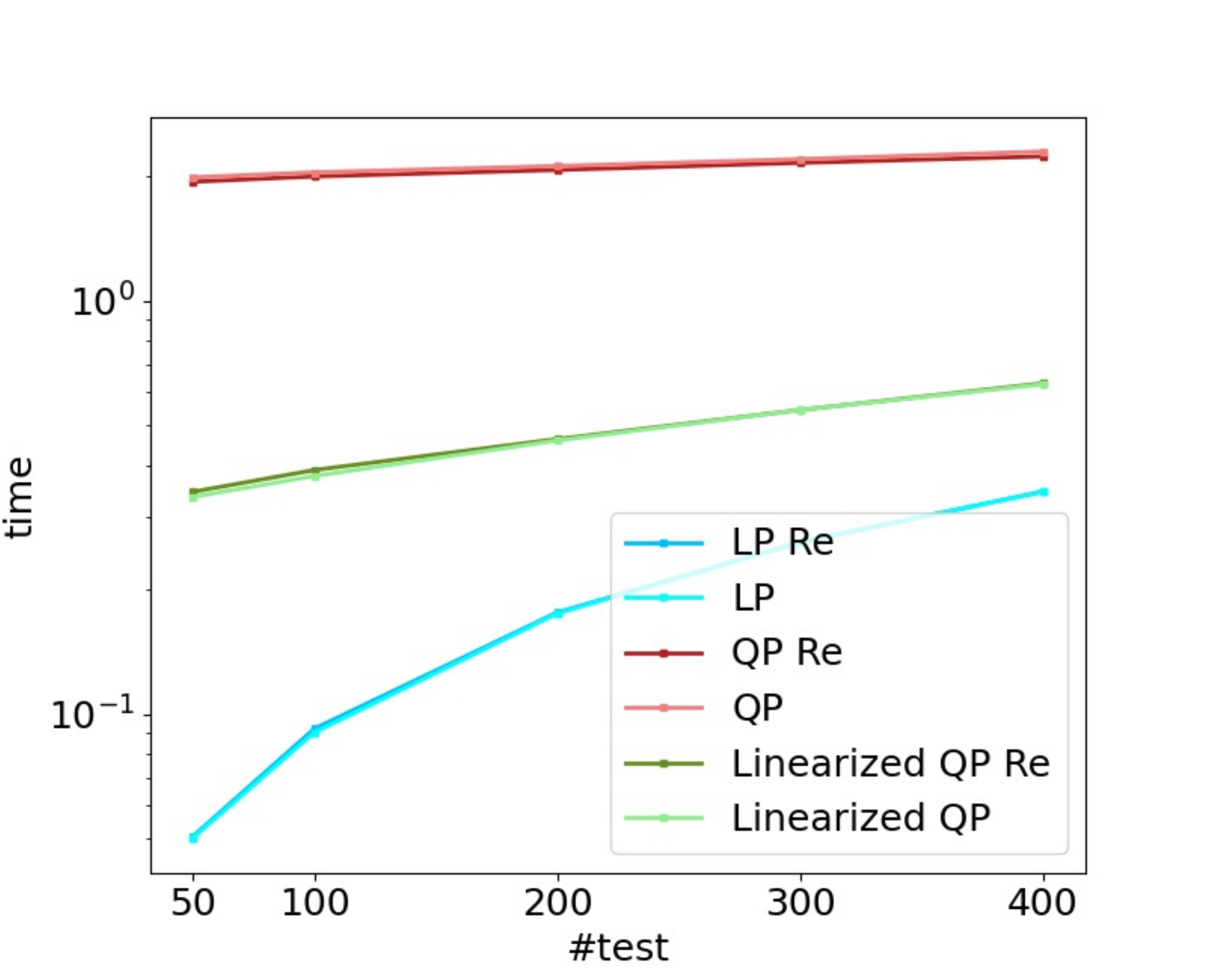}}
\end{minipage}

\begin{minipage}[b]{0.3\linewidth}
  \centering
  \centerline{\includegraphics[width=6.0cm]{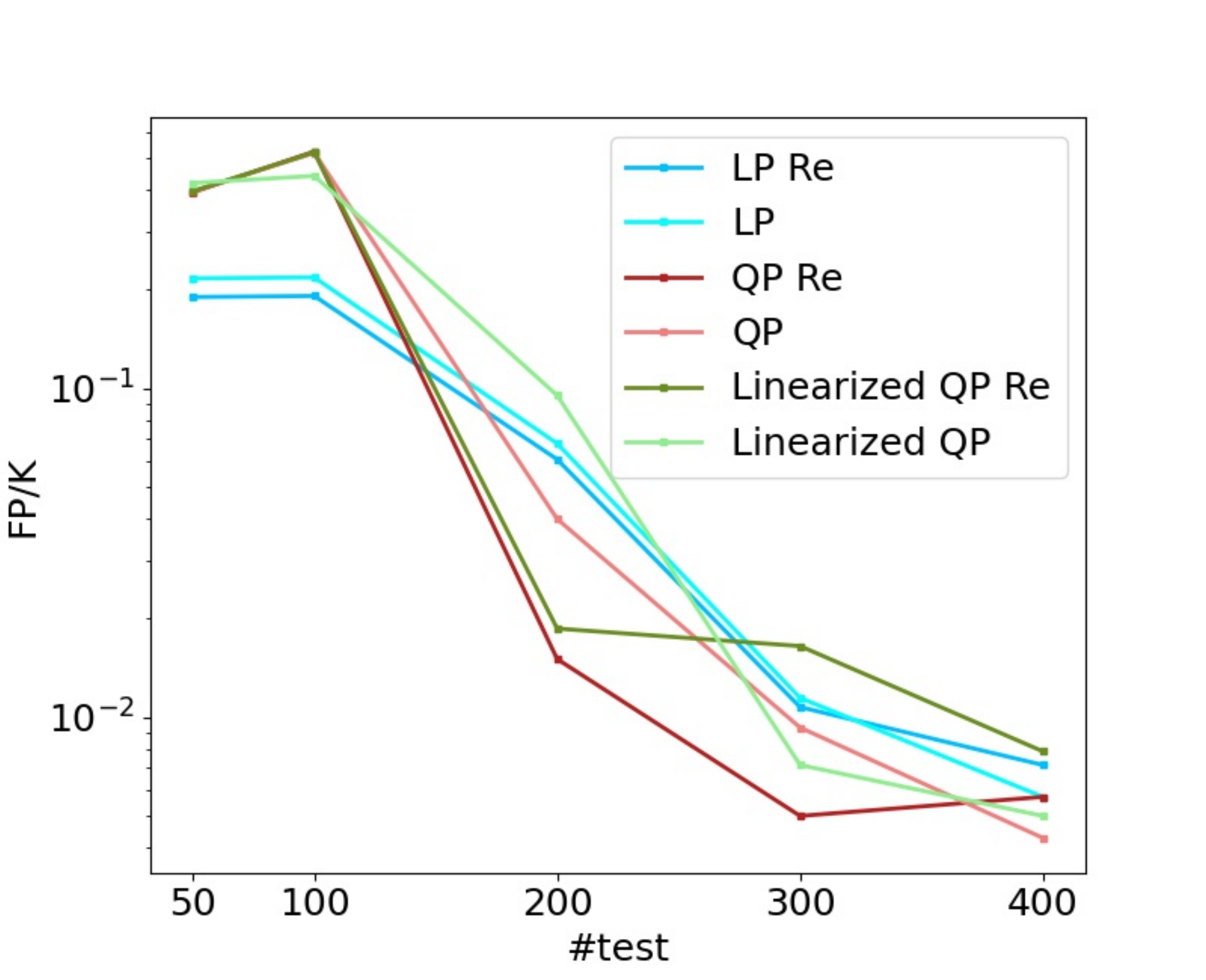}}
  \subfloat{$\mathrm{FP}/k$ (false positive rate)}
\end{minipage}
\hfill
\begin{minipage}[b]{0.3\linewidth}
  \centering
  \centerline{\includegraphics[width=6.0cm]{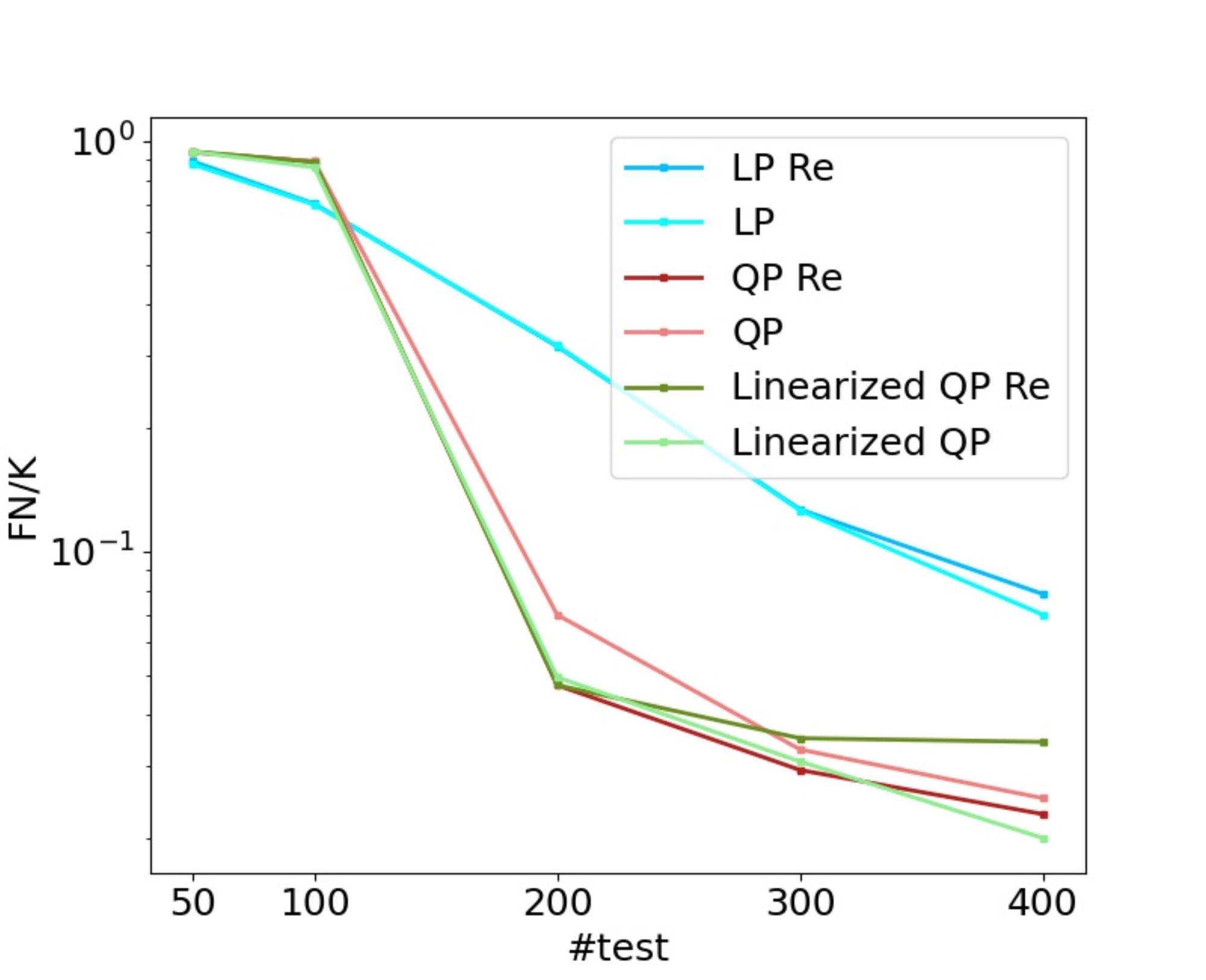}}
  \subfloat{$\mathrm{FN}/k$ (false negative rate)}
\end{minipage}
\hfill
\begin{minipage}[b]{0.3\linewidth}
  \centering
  \centerline{\includegraphics[width=6.0cm]{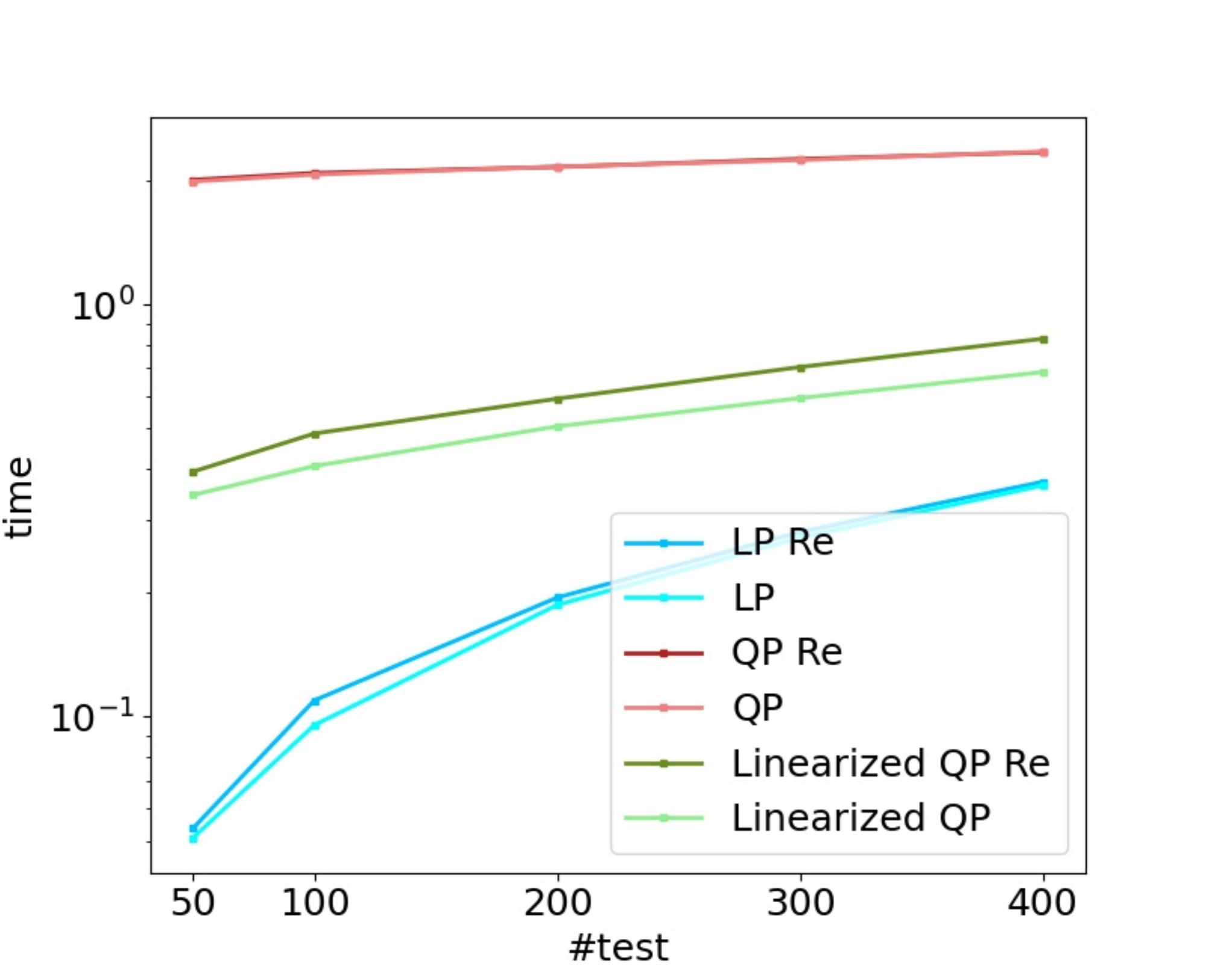}}
  \subfloat{computation time}
\end{minipage}

\caption{Results on Facebook graph with $\rho=0$ (top) and $\rho=0.01$ (bottom).}
\label{fb_decoding}
\end{figure*}

We apply group testing on this graph in the same way as the synthetic graphs, and present the results in Figure \ref{fb_decoding}.  Overall, we observe generally similar performance patterns to those of the simpler grid/block graphs (including computation time), particularly as the number of tests increases.

\section*{Acknowledgment} 

This work was supported by the Singapore National Research Foundation (NRF) under grant number R-252-000-A74-281.

\newpage
\bibliographystyle{IEEEtran}
\bibliography{main}

\clearpage
\appendices
% \appendix
\renewcommand\thesubsectiondis{\thesection-\Roman{subsection}}

\section*{\bf \large Supplementary Material (Model-Based and Graph-Based Priors for Group Testing)}
% \renewcommand{\thesubsection}{\Alph{subsection}}

% \subsection{Linear Programming and Quadratic Programming}
% \label{sect:appendix_0}
% \import{}{app0.tex}

\section{Proof of Theorem~\ref{thm:card_bound_achiev} (General Achievability Result)}
\label{app:general_achievability}
%!TEX root = main.tex

The analysis of \cite{scarlett2016phase} (see also \cite{Scarlett2017LimitsSupportRecovery}) considers the case that the 
space of defective sets, $\Sc$, is the entire space of $k$-sized subsets of $[n]$. 
In this appendix (achievability) and Appendix \ref{app:pf_converse} (converse), we adapt and generalize the analysis therein to the case where $\Sc$ is an arbitrary subset of the entire space of $k$-sized subsets of $[n]$.  This appendix and Appendix \ref{app:pf_converse} follow \cite{scarlett2016phase} quite closely, whereas the subsequent appendices deviate significantly from \cite{scarlett2016phase}.
% This generalization is useful for the refined achievability bound, and is also of interest in its own right.

\subsection{Notation} 
% Let $\Sc$ be an arbitrary subset of the entire space of $k$-sized subsets of $[n]$.
% % We let its size to be 
% % \[
%     % |\Sc| = 2^{\left(\beta \,  k \log_2 \frac{n}{k}\right)} \quad \text{where } \beta \in (0, 1)
% % \]    
% The unknown defective set, $S$, is uniform on $\Sc$.
% The binary test matrix is denoted by $\Xv \in \{0,1\}^{t \times n}$, and the notation $\Xv_s$ denotes the sub-matrix obtained by keeping only the columns indexed by the set $s \subseteq \{1, 2, \dotsc, n\}$.
% As in \cite{scarlett2016phase}, we consider non-adaptive i.i.d.~Bernoulli testing, where each item is placed in a given test with probability $\frac{\nu}{k}$ for some $\nu > 0$.  
% % In other words, independently over $i \in \{1,2, \dots, n\}$ and $t \in \{1,2,\dots, T\}$, we have $\PP(X_{ti} = 1) = \nu/k$ and $\PP(X_{ti} = 0) =  1 - \nu/k$.

% The analysis in \cite{scarlett2016phase} contains the combinatorial terms $\Ntaudefault$ in \eqref{eqn: combi term}, counting the number of ways that the correct defective set $S$ can have $\tau$ of its  $k$ items removed, and different $\tau$ items added (from $[n] \setminus S$), to produce some incorrect defective set $S'$.
% In our analysis, we replace these terms by more general values $\Ntau$ as defined in \eqref{eq:Ntau}, and we examine how the final bound on the number of tests changes 
% accordingly.

Given the test matrix $\Xv$, let $\Xv_s$ denote the sub-matrix obtained by keeping only the columns indexed by the set $s \subseteq \{1, 2, \dotsc, n\}$.  Recall also that the combinatorial quantities $\Ntau$ are defined in \eqref{eq:Ntau}.

The analysis utilizes the fact that the $t$ tests are independent, and so it is useful to have notations
for an arbitrary row of~$\Xv$ and its test outcome.
Let~$\mathbf{X}$ (resp., $\mathbf{X}_s$) be an arbitrary row of $\Xv$ (resp., $\Xv_s$), and let $Y \in \{0, 1\}$ be the corresponding entry of $\mathbf{Y}$.
Then $Y$ is generated according to some distribution $P_{Y|\mathbf{X}_s}$.  Initially, this may be any distribution that is symmetric in the sense that the dependence on $\mathbf{X}_s$ only enters through the number of defectives included (i.e., not the specific ordering); after the initial analysis, we will specialize to the noiseless model.  The $t$-fold product of $P_{Y|\mathbf{X}_s}$ gives the distribution of the overall
test vector $\mathbf{Y} = (Y_1, \dotsc, Y_t)$ given $\Xv_s$, and is denoted by $P_{\mathbf{Y}|\mathsf{X}_s}$.

Our analysis applies to any realization $s$ of $S$, and we can therefore set the realization arbitrarily throughout.
As in \cite{scarlett2016phase, Scarlett2017LimitsSupportRecovery}, we consider
% separate error events according to how much an incorrect defective set $\sbar$ overlaps with $s$. We 
the partition of the defective set $s$ into two sets $\seq$ and $\sdif$.
The set $\seq = s \cap \sbar$ is the intersection of the true defective set $s$ and 
some incorrect set $\sbar$; and the set  $\sdif = s \backslash \sbar$ is the set consisting the remaining elements of $s$. For a fixed defective set $s$, and a corresponding partition $(\sdif,\seq)$, we write
\begin{equation}
    P_{ Y | {\mathbf{X}_{\sdif} \mathbf{X}_{\seq}} }\left(y \mid \mathbf{x}_{\sdif}, \mathbf{x}_{\seq} \right) 
    \coloneqq 
    P_{ Y | \mathbf{X}_{s} } \left (y \mid \mathbf{x}_s \right), \label{eq:S0S1}
\end{equation}
where $\mathbf{X}_{\sdif}$ (resp. $\mathbf{X}_{\seq}$) is the subset of the test vector $\mathbf{X}$ indexed by $\sdif$ (resp. $\seq$). 
This allows us to define the marginal distribution
\begin{multline}
     P_{ Y | {\mathbf{X}_{\seq}} } \left(y \mid \mathbf{x}_{\seq} \right) \\
     \coloneqq 
     \sum_{\mathbf{X}_{\sdif}}
     P_{\mathbf{X}_{\sdif}} \left( \mathbf{x}_{\sdif} \right)
     P_{ Y | {\mathbf{X}_{\sdif} \mathbf{X}_{\seq}} } \left( y \mid \mathbf{x}_{\sdif},  \mathbf{x}_{\seq} \right),
    % P_{Y \mid {\vec X}_{S_1}}(y | {\vec x}_{S_1}) = \sum_{ {\vec x}_{S_0} } P_{{\vec X}_{S_0}}( {\vec x}_{S_0} ) P_{Y|{\vec X}_{S_0}, {\vec X}_{S_1}}(y \mid {\vec x}_{S_0}, {\vec x}_{S_1}),
\end{multline}
Using the preceding definitions, we define the {\em information density}
    \begin{equation}
        \imath\left( \Xsdif; Y \mid \Xseq \right) = 
        \log_2 
        \frac{ P_{Y| \Xsdif, \Xseq} \left(Y \mid \Xsdif, \Xseq \right) }
        { P_{Y| \Xseq} \left(Y \mid \Xseq \right) }, 
        \label{eq:info_dens}
    \end{equation}
    and let $\imath^t\left(\XXsdif; \mathbf{Y} \mid \XXseq \right)$ be the $t$-letter extension obtained by summing \eqref{eq:info_dens} over the $t$ tests.  Since the tests are independent, writing the sum of logarithms as the logarithm of a product yields
    \begin{equation}
        \imath^t\left(\XXsdif; \mathbf{Y} \mid \XXseq \right) = 
        \log_2 
        \frac{ P_{ \mathbf{Y}| \XXsdif, \XXseq} \left(\mathbf{Y} \mid \XXsdif, \XXseq \right) }
        { P_{ \mathbf{Y}| \XXseq} \left( \mathbf{Y} \mid \XXseq \right) }. \label{eq:info_dens_T}
    \end{equation}
    We also note that the expectation of \eqref{eq:info_dens} is a conditional mutual information
    $I \left( \Xsdif; Y \mid \Xseq \right)$.
    Since the mutual information for each $(\sdif,\seq)$ depends only on $\tau \coloneqq |\sdif|$
    by the assumed symmetry of $P_{Y|\mathbf{X}_s}$, we denote this conditional mutual information by
    \begin{equation}
        I_{\tau} \coloneqq I \left( \Xsdif; Y \mid \Xseq \right).
        \label{eq:Itau}
    \end{equation}
    It follows that the expectation of \eqref{eq:info_dens_T} is given by
     \begin{equation}
        \E \big[ \imath^t\left(\XXsdif; \mathbf{Y} \mid \XXseq \right)\big] = 
        % T \cdot I({\vec X}_{0,\tau}; Y \mid {\vec X}_{1,\tau}) 
        t \cdot I_\tau.
        \label{eq:I_tau}
    \end{equation}

\subsection{Choice of decoder and non-asymptotic error probability bound} 
\label{sect:inf_theoretic_decoder}

We use a similar information-theoretic threshold decoder as that in \cite[Appendix A]{scarlett2018noisy} (see also \cite[Sec. 4.2.3]{aldridge2019group}), except that it is constrained to only search over $\Sc$.  Specifically, the decoder searches for $s \in \Sc$ satisfying the following:
\begin{equation}
    \imath^t\left(\XXsdif; \mathbf{Y} \mid \XXseq \right)  \ge \gamma_{|\sdif|}, \quad \forall (\sdif,\seq)\text{ with } |\sdif| > \dmax,
    \label{eq:dec}
\end{equation}
where these pairs $(\sdif,\seq)$ are constrained to partition $s$ into two subsets, and the constants $\{ \gamma_{\tau} \}_{\tau = \dmax + 1}^{k}$ will be specified shortly.
% as in \eqref{eq:gamma_const}. 
If multiple such sets exist, or if no such set exists, then an error is declared.  
% We generalize the initial non-asymptotic bound of \cite[Equation (71)]{scarlett2018noisy}, by taking the defective space $\Sc$ to be an arbitrary subset of the entire space of $k$-sized subsets of~$[n]$. 

\begin{lemma}
\label{thm: non_asym_bound}
    Let $\delta > 0$ be an arbitrary real number, and set the constants $\gamma_{\tau}$ in \eqref{eq:dec} to be 
    \begin{equation}
    \label{eq:gamma_const}
        \gamma_{\tau} = \log_2 \frac{k \cdot N_{\tau}}{\delta} \quad 
        \text{for } \tau = \dmax + 1, \dotsc, k.
    \end{equation}
    Then, the error probability of the information-theoretic threshold decoder  above  satisfies 
    \begin{multline}
        \pe \le 
        \PP\Bigg[ 
        \bigcup_{(\sdif,\seq) \colon |\sdif| > \dmax} 
        \bigg\{ 
            \imath^t\left(\XXsdif; \mathbf{Y} \mid \XXseq \right) \\
            \le 
            \log_2 N_{\tau} 
                + 
            \log_2 \frac{k}{\delta} 
        \bigg\} 
        \Bigg] + \delta, \label{eq:p_init_bound}
    \end{multline}
    where we define $\tau = |\sdif|$ inside the union.
\end{lemma}

The only difference in \eqref{eq:p_init_bound} compared to \cite{scarlett2016phase} is the presence of $\log_2 \Ntau$ instead of $\log_2 \Ntaudefault$, since a union bound over the incorrect sets $\sbar$ with $|\sbar \setminus s| = \tau$ is now only over at most $N_{\tau}$ events (i.e., those with $\sbar \in \Sc$).

The non-asymptotic bound in Lemma \ref{thm: non_asym_bound} can be used to obtain conditions on the number of tests $t$, following \cite[Sec. 3.3]{scarlett2016phase}. 
    %  The idea is to use a concentration inequality. 
This gives the following analog of \cite[Thm. 6]{scarlett2016phase}.

\begin{lemma}
\label{lem: tests conditions}
    Let  $\delta > 0$ and $\{\deltatau \}_{\tau = \dmax+1}^k$ be
    arbitrary numbers in $(0, 1)$, and let $\{ \psi_{\tau} \colon \Z \times \R \to \R \}_{\tau = \dmax+1}^{k}$
    be some arbitrary functions. If the number of tests $t$ satisfies
    \begin{equation}
    \label{eq:T_first_Cond}
        t \ge 
        \max_{\tau \in \{\dmax+1, \dotsc, k \}} 
        \frac{ \log_2 N_{\tau} + \log_2\frac{k}{\delta} }
             { \left(1- \deltatau \right)\Itau},
    \end{equation}
    and if each information density satisfies a concentration bound of the form
    \begin{equation}
         \PP
         \bigg[ 
         \imath^t\left(\XXsdif; \mathbf{Y} \mid \XXseq \right) \le 
         t \big(1 - \deltatau \big) I_{\tau} 
         \bigg] \le 
         \psi_{\tau}(t, \delta'_{\tau}), \label{eq:psi_ach}
    \end{equation}
     then the error probability satisfies
    \begin{equation}
        \pe \le \sum_{\tau= \dmax + 1}^{k} \binom{k}{\tau} \psi_{\tau}(t, \delta'_{\tau}) + \delta. \label{eq:nonasymp_ach}
    \end{equation}
\end{lemma}

\subsection{Characterizing tail probabilities and the asymptotic number of tests}

We wish to find functions $\psi_{\tau}(t, \delta')$ in Lemma~\ref{lem: tests conditions} such that \eqref{eq:psi_ach} 	is satisfied, and the first term in the error probability \eqref{eq:nonasymp_ach} 
can be made arbitrarily small asymptotically.  For the noiseless model, the following result from
\cite[Proposition 3]{scarlett2016phase} provides such explicit choices.

\begin{lemma}[\cite{scarlett2016phase}]
\label{lem:functions_choice}
    Under the noiseless group testing model using i.i.d.~Bernoulli testing, for all 
    $(\sdif,\seq)$ with $|\sdif| = \tau \in \{\dmax+1, \dotsc, k \}$, and any constants $\{\deltatau \}_{\tau = \dmax+1}^k$ in $(0,1)$, we have
    \begin{multline}
        \PP\Big[ 
            \imath^t\left(\XXsdif; \mathbf{Y} \mid \XXseq \right) \le 
            t \big(1 - \deltatau \big) I_{\tau} 
        \Big] 
        \\ \le 
        2\exp 
        \left(- \frac{ \left(\deltatau I_{\tau} \right)^2 t}
               { 4 \left( 8 +  \deltatau I_{\tau} \right) } 
        \right). 
    \label{eq:conc_gen_disc}
    \end{multline}
\end{lemma}
    
We set the functions $\psi_{\tau}(t, \delta')$ in Lemma~\ref{lem: tests conditions}
to be the right-hand side of \eqref{eq:conc_gen_disc}. 
The analysis in \cite[Page 3657]{scarlett2018noisy} shows that the 
first term in the error probability \eqref{eq:nonasymp_ach} can be made arbitrarily small
asymptotically (see also \cite[Proposition 2]{scarlett2016phase}), as stated in the following.

\begin{lemma}[\cite{scarlett2016phase}]
\label{lem:error_bound}
    Let $k = \Theta\left(n^{\theta}\right)$ for some $\theta \in (0, 1)$, and
    $\dmax = \Theta(k)$.
     For any $\delta' \in (0, 1)$, as long as $t = \Omega(k)$, we have
     \begin{equation}
        \lim_{n \to \infty}
        \sum_{\ell= \dmax + 1}^{k} \binom{k}{\tau} \,
        2\exp 
            \left(- \frac{ \left(\delta' I_{\tau} \right)^2 t}
                   { 4 \left( 8 +  \delta' I_{\tau} \right) } 
            \right) = 0.
     \end{equation}
\end{lemma} 
     
Combining Lemmas~\ref{lem: tests conditions}, \ref{lem:functions_choice}, and \ref{lem:error_bound}
gives the following conditions on the number of tests $t$. 
 \begin{corollary}
\label{cor: tests conditions}
     Let $k = \Theta\left(n^{\theta}\right)$ for some $\theta \in (0, 1)$, and $\dmax = \Theta(k)$.  Fix $\delta > 0$ and $\delta' \in (0, 1)$.  
    % Let 
    % \[
    %     \psi_{\tau}(t, \delta') = 2\exp 
    %     \left(- \frac{ \left(\delta' I_{\tau} \right)^2 T}
    %           { 4 \left( 8 +  \delta' I_{\tau} \right) } 
    %     \right)
    %     \quad \text{for } \tau = \dmax+1, \dotsc, k.
    % \]    
    As $n \to \infty$, if the number of tests $t$ satisfies
    \begin{equation}
    \label{eq:bound_T}
        t \ge 
        \max_{\tau \in \{\dmax+1, \dotsc, k \}} 
        \frac{ \log_2 N_{\tau} + \log_2 \frac{k}{\delta} }
             { \left(1- \delta' \right)\Itau},
    \end{equation}
     then the error probability satisfies
    \begin{equation}
        \pe \le \delta + o(1). \label{eq:error_prob}
    \end{equation}
\end{corollary}

    We now analyze the requirement on the number of tests $t$ in \eqref{eq:T_first_Cond}. 
    We first characterize the conditional mutual information $I_\tau$ using the following known result \cite[Proposition 1]{scarlett2016phase}.
    
    \begin{lemma}[\cite{scarlett2016phase}] \label{lem:mi_asymp}
    	Under the noiseless group testing model using Bernoulli testing with probability~$\frac{\nu}{k}$, as $k \to \infty$, the conditional mutual information $I_{\tau}$ behaves as follows when $\tau/k \to \alpha \in (0,1]$:
    \begin{equation}
    \label{eq:gtn_I_const}
        I_{\tau} = e^{-(1-\alpha)\nu} H_2\big(e^{-\alpha \nu} \big)  (1+o(1)),
    \end{equation}
        where $H_2(\lambda) = \lambda \log_2\frac{1}{\lambda} + (1 - \lambda)\log_2\frac{1}{1-\lambda}$
        is the binary entropy function.
    \end{lemma}
    
    Finally, we characterize the asymptotic number of tests $t$ by following the analysis in \cite[Step 3]{scarlett2016phase}.
    By taking $\delta \to 0$ sufficiently slowly and noting that $\delta' > 0$ is arbitrarily small,
    the condition on the number of tests given in \eqref{eq:bound_T} gives
    \begin{equation}
         t \ge \max_{\alpha \in [\alpha^{\ast}, 1]}  
        \frac{\log_2 N_{\lceil \alpha k \rceil}  + 2 \log_2 k }
        { e^{-(1-\alpha)\nu} H_2\big(e^{-\alpha \nu} \big)} \left(1 + o(1) \right). \label{eq:t_final_proof}
        % t \ge \max_{\tau = d_{\text{max}}+1, \dotsc , k} 
        % \frac{\log_2 \left( N_{\tau} \right)  + 2 \log_2 (k) }
        % { e^{-(1-\alpha)\nu} H_2\big(e^{-\alpha \nu} \big)} \left(1 + o(1) \right).
    \end{equation}
    where the denominator follows from \eqref{eq:gtn_I_const},
    and the numerator follows from $\tau/k \to \alpha$.  Thus, we have established that \eqref{eq:t_final_proof} yields $\pe \to 0$, proving Theorem \ref{thm:card_bound_achiev}.
% \begin{theorem}
%     Let $n$ be the number of items, $k = \Theta\left(n^{\theta}\right)$ for some $\theta \in (0, 1)$ be the size of an unknown defective set $S$, and  $\Sc$ be an arbitrary subset of the entire space of $k$-sized subsets of $[n]$. Let $\alpha^{\ast} \in (0, 1)$ be arbitrary and $\dmax = \lfloor \alpha^{\ast} k \rfloor$.
%     For $\tau = \dmax +1, \dotsc, k$, let $\Ntau$ be the number of ways that the correct defective set
%     $S$ can have $\tau$ of its $k$ items removed, and different $\tau$ items added, to produce
%     some incorrect defective set $S' \in \Sc$.
%     Let $\mathsf{X} \in \{0,1\}^{T \times n}$ be a random binary matrix with i.i.d Bernoulli$\left(\frac{\nu}{k}\right)$ entries for some $\nu \in [0, 1]$. 
%     Under the non-adaptive, noiseless, and approximate recovery setting  group testing model,
%     % using Bernoulli testing with $p = \frac{\nu}{k}$, 
%     as $n \to \infty$, the error probability $\pe$
%     of the information-theoretic threshold decoder in Section~\ref{sect:inf_theoretic_decoder} vanishes, provided
%     \[
%         t \ge \max_{\tau = d_{\text{max}}+1, \dotsc , k} 
%         \frac{\log_2 \left( N_{\tau} \right)  + 2 \log_2 (k) }
%         { e^{-(1-\alpha)\nu} H_2\big(e^{-\alpha \nu} \big)} \left(1 + o(1) \right).
%     \]
% \end{theorem}

% \noindent Note that here we have renamed $\delta_2^{(1)}$ as $\delta_2$, as well as setting $\nu = \log 2$.

\section{Proof of Theorem \ref{thm:converse} (General Converse Result)} 
\label{app:pf_converse}
This result follows almost immediately from the analysis of \cite[Thm.~1]{scarlett2017little}, which we do not repeat here.  Specifically, in the case that $\mathcal{S}$ contains all $k$-sized subsets, the analysis therein shows that attaining $\pe \not\to 1$ requires\footnote{In \cite{scarlett2017little} the possibility of list decoding is considered, in which the decoder outputs a list of size $L \ge k$.  Equation \eqref{eq:t_lb_original} corresponds to the special case $L = k$.  For the approximate recovery criterion with distance measure \eqref{eq:dist}, it is known that having the decoder output a set of size $k$ is without loss of optimality \cite{Scarlett2017LimitsSupportRecovery,reeves2013approximate}, so it can be assumed here for the purpose of proving a converse.}
\begin{equation}
    t \ge \bigg( \log_2 \binom{n}{k} - \alpha^* k \log_2\frac{n}{k}\bigg)(1+o(1)), \label{eq:t_lb_original}
\end{equation}
where the first term represents the number of defective sets, and the second term represents the (asymptotic) logarithm of how many possible defective sets are within distance $\alpha^* k$ of a fixed size-$k$ set, according to the notion of distance defined in \eqref{eq:dist}.

In our setting, $\log_2 \binom{n}{k}$ naturally generalizes to $\log_2 |\mathcal{S}|$.  Regarding the second term, we still consider possible defective sets within distance $\alpha^* k$ of some fixed $\Shat$ that the decoder may output, and the term naturally generalizes to $\log_2 \Ntil_{\lceil \alpha^* k \rceil}$ according to the definition in \eqref{eq:Ntilde}. 
%\footnote{When $\Sc_{\rm dec}$ equals $\{S \,:\,|S|=k\}$ in \eqref{eq:Sdec}, we use the fact that outputting a size-$k$ set is without loss of optimality \cite{reeves2013approximate,Scarlett2017LimitsSupportRecovery}, so it can be assumed for the purpose of proving a converse.}  
Thus, we obtain \eqref{eq:converse_bound_T_Ntau} in Theorem~\ref{thm:converse}.

% Moreover, the second term in \eqref{eq:t_lb_original} still serves as a valid \emph{upper bound} on the number of possible defective sets within distance $\alpha^* k$ of any fixed one.  This trivially follows from the fact that our $\mathcal{S}$ is a subset of the ``full set'' used to derive \eqref{eq:t_lb_original}.  Thus, Corollary \ref{cor:converse_card} immediately follows.
% {\color{magenta} See Appendix~\ref{sect:app_conv_card} for Ivan's proof of Corollary~\ref{cor:converse_card}: }

\section{Proofs of Corollaries~\ref{cor:card_bound_achiev} and \ref{cor:converse_card} (Cardinality-Bounded Structure)}
\label{app:cardinality_bound}
%!TEX root = main.tex

\subsection{Achievability}
\label{sect:app_achiev_card}
Since $|\Sc| = 2^{\left(\beta \,  k \log_2 \frac{n}{k}\right)(1+o(1))}$, we have
\begin{equation}
    \Ntau \le 
    \min
    \left\{\Ntau^{\text{default}}, 2^{\left(\beta \,  k \log_2 \frac{n}{k}\right)(1+o(1))} 
    \right\}, 
\end{equation}
and substituting \eqref{eq:size_of_def_spac} and \eqref{eqn: combi term} gives
\begin{multline}
\label{eqn: trivial bound}
    \log_2 \Ntau \le 
    \min \bigg\{\log_2 \binom{k}{\tau} + \log_2 \binom{n-k}{\tau}, \\
                \Big(\beta \,  k \log_2 \frac{n}{k} \Big)(1+o(1))
         \bigg\}.
\end{multline}
Using Theorem~\ref{thm:card_bound_achiev}, we have the following condition on the number of tests $t$:
\begin{equation}
\label{eq:card_bound_achievability}
    t  = \max_{\alpha \in [\alpha^{\ast}, 1]}  
        \frac{\log_2 N_{\lceil \alpha k \rceil}  + 2 \log_2 k }
        { e^{-(1-\alpha)\nu} H_2\big(e^{-\alpha \nu} \big)} \left(1 + o(1) \right).
    % \max_{\tau = \dmax+1, \dotsc , k} 
        % \frac{\log_2 \left( N_{\tau} \right)  + 2 \log_2 (k) }
            %   { e^{-(1-\alpha)\nu} H_2\big(e^{-\alpha \nu} \big)} \left(1 + o(1) \right)
\end{equation}

Using the bound \eqref{eqn: trivial bound} and the identity $\log_2 \binom{n-k}{\tau} = \left(\tau \log_2 \frac{n}{\tau} \right)(1 + o(1) )$ for $k = o(n)$, and retaining only the
dominant terms, we obtain the following condition on the number of tests $t$:
% \[
%      T = \max_{\tau = \dmax+1, \dotsc , k} 
%      \frac{\min\left\{ \tau \log_2 \frac{n}{\tau} , \beta \,  k \log_2 \frac{n}{k} \right\}}
%      { e^{-(1-\alpha)\nu} H_2\big(e^{-\alpha \nu} \big)}  \left(1 +o(1) \right).
%     \]
% Writing $\frac{\tau}{k} \to \alpha$ and $\dmax + 1 = \alpha^{\ast} k$,  we get the condition
\begin{align}
     t&= \max_{\alpha \in [\alpha^{\ast}, 1]} 
    \frac{\min\left\{\alpha k \log_2 \frac{n}{k} , \beta \,  k \log_2 \frac{n}{k} \right\}}{e^{-(1-\alpha)\nu} H_2(e^{-\alpha \nu})}  \left(1 +o(1) \right) \\
    &= \max_{\alpha \in [\alpha^{\ast}, 1]} 
    \frac{\min\left\{\alpha , \beta \right\}}{e^{-(1-\alpha)\nu} H_2(e^{-\alpha \nu})}  \left(k \log_2 \frac{n}{k}\right) \left(1 +o(1) \right).
    % \numberthis 
    \label{ref: final}
\end{align}
From \eqref{ref: final}, it remains to determine the coefficient
\begin{equation}
\label{eqn: coeff}
    \max_{\alpha \in [\alpha^{\ast}, 1]} 
    \frac{\min\left\{\alpha , \beta \right\}}{e^{-(1-\alpha)\nu} H_2(e^{-\alpha \nu})}= 
    e^{\nu} \max_{\alpha \in [\alpha^{\ast}, 1]} 
    \frac{ e^{-\alpha\nu} \min\left\{\alpha , \beta \right\}}{H_2(e^{-\alpha \nu})}.
\end{equation}
For fixed $\nu > 0$, a change of variable $\lambda = e^{-\alpha \nu}$ allows us to verify easily through direct differentiation that
% gives the following: 
\begin{equation}
    \diff{}{\alpha} \left(\frac{\alpha e^{-\alpha \nu}}{H_2(e^{-\alpha \nu})}  \right)  
    % = - \log 2 \
    %   \frac{\log\left(e^{-\alpha \nu} \right) + 
    %          \log\left(1- e^{-\alpha \nu} \right)\left( e^{\alpha \nu} - 1 - \alpha \nu e^{\alpha \nu} \right)}
    %   {\left(
    %     \log\left(e^{-\alpha \nu} \right) - \log\left(1 - e^{-\alpha \nu} \right)
    %     + e^{\alpha \nu} \log\left(1 - e^{-\alpha \nu} \right)
    %     \right)^2}
        \ge 0,
\end{equation}  
i.e., $\frac{\alpha e^{-\alpha \nu}}{H_2(e^{-\alpha \nu})}$ is increasing for $\alpha \in [0, 1]$, whereas
\begin{equation}
    \diff{}{\alpha} \left(\frac{ e^{-\alpha \nu}}{H_2(e^{-\alpha \nu})}  \right)  
    % =
    % \log 2 \
    % \frac{ \nu e^{\alpha \nu} \log\left(1- e^{-\alpha \nu} \right)}
    %     {\left(
    %     \log\left(e^{-\alpha \nu} \right) - \log\left(1 - e^{-\alpha \nu} \right)
    %     + e^{\alpha \nu} \log\left(1 - e^{-\alpha \nu} \right)
    %     \right)^2}
        \le 0
\end{equation} 
i.e., $\frac{e^{-\alpha \nu}}{H_2(e^{-\alpha \nu})}$ is decreasing for $\alpha \in [0, 1]$. 
It follows that the objective function \eqref{eqn: coeff} increases in~$\alpha$ until $\alpha \ge \beta$, after which it decreases.
This implies that the maximizing $\alpha$ in the objective function in \eqref{eqn: coeff} 
is always $\max \{\alpha^{\ast}, \beta \}$, and hence, the coefficient to $k\log_2\frac{n}{k}$ is given by
% for $\beta \in (0, 1]$ we have
% \begin{equation}
%     \max_{\alpha \in [\alpha^{\ast}, 1]} 
%     \frac{ e^{-\alpha\nu} \min\left\{\alpha , \beta \right\}}{H_2(e^{-\alpha \nu})}
%     =
%     \begin{cases}
%     \frac{\beta \, \mathrm{exp}\left(-\nu \alpha^{\ast}\right)}
%     {H_2\left(\mathrm{exp}(-\nu \alpha^{\ast}  )\right)} \quad & \text{if } \alpha^{\ast} > \beta;
%     \\
%     \\
%     \frac{\beta \, \mathrm{exp}\left(-\nu \beta \right)}
%     {H_2\left(\mathrm{exp}(-\nu  \beta  )\right)} \quad &\text{otherwise}
%     \end{cases}.
% \end{equation}
% Combining with \eqref{eqn: coeff}, we have the coefficient
\begin{equation}
    % \label{eqn: simplified coeff}
    \frac{\beta \, \mathrm{exp}\left(\nu \left(1 - \max \left\{ \alpha^{\ast}, \beta \right\} \right)\right)}
    {H_2\left(\mathrm{exp}(-\nu  \max \left\{ \alpha^{\ast}, \beta \right\} )\right)}. 
\end{equation}

\subsection{Converse}
\label{sect:app_conv_card}

For $\Sc_{\rm dec} = \{S \,:\, |S| = k \}$, we have the upper bound
\begin{align}
      \Ntil_{\dmax} 
    &= \max_{S \in \Sc_{\rm dec}} \big| S' \in \Sc : d(S,S') \le d_{\max} \big| \\
    &\le \min \left\{\Sc, \, (\dmax + 1) \max_{\tau = 0, 1, \dotsc, \dmax } \Ntau^{\text{default}} \right\} \\
    &= \min \left\{ \Sc, \, (\dmax + 1) \max_{\tau = 0, 1, \dotsc, \dmax } \binom{k}{\tau} \binom{n-k}{\tau} \right\}.
    \label{eq:pre_log}
\end{align}
% we have 
% \begin{equation}
% \label{eq:logNtilde}
%     % \max_{\tau = 0, 1, \dotsc, \dmax } \log_2 \left(\Ntau\right) \le 
%     \log_2 \Ntil_{\dmax} \le 
%     \log_2 (\dmax + 1) + \max_{\tau = 0, 1, \dotsc, \dmax } \log_2 \Ntau.
% \end{equation}
Taking the logarithm, using the bound \eqref{eqn: trivial bound} and the identity $\log_2 \binom{n-k}{\tau} = \left(\tau \log_2 \frac{n}{\tau} \right)(1 + o(1) )$ for $k = o(n)$, and retaining only the dominant terms, we obtain the asymptotic bound
\begin{align}
\label{eq:bound_Ntilde_card}
    \log_2 \Ntil_{\dmax} &\le 
    \min\left\{  \beta \,  k \log_2 \frac{n}{k} ,  \max_{0 \le \tau\le \dmax } \tau \log_2 \frac{n}{\tau} \right\} (1+o(1)) \nonumber \\ & \le
    % \min\left\{  \max_{0 \le \tau\le \dmax} \tau \log_2 \frac{n}{\tau} , \beta \,  k \log_2 \frac{n}{k} \right\} =
    \min \{\beta, \alpha^* \} \left( k \log_2 \frac{n}{k} \right) (1+o(1)).
\end{align}
Combining \eqref{eq:bound_Ntilde_card} and \eqref{eq:converse_bound_T_Ntau}, along with $\log_2|\Sc| = \big( \beta k \log_2\frac{n}{k} \big)(1+o(1))$, we obtain the condition
\begin{align}
    t &\ge 
    \left( \beta  k \log_2 \frac{n}{k} - \log_2 \Ntil_{\dmax} \right) (1+o(1)) \nonumber \\ &\ge
    % \ge \left(\beta - \min \{\alpha^* , \beta \} \right)  \left( k \log_2 \frac{n}{k} \right) (1+o(1)) =
    \max\{ \beta  - \alpha^* , 0 \} \left( k \log_2 \frac{n}{k} \right) (1+o(1)),
\end{align}
which yields Corollary \ref{cor:converse_card}.

% \magenta{Jon's proof:}
% Since the space of defective sets $\Sc$  is a subset of all size-$k$ sets, the second term in \eqref{eq:t_lb_original} still serves as a valid \emph{upper bound} on the number of possible defective sets within distance $\alpha^* k$ of any fixed decoder output $\Shat$. Thus, Corollary \ref{cor:converse_card} immediately follows.

\section{Proofs of Corollaries~\ref{cor: block_graph_achiev} and \ref{cor:converse_block} (Block Structure)}
\label{app:block_structure}
%!TEX root = main.tex

% Since 
% \[
%     \Ntau = \binom{k/q}{\tau/q} \binom{\frac{n-k}{q}}{\tau/q},
% \]
% we have

The analysis for the block-structured case is similar to Appendix \ref{app:cardinality_bound}, but is relatively simpler and has more in common with \cite{scarlett2016phase}.  Accordingly, we omit some of the details.

Recall that $N_{\tau}$ is only non-zero when $\tau$ is divisible by $q$.  For such values, it follows from \eqref{eqn: block graph Nl bound} and $\tau = o(n)$ that
\begin{align}
\label{eq:blockgraph_log2Ntau}    
    \log_2 \Ntau &= \log_2\binom{k/q}{\tau/q} + \log_2 \binom{\frac{n-k}{q}}{\tau/q}
     \nonumber \\ &= \frac{1}{q} \left(\tau \log_2 \frac{n}{\tau} \right)(1 + o(1) ).
\end{align}
Then, using similar steps to those in Appendix~\hyperref[sect:app_achiev_card]{C-I} 
with \eqref{eq:blockgraph_log2Ntau}, we obtain the following condition on the number of tests $t$:
\begin{align}
    t &\ge \frac{1}{q} 
        \left( 
            \max_{\alpha \in [\alpha^{\ast}, 1]} 
            \frac{\alpha k \log_2 \frac{n}{k}}{e^{-(1-\alpha)\nu} H_2(e^{-\alpha \nu})} 
        \right)
        \left(1 +o(1) \right) \nonumber \\
      &= \frac{1}{q} \frac{k \log_2 \frac{n}{k}}{H_2(e^{-\nu})} \left(1 +o(1) \right),
\end{align}
with $\alpha = 1$ achieving the maximum as already shown in \cite[p. 48]{scarlett2016phase}.

For the converse bound, we follow similar steps to those in Appendix~\hyperref[sect:app_conv_card]{C-II}, but now using
$\Sc_{\rm dec} = \Sc$ for $\Ntil_{\dmax}$. The corresponding upper bound is 
\begin{equation}
    \Ntil_{\dmax} \le (\dmax + 1) \max_{0 \le \tau \le \dmax } \Ntau.
\end{equation}
Combining the upper bound with \eqref{eq:blockgraph_log2Ntau} and $\beta = \frac{1}{q}(1+o(1))$ yields 
\begin{align}
\label{eq:bound_Ntilde_block}
    \log_2 \Ntil_{\dmax} &\le 
    \frac{1}{q}
    \left( \max_{0 \le \tau\le \dmax }
      \tau \log_2 \frac{n}{\tau} \right)  (1+o(1)) \nonumber \\
      &=
    \beta \left( \alpha^*  k \log_2 \frac{n}{k} \right) (1+o(1)) \nonumber \\
    &=
    \alpha^* \left( \log_2 |\Sc| \right)  (1+o(1)).
\end{align}
Combining \eqref{eq:bound_Ntilde_block} and \eqref{eq:converse_bound_T_Ntau} 
gives
\begin{align}
     t &\ge \left( \log_2 |\Sc|  - \log_2 \Ntil_{\dmax} \right)(1+o(1)) \nonumber \\
    %   = \left( \beta - \frac{1}{q} \alpha^* \right) \left( k \log_2 \frac{n}{k} \right) (1+o(1)) 
    %   &= \left( \beta - \alpha^* \frac{c-\theta}{1-\theta}\right)
    %         \left(  k \log_2 \frac{n}{k} \right)(1+o(1)) \\
    %   \ge   \left( \beta - \frac{\alpha^*}{q} \right) \left( k \log_2 \frac{n}{k} \right) (1+o(1))
        &\ge   (1-\alpha^*) \left( \log_2 |\Sc| \right) (1+o(1)),
\end{align}
which yields Corollary \ref{cor:converse_block}.
\section{Proofs of Corollaries~\ref{cor: imbal_achiev_linear_k2} -- \ref{cor:converse_imbalanced_linear_k2} (Imbalanced Structure)}
\label{app:imbalanced_structure}
%!TEX root = main.tex

The size of the defective space $\Sc$ is given by
\begin{equation}
    \left| \Sc \right| = \binom{n_1}{k_1} \binom{n_2}{k_2}.
\end{equation}
We proceed to give a bound on $\Ntau$. Observe that the $\tau$ removals and additions of items 
can be decomposed into $\tau_1$ removals and additions of items in the first block, 
and $\tau - \tau_1$ removals and addition of items in the second block, 
for some integer $\tau_1 \in [0, \tau]$.
For fixed $\tau_1$, there are $\binom{k_1}{\tau_1}\binom{n_1-k_1}{\tau_1}$ ways to remove $\tau_1$ of the items from the $k_1$
defective items, the analogous term is $\binom{k_2}{\tau-\tau_1}\binom{n_2-k_2}{\tau-\tau_1}$ in the second block.
%and then replace these $\tau_1$ items with items from the remaining $n_1-k_1$ items; and there are $\binom{k_2}{\tau-\tau_1}\binom{n_2-k_2}{\tau-\tau_1}$ ways to remove $\tau - \tau_1$ of the items from the $k_2$defective items, and then replace these $\tau- \tau_1$ items with items from the remaining $n_2-k_2$ items. 
Multiplying these terms and summing over all values of $\tau_1$, we obtain
% The corresponding $\Ntau$ are given by
\begin{equation}
\label{eq:imb_combi_term}
    \Ntau = \sum_{\tau_1 = \max(0, \tau-k_2)}^{\min(k_1, \tau)}
            \binom{k_1}{\tau_1} \binom{n_1-k_1}{\tau_1} 
            \binom{k_2}{\tau - \tau_1} \binom{n_2-k_2}{\tau - \tau_1} , 
    % \quad \text{for } \tau = \dmax+1, \dotsc, k,
\end{equation}
where the upper limit of the summation follows since $\tau_1 > k_1  \Rightarrow \binom{k_1}{\tau_1} = 0$,
and the lower limit follows since $\tau - \tau_1 > k_2  \Rightarrow \binom{k_2}{\tau - \tau_1} = 0$.
Then, we can bound $\Ntau$ by
\begin{multline}
\label{eq:imb_relaxed_bound}
    \Ntau \le 
    \min 
    \bigg\{
     \left|\Sc\right|, \, \\
    (\tau + 1) 
    \left( 
    \max_{\tau_1 }
    \binom{k_1}{\tau_1} \binom{n_1-k_1}{\tau_1} 
    \binom{k_2}{\tau - \tau_1} \binom{n_2-k_2}{\tau - \tau_1}  
    \right)
    \bigg\},
\end{multline}
where $\tau_1$ is implicitly constrained according to \eqref{eq:imb_combi_term}.
We now take the logarithm of the two different terms in the right-hand side of \eqref{eq:imb_relaxed_bound}, which gives
\begin{equation}
\label{eq:imb_spars_log_2_S_bound}
    \log_2 \Ntau \le 
    \log_2  \left|\Sc\right| =
    \log_2 \binom{n_1}{k_1} + \log_2 \binom{n_2}{k_2},
\end{equation}
and
\begin{multline}
\label{eq:imb_spars_relaxed_log_2_bound}
    \log_2 \Ntau \le 
    \log_2 \left(\tau+1\right)
    +
    \max_{\tau_1}
    \bigg\{
    \log_2 \binom{k_1}{\tau_1} \\ +
    \log_2 \binom{n_1-k_1}{\tau_1} +
    \log_2 \binom{k_2}{\tau - \tau_1} +
    \log_2 \binom{n_2-k_2}{\tau - \tau_1}  
    \bigg\}.
\end{multline}
% $n_1 \sim n^c$ where $c \in (0, 1)$, \\
% $n_2 = n - n_1 = \Theta(n)$, \\
% $k \sim n^{\theta}$ where $\theta \in (0, c)$, \\
% $k_2 \sim k^d = n^{d\theta} = o(n_2)$ where $d \in (0, 1)$, \\
% $k_1 = k - k_2 = k(1- o(1)) = o(n_1)$, \\
% $\tau \in \{\alpha^{\ast} k + 1, \dotsc k \}$
Since $k_1 = o(n_1)$ and $k_2 = o(n_2)$, the asymptotic equivalences 
$\log_2 \binom{n_1}{k_1} = 
    \big( k_1 \log_2 \big(\frac{n_1}{k_1}\big) \big) (1+o(1))$
 and $\log_2 \binom{n_2}{k_2} = 
    \big( k_2 \log_2 \big(\frac{n_2}{k_2}\big)  \big) (1+o(1))$ yield
% \[
%     \log_2 \binom{n_1-k_1}{\tau_1} = 
%     \log_2 \binom{n_1}{\tau_1} 
%     \left( \tau_1 \log_2 \left(\frac{n_1}{\tau_1}\right) \right) \left(1+o(1)\right)
% \]    
% and
% \[
%     \log_2 \binom{n_2-k_2}{\tau - \tau_1} = 
%     \left( (\tau - \tau_1) \log_2 \left(\frac{n_2}{\tau - \tau_1}\right)  \right) \left(1+o(1)\right)
% \]
\begin{equation}
\label{eq:imb_spars_log_2_S_asym_bound}
     \log_2  \left|\Sc\right| = 
    \left( 
    k_1 \log_2 \left(\frac{n_1}{k_1}\right) + 
    k_2 \log_2 \left(\frac{n_2}{k_2}\right) 
    \right) 
    \left(1+o(1)\right).
    % =
    % \left( 
    % k_1 \log_2 \left(\frac{n_1}{k_1}\right)
    % \right) 
    % \left(1+o(1)\right),
\end{equation}
We can apply a similar analysis to derive the asymptotic value of the right-hand side of \eqref{eq:imb_spars_relaxed_log_2_bound}:
\begin{equation}
\label{eq:imb_spars_relaxed_log_2_asymp_bound}
    \max_{\tau_1}
    % \max_{\tau_1 \in \{\tau-k_2, \dotsc, \min(k_1, \tau) \}} 
    \left\{ 
        \tau_1 \log_2 \left(\frac{n_1}{\tau_1}\right)+ 
        \left(\tau - \tau_1\right) \log_2 \left(\frac{n_2}{\tau - \tau_1} \right)
    \right\}  \left(1+o(1)\right).
    % =
    % \left( \tau \log_2\left(\frac{n_2}{\tau} \right) \right) \left(1+o(1)\right),
\end{equation}
% where the range of $\tau_1$ is as in \eqref{eq:imb_combi_term}.
% where the range of $\tau_1$ follows from $\tau = \Theta(k)$ and $k_2 = o(k)$.
Since the first derivative of the objective function 
in \eqref{eq:imb_spars_relaxed_log_2_asymp_bound} (with respect to $\tau_1$) is
\begin{multline}
    % \diff{}{\tau_1} 
    % \left( 
    % \tau_1 \log_2 \left(\frac{n_1}{\tau_1}\right)+ 
    %     \left(\tau - \tau_1\right) \log_2 \left(\frac{n_2}{\tau - \tau_1} \right)
    % \right) = 
    \log_2\left(\frac{n_1}{\tau_1}\right) - \log_2\left(\frac{n_2}{\tau - \tau_1}\right) =
    \log_2\left(\frac{\tau - \tau_1 }{\tau_1}\right) - \log_2\left(\frac{n_2}{n_1}\right) \\ =
    \log_2\left(\frac{\tau}{\tau_1} - 1\right) - \log_2\left(\frac{n}{n_1} - 1\right),   
\end{multline}
and the second derivative is 
\begin{equation}
    -\frac{1}{\log 2} 
    \left( \frac{1}{\tau-\tau_1} + \frac{1}{\tau_1}\right) < 0,
\end{equation}
the objective function in \eqref{eq:imb_spars_relaxed_log_2_asymp_bound}
increases on $\left(0, \frac{n_1}{n} \tau \right)$ and decreases on 
$\left(\frac{n_1}{n} \tau, \infty\right)$,
% \[
%     \tau_1^{\rm opt} \coloneqq \frac{n_1}{n} \tau = o(1),
% \]    
and so the maximizing $\tau_1$ in \eqref{eq:imb_spars_relaxed_log_2_asymp_bound} (subject to the constraints in \eqref{eq:imb_combi_term}) is 
\begin{equation}
\label{eq:maximizing_tau1}
    %\argmax_{\tau_1}
    % \max_{\tau_1 \in \{\tau-k_2, \dotsc, \min(k_1, \tau) \}} 
    %\left\{ 
    %    \tau_1 \log_2 \left(\frac{n_1}{\tau_1}\right)+ 
    %    \left(\tau - \tau_1\right) \log_2 \left(\frac{n_2}{\tau - \tau_1} \right)
    %\right\}
    % \tau_1^{\rm opt} 
    \tau_1^{\rm opt} = \max\left\{ \tau-k_2, \frac{n_1}{n} \tau \right\}.
\end{equation}
We are now in a position to prove our main results for imbalanced structure.

Since we have $k = \Theta(n^{\theta})$ and $n_1 = \Theta( n^c )$, as well as $k_1 = \gamma k$ and $k_2 = (1-\gamma) k$ with $\gamma \in (0,1)$, it follows that  \eqref{eq:imb_spars_log_2_S_asym_bound} can be simplified as
\begin{align}
    &\log_2  \left|\Sc\right| \nonumber \\ &= 
     \left( 
    \gamma k \log_2 \frac{n^c}{n^{\theta}} + 
    (1-\gamma)k \log_2 \frac{n}{n^{\theta}}
    \right) 
    \left(1+o(1)\right)  \\
    &= \left( 
    (c -\theta) \gamma k \log_2 n  + 
    (1-\gamma)(1-\theta) k \log_2 n
    \right) 
    \left(1+o(1)\right) 
    \\ &=
    % \left( \left(1 -\gamma\right)\left(1-c\right) + c -\theta \right) 
    % \left(1 -\gamma\left(1-c\right) -\theta \right) 
    \left( \left(1 - \theta\right) - \gamma\left(1-c\right)  \right)\left( k \log_2 n \right)
    \left(1+o(1)\right) 
        % \left(1 - \theta\right) - \gamma\left(1-c\right)  - (1-\alpha)\left(c-\theta\right)
    \label{eq:imb_spars_log_2_S_asym_bound_with_k2}.
\end{align}
% \begin{equation}
% \label{eq:imb_spars_log_2_S_asym_bound_with_k2}
%   \log_2  \left|\Sc\right| = 
%     \left( 
%     c \gamma k \log_2 \left(\frac{n}{\gamma k}\right) + 
%     (1-\gamma)k \log_2 \left(\frac{n}{(1-\gamma)k}\right) 
%     \right) 
%     \left(1+o(1)\right) =
%     \left( 
%     (c \gamma + 1-\gamma)k  \log_2 \left(\frac{n}{ k}\right)
%     \right) 
%     \left(1+o(1)\right).
% \end{equation}
Writing $\tau = \alpha k$ for $\alpha \in (0, 1]$, by \eqref{eq:maximizing_tau1}, 
the maximizing~$\tau_1$ of the objective function \eqref{eq:imb_spars_relaxed_log_2_asymp_bound} is
% objective function in \eqref{eq:imb_spars_relaxed_log_2_asymp_bound} achieves its maximum at 
\begin{equation}
    % \tau_1^{\rm opt} = 
    %\argmax_{\tau_1}
    % \max_{\tau_1 \in \{\tau-k_2, \dotsc, \min(k_1, \tau) \}} 
    % \left\{ 
    %     \tau_1 \log_2 \left(\frac{n_1}{\tau_1}\right)+ 
    %     \left(\tau - \tau_1\right) \log_2 \left(\frac{n_2}{\tau - \tau_1} \right)
    % \right\} =
    \tau_1^{\rm opt} =
    \begin{cases}
        \tau - k_2 = (\alpha + \gamma -1)k & \text{if } \alpha > 1 - \gamma;\\
        \frac{n_1}{n} \alpha k & \text{otherwise.}
    \end{cases}
    % \max\left\{ (\alpha + \gamma -1)k , \frac{n_1}{n} \alpha k \right\}.
\end{equation}   
% It follows that the maximum value of the objective function \eqref{eq:imb_spars_relaxed_log_2_asymp_bound}
% decreases after
% \[
%     \tau_1 \ge \frac{n_1}{n} \tau = o(1),
% \]    
% and so it achieves its maximum at 
% $\tau_1 = \max\left\{ \tau-k_2, \frac{n_1}{n} \tau \right\} = \tau-k_2$, 
If $\alpha > 1 - \gamma$, this gives the following asymptotic value of the objective in \eqref{eq:imb_spars_relaxed_log_2_asymp_bound}:
\begin{align}
    &\left( 
        (\alpha + \gamma -1)k \log_2 \frac{n^c}{(\alpha + \gamma -1)k} + 
        \left(1-  \gamma \right) k \log_2 \frac{n}{\left(1-  \gamma \right) k}
    \right) \nonumber \\
    &\quad \sim \left( 
        (c-\theta)(\alpha + \gamma -1) k \log_2 n + 
        (1-\theta)\left(1-  \gamma \right) k \log_2 n
    \right)  \\
    &\quad \sim \left(
        \left(1 - \theta\right) - \gamma\left(1-c\right)  - (1-\alpha)\left(c-\theta\right)
        % \left(1 - \gamma\right)\left(1-c\right) + \alpha \left(c-\theta\right)  
        \right)
         \left( k \log_2 n  \right) 
        \label{eq:log_2Ntau_big_alpha},
\end{align}
where here $\sim$ denotes equality up to multiplication by $1+o(1)$.  On the other hand, if $\alpha \le 1 - \gamma$, then the asymptotic value of the objective is as follows (using $\frac{n_1}{n} \to 0$):
\begin{multline}
\label{eq:log_2Ntau_small_alpha}
     \left( 
        \frac{n_1}{n} \alpha k \log_2 \frac{n}{\alpha k}+ 
        \alpha k \log_2 \frac{n}{\alpha k}
    \right)  \left(1+o(1)\right) \\ = 
    \alpha (1- \theta) \left( k \log_2 n \right) \left(1+o(1)\right).
\end{multline}
Taking the minimum of \eqref{eq:imb_spars_log_2_S_asym_bound_with_k2} and
\eqref{eq:log_2Ntau_big_alpha}, as well as the minimum of 
\eqref{eq:imb_spars_log_2_S_asym_bound_with_k2} and  \eqref{eq:log_2Ntau_small_alpha}, we have 
% for $\alpha > 1 - \gamma$ that
\begin{multline}
\label{eq:logNtau_imbal_alpha_big}
    \text{if $\alpha > 1 - \gamma$, then } 
    \log_2 N_{\alpha k} \le 
    \big(
    \left(1 - \theta\right) - \gamma\left(1-c\right) \\  - (1-\alpha)\left(c-\theta\right)
        % \left(1 - \gamma\right)\left(1-c\right) + \alpha \left(c-\theta\right) 
        \big)
         \left( k \log_2 n  \right)
        \left(1+o(1)\right) 
\end{multline}
and
\begin{equation}
\label{eq:logNtau_imbal_alpha_small}
    \text{if $\alpha \le 1 - \gamma$, then } 
    \log_2 N_{\alpha k} \le 
    \alpha (1- \theta) \left( k \log_2 n \right) \left(1+o(1)\right).
\end{equation}
Continuing with the case $\alpha^* \le 1-\gamma$, we deduce the following condition on the number of tests:
\begin{multline}
    t \ge
    \max
    \Bigg\{
        \max_{\alpha \in [\alpha^*, 1- \gamma]}
    \frac{\alpha (1- \theta)}
    {e^{-(1-\alpha)\nu} H_2(e^{-\alpha \nu})}, \\ \,
    \max_{\alpha \in (1- \gamma, 1]}
    \frac{ \left(1 - \theta\right) - \gamma\left(1-c\right)  - (1-\alpha)\left(c-\theta\right) }
    {e^{-(1-\alpha)\nu} H_2(e^{-\alpha \nu})}  
    \Bigg\}
    \left(k \log_2 n \right) \\ \times \left(1 +o(1) \right). \label{eq:two_terms}
\end{multline}
% When $\alpha^* > 1-\gamma$, the first term is absent, and the maximization in the second term becomes over $(\alpha^*,1]$; subsequently, we omit the details for $\alpha^* > 1-\gamma$, which is the simpler of the two cases.
Since the first term in \eqref{eq:two_terms} is increasing in $\alpha$ (see \cite[p.~48]{scarlett2016phase}) and the limits of both terms at $\alpha = 1 - \gamma$ coincide, we have the following condition on the number of tests $t$ given in \eqref{eq:imbal_achiev_linear_k_2} of
Corollary~\ref{cor: imbal_achiev_linear_k2}:
\begin{multline}
     t \ge
        \max_{\alpha \in [1- \gamma, 1]}
        \frac{ \left(1 - \theta\right) - \gamma\left(1-c\right)  - (1-\alpha)\left(c-\theta\right) }
        {e^{-(1-\alpha)\nu} H_2(e^{-\alpha \nu})} \\ \times
        \left( k \log_2 n \right) 
        \left(1 +o(1) \right). \label{eq:t_imb_final}
        % \\
        % &= e^{\nu} \max_{\alpha \in (1- \gamma, 1]}
        % \frac{\left(1 - \gamma\right)\left(1-c\right) + \alpha \left(c-\theta\right) }
        % {e^{-(1-\alpha)\nu} H_2(e^{-\alpha \nu})}  
        % \left( k \log_2 n \right) 
        % \left(1 +o(1) \right)
        % &=  \left(\frac{1 + c\gamma   -  \gamma}
        %     {H_2(e^{- \nu})}\right)
        %     \left( k \log_2 \left(\frac{n}{k}\right) 
        %     \right) \left(1 +o(1) \right). 
\end{multline}
The case $\alpha^* > 1-\gamma$ is similar but simpler, so we omit its remaining details.

We now derive the condition given in Corollary~\ref{cor: imbal_achiev_linear_k2_convexity}, which assumes $\alpha^* \le 1-\gamma$.  We rewrite the coefficient to $\left( k \log_2 n \right) \left(1 +o(1) \right)$ in \eqref{eq:t_imb_final} 
as follows:
\begin{multline}
        \frac{ \left(1 - \theta\right) - \gamma\left(1-c\right)  - (1-\alpha)\left(c-\theta\right) }
        {e^{-(1-\alpha)\nu} H_2(e^{-\alpha \nu})} \\  =
        e^{\nu} 
        \left(
        \left(1 - \gamma\right)\left(1-c\right) \frac{e^{-\alpha \nu}}{H_2(e^{-\alpha \nu})} +
        (c- \theta) \frac{\alpha e^{-\alpha \nu}}{H_2(e^{-\alpha \nu})}
        \right),
\end{multline}
% Derivative tests show that for $\alpha \in [0, 1]$, we have
which gives an optimizing $\alpha$ value of
\begin{equation}
    \label{eq:argmaxalpha}
    %\argmax_{\alpha \in [1- \gamma, 1]} 
    %\frac{ \left(1 - \theta\right) - \gamma\left(1-c\right)  - (1-\alpha)\left(c-\theta\right) }
    %    {e^{-(1-\alpha)\nu} H_2(e^{-\alpha \nu})} =
    \argmax_{\alpha \in [1- \gamma, 1]} 
        \left(
        \frac{\left(1 - \gamma\right)\left(1-c\right)}{c- \theta}        \frac{e^{-\alpha \nu}}{H_2(e^{-\alpha \nu})} +
        \frac{\alpha e^{-\alpha \nu}}{H_2(e^{-\alpha \nu})}
        \right).     
\end{equation}
By a change of variable $\lambda = e^{-\alpha \nu}$, this is equivalent to determining
\begin{equation}
\label{eq:argmaxlambda}
    \argmax_{\lambda \in [e^{-\nu}, e^{-(1-\gamma)\nu}]} 
        \left(
        \frac{\nu \left(1 - \gamma\right)\left(1-c\right)}{c- \theta}        
        \frac{\lambda}{H_2(\lambda)} -
        \frac{\lambda \log \lambda}{H_2(\lambda)}
        \right).
    %     =
    % \argmax_{\lambda \in [e^{-\nu}, e^{(1-\gamma)\nu}]} 
    %     \left(
    %     \frac{\nu \left(1 - \gamma\right)\left(1-c\right)}{c- \theta}        
    %     \frac{\lambda}{H_2(\lambda)} - \frac{\lambda \log \lambda}{H_2(\lambda)}
    %     \right).       
\end{equation}
An evaluation of the second derivative reveals that the objective function \eqref{eq:argmaxlambda}
is a convex function of $\lambda$ if
\begin{equation}
    \frac{\nu \left(1 - \gamma\right)\left(1-c\right)}{c- \theta}  > 0.029.
\end{equation}
It follows that for parameters in this regime, the maximizing $\lambda$ (or equivalently, the maximizing~$\alpha)$ is at one of the two end points, yielding Corollary~\ref{cor: imbal_achiev_linear_k2_convexity}.

We now derive the converse bound for the 
case $\alpha^* > 1 - \gamma$ in Corollary~\ref{cor:converse_imbalanced_linear_k2}.  Again applying similar steps to those in Appendix~\ref{app:block_structure}, and substituting the expressions in \eqref{eq:imb_spars_log_2_S_asym_bound_with_k2} and \eqref{eq:logNtau_imbal_alpha_big}, we obtain that $\log_2 \Ntil_{\dmax}$ is upper bounded by the following up to multiplication by $1+o(1)$:
\begin{align}
    &
    \left( \max_{0 \le \alpha \le \alpha^* }
      \left(1 - \theta\right) - \gamma\left(1-c\right)  
            - (1-\alpha)\left(c-\theta\right)
        % \left(1 - \gamma\right)\left(1-c\right) + \alpha \left(c-\theta\right) 
       \right)  \left( k \log_2 n  \right)  \\
     &\sim
     \left(  \left(1 - \theta\right) - \gamma\left(1-c\right)  
            - (1-\alpha^*)\left(c-\theta\right) \right)  \left( k \log_2 n  \right)  \\
    &\sim 
       \log_2 |\Sc| - 
       (1-\alpha^*)\left(c-\theta\right) \left( k \log_2 n  \right)
       \label{eq:bound_Ntilde_less_imbal}.
\end{align}
Combining \eqref{eq:bound_Ntilde_less_imbal} and \eqref{eq:converse_bound_T_Ntau} 
gives 
\begin{align}
     t &\ge \big( \log_2 |\Sc|  - \log_2 \Ntil_{\dmax} \big)(1+o(1)) \nonumber \\
       &\ge (1-\alpha^*)\left(c-\theta\right) \left( k \log_2 n  \right) (1+o(1)).
    %   =  (1-\alpha^*)\frac{c-\theta}{1-\theta} \left( k \log_2 \frac{n}{k}  \right) (1+o(1)) 
\end{align}
The converse bound for the case $\alpha^* \le 1 - \gamma$ in Corollary~\ref{cor:converse_imbalanced_linear_k2} follows directly from Corollary \ref{cor:converse_card}.

\section{Proof of Theorem \ref{thm:qp_map}} \label{sec:map_proof}

For the noiseless case, we express the posterior probability $P(\uv|\mathsf{X},\Yv)$ as
\begin{align}
    &P(\uv|\mathsf{X},\Yv) \nonumber \\
    &= \frac{P(\uv,\Yv|\mathsf{X})}{P(\Yv|\mathsf{X})} \\
    &\propto P(\uv) P(\Yv|\mathsf{X},\uv) \\ 
    &= P(\uv) \bone\{ \uv \in \Cless \} \\ 
    % &= P_{U}(\uv)\prod_{i=1}^{t} P(Y_{i}|\mathbf{X}_{i}) \\ 
    % &\propto \exp{\Big(\lambda \sum_{(j,j') \in E} (2u_{j}-1)(2u_{j'}-1) - \phi\sum_{j \in V}(2u_{j}-1)\Big)} \times 1 \\
    &\propto \exp{\Big(\lambda \sum_{(j,j') \in E} (2u_{j}-1)(2u_{j'}-1) - \phi\sum_{j \in V}(2u_{j}-1)\Big)} \nonumber \\
        & \hspace*{4.5cm} \times \bone\{ \uv \in \Cless \}
    % \label{posterior}
\end{align}
where we substituted in the Ising model from \eqref{eq:Ising01}.  Since maximizing the probability is equivalent to minimizing the negative log-probability, it follows that
\begin{multline}
    \hat{\uv} 
    %= &\argmax P(\uv|\mathsf{X},\Yv) \\
    %= &\argmax_{\uv} \exp{\Big(\lambda \sum_{(j,j') \in E} (2u_{j}-1)(2u_{j'}-1) - \phi\sum_{j \in V}(2u_{j}-1)\Big)} \\
    = \argmin_{\uv \in \Cless} -  \Big(\lambda \sum_{(j,j') \in E} (2u_{j}-1)(2u_{j'}-1) \\ - \phi\sum_{j \in V}(2u_{j}-1)\Big), 
    % \label{qp_obj}
\end{multline}
as desired.

% Considering the linear constraints \cancel{and the relaxation}, we have the objective function in the noiseless setting as
% \begin{equation}
%     \begin{aligned}
%         &\min_{\uv,\boldsymbol{\xi}} - \Big(\exp{\Big(\lambda \sum_{(j,j') \in E} (2u_{j}-1)(2u_{j'}-1) - \phi\sum_{j \in V}(2u_{j}-1)\Big)} \Big) \\
%         &\text{subject to } \uv \in \Cless \cancel{\text{ (or $\Clessr$ if relaxed)}} \\
%     \end{aligned}
%     \label{qp_noiseless}
% \end{equation}
% Observe that when $\rho=0$, performing \emph{maximum a posteriori} on $P(\uv|\mathsf{X},\Yv)$ is just equivalent to maximizing the prior probability $P_{U}(\uv)$.

In the noisy setting, we let $\xi_i$ be a binary random variable equaling $1$ if and only if the $i$-th test is flipped, and write
\begin{align}
    &P(\uv|\mathsf{X},\Yv) \nonumber \\
    &\propto P(\uv) P(\Yv|\mathsf{X},\uv) \\
    &= P(\uv)\prod_{i=1}^{t} P(Y_{i}|\mathbf{X}_{i},\uv) \\ 
    &\propto \exp\Big(\lambda \sum_{(j,j') \in E} (2u_{j}-1)(2u_{j'}-1) \nonumber \\
            &\hspace*{1cm} - \phi\sum_{j \in V}(2u_{j}-1)\Big) \rho^{\sum_{i=1}^{t}\xi_{i}} (1-\rho)^{t-\sum_{i=1}^{t}\xi_{i}} \\
    &= \exp\Big(\lambda \sum_{(j,j') \in E} (2u_{j}-1)(2u_{j'}-1) \nonumber \\
            &\hspace{1cm}- \phi\sum_{j \in V}(2u_{j}-1)\Big) \Big(\frac{\rho}{1-\rho}\Big)^{\sum_{i=1}^{t}\xi_{i}} .
    % \label{posterior2}
\end{align}
Taking the negative logarithm gives the objective function
\begin{equation}
    - \Big(\lambda \sum_{(j,j') \in E} (2u_{j}-1)(2u_{j'}-1) - \phi\sum_{j \in V}(2u_{j}-1) \Big) - \eta\sum_{i=1}^{t} \xi_{i},
\end{equation}
where $\eta = \log\frac{\rho}{1-\rho}$.  Finally, by the above definition of $\xi_i$ and the definition of $\Cnoisy$ in \eqref{eq:C_noisy}, minimizing this expression with respect to $\uv$ (with $\boldsymbol{\xi}$ implicitly depending on $(\uv,\mathsf{X},\Yv)$) is equivalent to minimizing jointly over  $(\uv,\boldsymbol{\xi}) \in \Cnoisy$.  This completes the proof.
% \olive{With the maximization of posterior probability and considering the constraints we have the objective function in noisy setting as}
% \begin{equation}
%     \begin{aligned}
%         &\min_{\uv,\boldsymbol{\xi}} - \Big(\lambda \sum_{(j,j') \in E} (2u_{j}-1)(2u_{j'}-1) - \phi\sum_{j \in V}(2u_{j}-1) \Big) - \eta\sum_{i=1}^{t} \xi_{i} \\
%         &\text{subject to } (\uv,\boldsymbol{\xi}) \in \Cnoisy  \cancel{\text{ (or $\Cnoisyr$ if relaxed)}} \\
%     \end{aligned}
%     \label{qp}
% \end{equation}
% where $\eta = \log(\frac{\rho}{1-\rho})$.

\section{Computation Times on Larger Graphs}

\begin{figure*}
\centering
\begin{minipage}[b]{0.4\linewidth}
  \centering
  \centerline{\includegraphics[width=7.2cm]{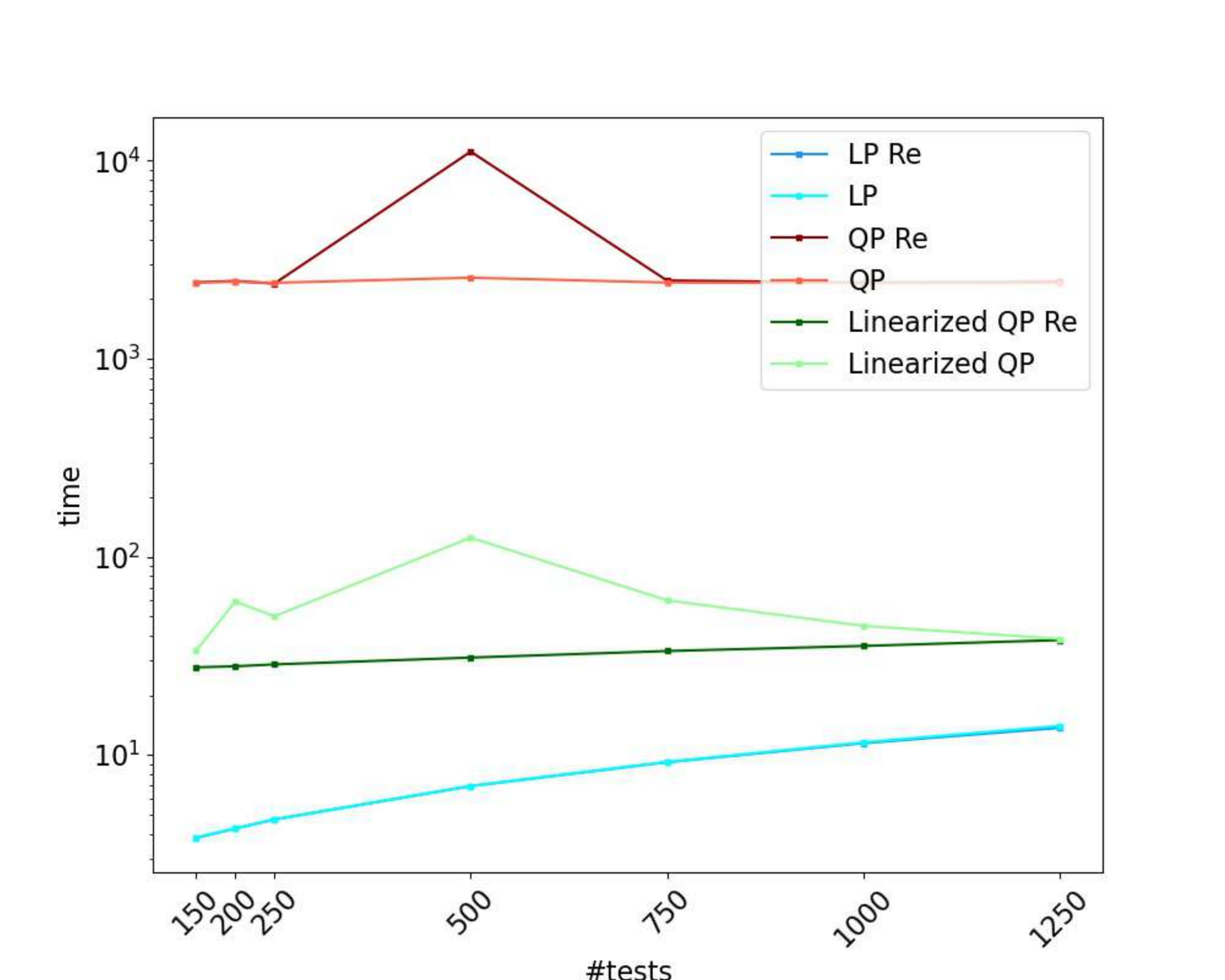}}
  \centerline{Computation time of grid graph}
  \centerline{($72 \times 72$)}\medskip
\end{minipage}
\quad
\begin{minipage}[b]{0.4\linewidth}
  \centering
  \centerline{\includegraphics[width=7.2cm]{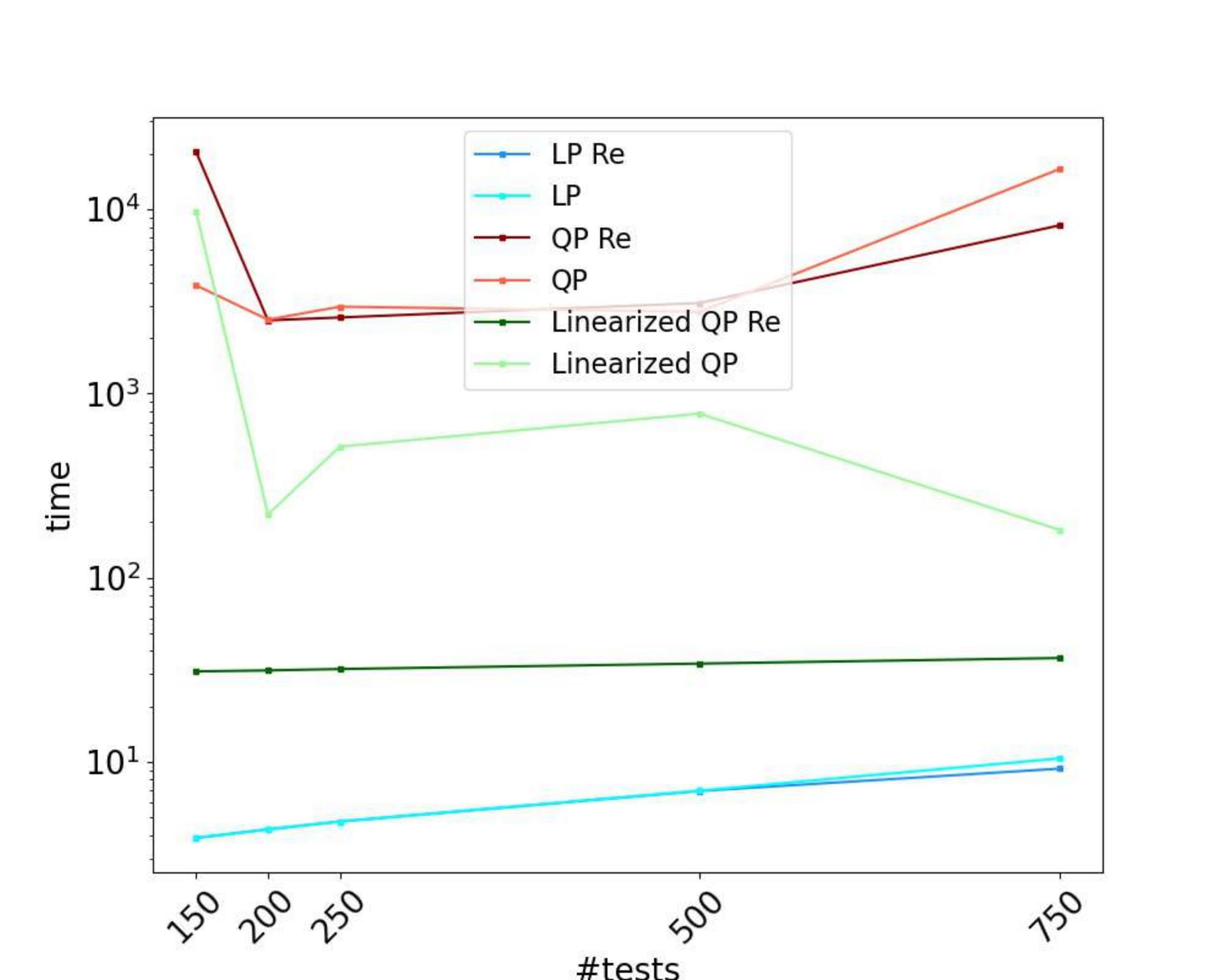}}
  \centerline{Computation time of block graph}
  \centerline{($4 \times 4$ blocks of size $18 \times 18$ each)}\medskip
\end{minipage}
\caption{Computation times for larger graphs in the noisy setting.} \label{fig:comp_large}
\end{figure*}

The computation times for lager graphs (discussed in Section \ref{sec:experiments}) are shown in Figure \ref{fig:comp_large}.

\end{document}